\documentclass[11pt,a4paper]{article}
\usepackage{jheppub}
\usepackage{amssymb,xcolor}
\usepackage[T1]{fontenc}
\usepackage[utf8]{inputenc}
\usepackage{lmodern,mathtools,amsthm,mathrsfs,booktabs,longtable,array,microtype}
\usepackage{hyperref}
\usepackage{enumitem,placeins,needspace}
\setlist{itemsep=3pt,topsep=5pt}
\newcommand{\C}{\mathbb C}
\newcommand{\R}{\mathbb R}
\newcommand{\Z}{\mathbb Z}
\newcommand{\PP}{\mathbb P}
\newcommand{\eps}{\epsilon}
\newcommand{\dd}{\mathrm d}

\newcolumntype{L}[1]{>{\raggedright\arraybackslash}p{#1}}

\numberwithin{equation}{section}
\renewcommand{\beforetochook}{\pagestyle{myplain}\pagenumbering{roman}\begingroup\small\renewcommand{\baselinestretch}{1}\selectfont}
\hypersetup{pdftitle={Instantons from geometry in M2-brane sphere partition functions},pdfauthor={Kiril Hristov and Ali Mert Yetkin}}
\title{Instantons from geometry in\\ M2-brane sphere partition functions}
\author[a,b,c]{Kiril Hristov\footnote{Visiting Member at the Institute for Advanced Study; on leave from Sofia University.}}
\author[d]{and Ali Mert Yetkin}
\affiliation[a]{School of Natural Sciences, Institute for Advanced Study,\\ 1 Einstein Drive, Princeton, NJ 08540, USA}
\affiliation[b]{Faculty of Physics, Sofia University ``St.\ Kliment Ohridski'',\\ J. Bourchier Blvd.\ 5, 1164 Sofia, Bulgaria}
\affiliation[c]{INRNE, Bulgarian Academy of Sciences,\\ Tsarigradsko Chaussee 72, 1784 Sofia, Bulgaria}
\affiliation[d]{Department of Physics and Astronomy, University of Kentucky,\\ Lexington, KY 40506-0055, USA}
\date{21 October 2026}
\preprint{}
\abstract{Can the geometry transverse to M2-branes capture effects that are exponentially small in their three-sphere ($S^3$) partition function? We investigate this question by extending beyond constant maps the proposal that M2-brane observables are encoded by equivariant topological strings on the transverse Calabi–Yau fourfold (CY$_4$). Our central object is the disk partition function of the CY$_4$ gauged linear sigma model, a quantum generalization of the equivariant volume. After a Gamma-function normalization and a Mellin transform, it reproduces the complete leading WKB grand potential in the examples admitting a suitable Fermi-gas description, including the R-charge and mass deformations compatible with that description. In particular, the transform recovers membrane-instanton series even for $\mathbb C^4$, which has no compact curves. For resolved CY$_4$ cones with compact curves, the match requires a quantum refinement of the classical mesonic constraint. We establish the exact equalities of the corresponding dressed disk partition functions with the classical spectral traces for ABJM, ADHM and a representative set of flavored M2-brane quiver theories. Away from the deformation loci where this match can be established, the same construction yields geometric predictions for the membrane-instanton sector that remain to be tested.}
\begin{document}
\maketitle
\endgroup
\normalsize
\flushbottom
\section{Introduction and summary of results}
\label{sec:intro}

The supersymmetric partition function on the round $S^3$ offers a precise setting in which to explore how M-theory encodes quantum field theory. For three-dimensional theories describing M2-branes at a Calabi--Yau fourfold singularity, the same quantity can be studied as a gauge-theory path integral or as a gravitational observable with a prescribed conformal boundary. Supersymmetric localization expresses the finite-rank answer as a matrix integral, while the gravitational description makes its large-rank behaviour and its relation to the transverse geometry more transparent \cite{Maldacena:1997re,Aharony:2008ug,Kapustin:2009kz,Drukker:2010nc,Herzog:2010hf}. We denote this partition function by $Z_{S^3}(N;\Delta)$, where $N$ is the rank and $\Delta$ collectively denotes the complexified R-charge parameters, including real-mass deformations. In the charge parametrizations used below,
$\sum_i\Delta_i=2$. We keep these parameters arbitrary within the supersymmetric constraints, since their dependence provides a more detailed comparison with geometry than the extremized answer alone. Our interest here is in how this comparison extends to effects that are exponentially small at large rank and hence invisible in any finite expansion in inverse powers of $N$.

On the field-theory side, localization enabled the direct derivation of the characteristic $N^{3/2}$ growth of the M2-brane free energy, $-\log Z_{S^3}$, first in ABJM and subsequently in broader classes of quiver theories \cite{Drukker:2010nc,Herzog:2010hf,Santamaria:2010dm,Martelli:2011qj}. Extremizing the sphere free energy over trial R-charges determines the superconformal R-symmetry through $F$-maximization \cite{Jafferis:2010un,Jafferis:2011zi,Closset:2012vg}. This dependence also contains physical information beyond the extremum: derivatives of the sphere free energy compute integrated correlation functions and constrain interactions in the holographic effective action \cite{Binder:2018yvd,Chester:2021gdw}. Squashing the sphere, together with suitable background fields, provides a further supersymmetric deformation with a known holographic realization, although we restrict attention here to the round sphere \cite{Hama:2011ea,Martelli:2011fu}. Beyond the leading large-$N$ saddle, the Fermi-gas formulation recasts certain localization integrals as spectral problems, providing a natural framework for the Airy-function resummation and the systematic study of exponentially suppressed contributions \cite{Fuji:2011km,Marino:2011eh,Hatsuda:2012dt,Calvo:2012du,Putrov:2012zi,Hatsuda:2012hm,Hatsuda:2014vsa,Nosaka:2015iiw,Hatsuda:2015lpa,Moriyama:2014gxa,Bobev:2025ltz} (see \cite{Hosseini:2026lvj} for a related observable). We use the deformed round-sphere reductions and conventions assembled in \cite{Nosaka:2015iiw,Hatsuda:2015lpa,Moriyama:2014gxa,Hristov:2026zjh,Kubo:2024qhq,Kubo:2025dot}.

On the gravitational side, the holographic description identifies the sphere partition function with a bulk path integral with prescribed supersymmetric boundary conditions \cite{Gubser:1998bc,Witten:1998qj}. The starting point is the near-horizon geometry AdS$_4\times Y_7$, where $Y_7$ is the link of the transverse Calabi--Yau cone \cite{Martelli:2008rt,Jafferis:2008qz}. For the undeformed round-sphere vacuum, Euclidean AdS$_4$ fills the boundary $S^3$, and its renormalized on-shell action gives the leading free energy. R-charge and mass deformations are encoded in boundary conditions for the bulk fields, as illustrated by the supergravity realization of $F$-maximization in ABJM \cite{Freedman:2013oja}. Equivariant localization allows supersymmetric on-shell actions to be evaluated from fixed-point data, often without constructing the full solution; see \cite{BenettiGenolini:2023kxp} and subsequent developments. Higher-derivative interactions and quantum fluctuations account for further terms in the large-$N$ expansion, including logarithmic corrections \cite{Bobev:2020egg,Bobev:2021oku,Bhattacharyya:2012ye,Liu:2017vbl,Hristov:2021zai,Bobev:2023dwx}. The relative equivariant localization proposal of
\cite{BenettiGenolini:2026cyc} recovers the Airy-function
structure predicted by equivariant topological strings
\cite{Cassia:2025aus,Cassia:2025jkr}.
The agreement of the coefficients controlling this structure
is explained by their common origin in the twelve-dimensional
M-theory anomaly polynomial \cite{Hristov:2026uxs}.
The remaining $N$-independent normalization is addressed
by a subsequent proposal for the bulk one-loop determinant
in terms of a tangent--cotangent character
\cite{BenettiGenolini:2026kjy}.
Beyond the perturbative answer, wrapped-string and M2-brane
calculations reproduce worldsheet-instanton contributions,
including fluctuation prefactors and, more recently,
higher-loop corrections
\cite{Gautason:2023igo,Beccaria:2023ujc,Gautason:2025per,
Kurlyand:2026yke,Tseytlin:2026dhy,Kurlyand:2026gxi}.
In ABJM, the relevant M2-brane wraps $S^3/\mathbb Z_k$
and lifts the fundamental-string instanton wrapping
$\mathbb{CP}^1\subset\mathbb{CP}^3$
\cite{Cagnazzo:2009zh,Park:2020hgt}.
These calculations concern the worldsheet sector in the
Fermi-gas terminology, rather than the membrane sector
studied here; we explain the distinction in
Section~\ref{sec:intro-qc}.

The geometric approach pursued here starts directly from the transverse cone $C(Y_7)$, or a smooth resolution $X$, and constructs equivariant integrals from its torus action and characteristic classes.\footnote{We distinguish these geometric constructions from calculations in the gravitational effective theory. Their relation is established at the classical level in the settings relevant here through equivariant localization and anomaly-based constructions \cite{BenettiGenolini:2023kxp,Martelli:2023oqk,Hristov:2026uxs}. We do not assume that this classical agreement extends to a general equivalence of their quantum completions.} The equivariant volume regulates the noncompact geometry while retaining its dependence on torus weights, which are identified with trial R-charges. It thus extends the familiar relation between the leading sphere free energy and the Sasaki--Einstein volume to a function of the deformation parameters \cite{Martelli:2005tp,Martelli:2006yb,Martelli:2011qj}. Geometric constructions based on equivariant volumes and characteristic classes provide a framework for organizing holographic observables \cite{Hosseini:2019ddy,Martelli:2023oqk,Colombo:2023fhu}. Building on this framework, the proposal of \cite{Cassia:2025aus,Cassia:2025jkr} identifies equivariant topological strings on the transverse fourfold as the geometric description of M2-brane partition functions. Its starting point is the constant-map sector, in which the
worldsheet maps to a point of $X$. This sector organizes the
perturbative large-$N$ contributions. We ask how to extend
this description to exponentially small terms, which are
invisible in the perturbative large-$N$ expansion.

The simplest transverse geometry already exposes the central puzzle we address here. The fourfold $\C^4$ has no compact curves, yet the associated Fermi-gas grand potential contains membrane-instanton contributions even at leading semiclassical order \cite{Marino:2011eh,Calvo:2012du,Nosaka:2015iiw,Hatsuda:2015lpa,Moriyama:2014gxa}. How can the transverse geometry encode these exponentially small terms when there is no sum over positive-degree maps to compact curves? The puzzle concerns their geometric encoding on the transverse fourfold, rather than the existence of instanton configurations in the gravitational background.

Our main observation is that this information can be recovered from the equivariant disk partition function of the gauged linear sigma model (GLSM) for $X$, through a suitable Gamma normalization and Mellin transform. The disk partition function is an observable of an auxiliary two-dimensional theory whose target is the transverse fourfold \cite{Honda:2013uca,Hori:2013ika,Sugishita:2013jca,Benini:2012ui,Doroud:2012xw,Jockers:2012dk,Cassia:2022lfj}. For the Fermi-gas descriptions considered here, the transform reproduces the complete leading semiclassical, or WKB, grand potential, including its membrane-instanton series. The same geometric construction remains available beyond the parameter loci where we establish the Fermi-gas match, providing predictions for the membrane-instanton sector that remain to be tested against field theory.

\subsection{Field-theory results on quantum-curve loci}
\label{sec:intro-qc}
\label{sec:intro-match}
\begin{figure}[t]
 \centering
 \includegraphics[width=0.9\textwidth]{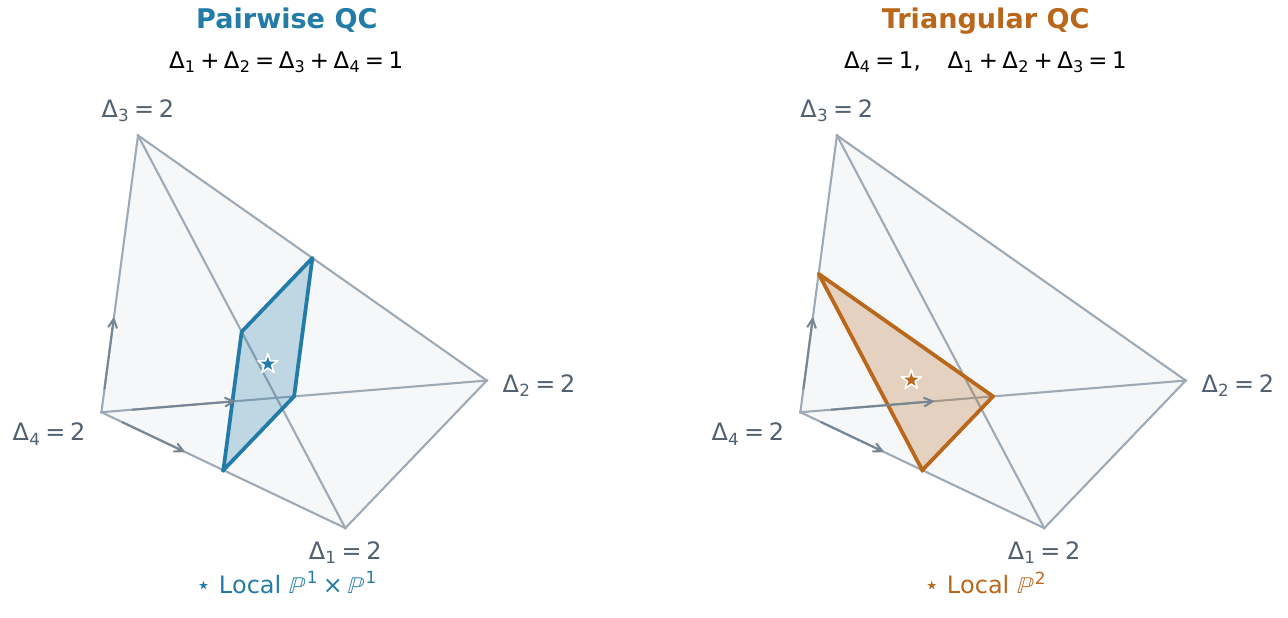}
 \caption{ABJM QC loci in the charge simplex $\Delta_i\geq0$, $\sum_i\Delta_i=2$. Stars mark the indicated local-CY$_3$ specializations. The pairwise locus applies at arbitrary level $k$, and the triangular locus at level one.}
 \label{fig:parameter-space}
\end{figure}

To state the match precisely, we first introduce the field-theory quantities that enter it. We pass to the grand-canonical ensemble by summing over the rank $N$, with chemical potential $\mu$, while retaining the dependence on the complexified R-charges $\Delta$. The grand partition function and grand potential are
\begin{equation}
\Xi(\mu;\Delta):=\sum_{N\geq0}e^{N\mu}Z_{S^3}(N;\Delta),
 \qquad
 J(\mu;\Delta)=\log\Xi(\mu;\Delta),
 \qquad
 Z_{S^3}(0;\Delta)=1.
 \label{eq:intro-Xi}
\end{equation}
For certain choices of $\Delta$, determinant identities recast the localized matrix integral exactly as the partition function of an ideal Fermi gas. In this description,
\begin{equation}
 \Xi(\mu;\Delta)
 =\det\!\left(1+e^\mu\widehat\rho(\Delta)\right),
 \qquad
 \widehat O(\Delta):=\widehat\rho(\Delta)^{-1},
 \qquad
 [\widehat x,\widehat p]=2\pi i\hbar,
 \label{eq:intro-QC}
\end{equation}
where $\widehat\rho(\Delta)$ is the one-particle density operator and $\hbar$ controls its noncommutativity \cite{Marino:2011eh,Nosaka:2015iiw,Hatsuda:2015lpa,Moriyama:2014gxa}. Its inverse $\widehat O(\Delta)$ is built from functions of the exponentiated conjugate operators $e^{\widehat x}$ and $e^{\widehat p}$, with a specified ordering. We refer to it as the quantum-curve operator and call the restrictions on $\Delta$ that permit this reformulation the quantum-curve (QC) loci.~\footnote{At further special choices of the complexified R-charges, these operators become quantized mirror curves of toric Calabi--Yau threefolds. The topological-string/spectral-theory (TS/ST) correspondence then proposes a nonperturbative description of their spectral determinants \cite{Marino:2009jd,Hatsuda:2013oxa,Grassi:2014zfa,Kashaev:2015kha}. This threefold encodes the spectral problem and differs from the transverse fourfold. Identifying it requires the powers appearing in the curve, their integral lattice and the monomial carrying the spectral parameter. General charge deformations change these powers, so they need not describe K\"ahler deformations of one fixed threefold \cite{Grassi:2014zfa,Kashaev:2015kha}. Figure~\ref{fig:parameter-space} shows the two QC loci of ABJM theory and their local-threefold specializations. }

The Fermi-gas reformulation is exact; the semiclassical
approximation enters when we expand in $\hbar$ at fixed
$\mu$ and $\Delta$. For the operator presentations considered here, the formal
semiclassical expansion contains only even powers of $\hbar$
after extracting the phase-space factor $1/\hbar$.
This follows from reversal of the Moyal product and cyclicity
of the spectral trace, as explained in
Appendix~\ref{app:wigner-calculus}.
The Wentzel--Kramers--Brillouin (WKB) expansion therefore takes
the form
\begin{equation}
 J(\mu;\Delta,\hbar)\sim
 \frac{1}{\hbar}\sum_{n=0}^{\infty}
 \hbar^{2n}J_n(\mu;\Delta)
 =
 \frac{1}{\hbar}J_0(\mu;\Delta)
 +\hbar J_1(\mu;\Delta)+\cdots,
 \qquad \hbar\to0.
 \label{eq:intro-WKB}
\end{equation}
Only the leading coefficient $J_0$ is needed here. The index $n$ counts orders in the spectral semiclassical
expansion; it is not a worldsheet genus. Contributions nonperturbative in $\hbar$ are not captured at any finite WKB order.

The WKB expansion and the large-rank expansion organize different
features of the answer. Each coefficient $J_n(\mu;\Delta)$ has its
own large-$\mu$ expansion, consisting of polynomial terms and
exponentially suppressed corrections. In ABJM and a broader class
of deformed quiver Fermi gases, these terms follow the hierarchy
\cite{Marino:2011eh,Nosaka:2015iiw,Hatsuda:2015lpa,Moriyama:2014gxa}
\begin{equation}
 \begin{aligned}
 J_0(\mu;\Delta)
 &=\frac{C_0(\Delta)}{3}\mu^3
   +B_0(\Delta)\mu+A_0(\Delta)
   +J_{0,\mathrm{exp}}(\mu;\Delta),\\
 J_1(\mu;\Delta)
 &=B_1(\Delta)\mu+A_1(\Delta)
   +J_{1,\mathrm{exp}}(\mu;\Delta),\\
 J_n(\mu;\Delta)
 &=A_n(\Delta)+J_{n,\mathrm{exp}}(\mu;\Delta),
 \qquad n\geq2.
 \end{aligned}
 \label{eq:intro-WKB-large-mu}
\end{equation}
Here $J_{n,\mathrm{exp}}$ denotes exponentially small terms,
possibly multiplied by powers of $\mu$. Our comparison concerns the complete leading coefficient
$J_0$, and does not require the higher-order truncation in
\eqref{eq:intro-WKB-large-mu} to hold for every theory considered below.

Collecting the polynomial terms gives
\begin{equation}
 J_{\mathrm{pert}}(\mu;\Delta,\hbar)
 =\frac{C(\Delta,\hbar)}{3}\mu^3
  +B(\Delta,\hbar)\mu+A(\Delta,\hbar).
 \label{eq:intro-Jpert}
\end{equation}
For the hierarchy above, the coefficients satisfy
\begin{equation}
 \begin{aligned}
 C(\Delta,\hbar)&=\frac{C_0(\Delta)}{\hbar},\qquad
 B(\Delta,\hbar)=\frac{B_0(\Delta)}{\hbar}
                  +\hbar B_1(\Delta),\\
 A(\Delta,\hbar)&\sim
 \sum_{n=0}^{\infty}\hbar^{2n-1}A_n(\Delta).
 \end{aligned}
 \label{eq:intro-ABC}
\end{equation}
The last relation specifies the semiclassical expansion of $A$,
rather than its full finite-$\hbar$ value.
Here ``perturbative'' refers to the large-rank expansion:
the inverse Laplace transform of $\exp J_{\mathrm{pert}}$ gives
\begin{equation}
 Z_{\mathrm{pert}}(N;\Delta,\hbar)
 =e^{A(\Delta,\hbar)}C(\Delta,\hbar)^{-1/3}
 \operatorname{Ai}\!\left[
 C(\Delta,\hbar)^{-1/3}
 \bigl(N-B(\Delta,\hbar)\bigr)
 \right],
 \label{eq:intro-airy}
\end{equation}
which resums the perturbative large-$N$ answer
\cite{Fuji:2011km,Marino:2011eh,Bobev:2025ltz,Nosaka:2015iiw,Hatsuda:2015lpa,Moriyama:2014gxa}.
At fixed $\Delta$ and $\hbar$, the saddle obeys
$\mu_*\simeq\sqrt{(N-B)/C}$.
Consequently, terms exponentially small in $\mu$ give corrections
exponentially small in $\sqrt N$, beyond the Airy approximation.

The distinction between the instanton sectors is most transparent
in undeformed ABJM, where $\hbar=k$ in our convention
$[\widehat x,\widehat p]=2\pi i\hbar$.
The two basic families have weights
\begin{equation}
 e^{-2\ell\mu}
 \quad\text{(membrane)},\qquad
 e^{-4m\mu/k}
 \quad\text{(worldsheet)},\qquad
 \ell,m\geq1.
 \label{eq:intro-instanton-weights}
\end{equation}
Their names reflect their gravitational origin.
In the type-IIA description, the worldsheet family arises from
fundamental strings wrapping $\mathbb{CP}^1\subset\mathbb{CP}^3$,
while the membrane family is associated with Euclidean D2-branes
wrapping $\mathbb{RP}^3\subset\mathbb{CP}^3$
\cite{Drukker:2011zy,Cagnazzo:2009zh}.
Both lift to wrapped M2-branes in eleven dimensions
\cite{Park:2020hgt}.
The membrane weights are independent of $k$, and their prefactors
admit a WKB expansion; they therefore contribute to
$J_{n,\mathrm{exp}}$, already at $n=0$
\cite{Calvo:2012du,Hatsuda:2012dt}.
The worldsheet weights are instead nonperturbative in $k$
at fixed positive $\mu$, and are absent at every finite WKB order.
The full answer also contains bound states of the two families
\cite{Hatsuda:2013gj}.\footnote{
These instanton families are conventionally organized using the
modified grand potential. Throughout this work, $J$ denotes
$\log\Xi$; the relation between the two objects, which share the
same perturbative cubic polynomial, is explained in
Section~\ref{sec:QC} \cite{Hatsuda:2012dt,Grassi:2014zfa}.}
With deformations, the actions can depend on $\Delta$.
We use the ``membrane-instanton'' terminology for the corresponding
exponentially small terms captured by the spectral WKB expansion
\cite{Nosaka:2015iiw,Hatsuda:2015lpa,Moriyama:2014gxa}.

Section~\ref{sec:QC} computes $J_0$ from the classical phase-space
trace. Its Mellin--Barnes representation determines the polynomial,
constant and exponential terms from a single meromorphic function
\cite{Hatsuda:2015oaa,Nosaka:2015iiw,Hatsuda:2015lpa,Moriyama:2014gxa}.\footnote{Higher WKB terms,
the all-orders constant functions and squashing are studied in a
follow-up work \cite{CompanionII}.} We evaluate this integral for ABJM, ADHM, grouped necklace
quivers, and flavored quiver theories associated with the
conifold, suspended pinch point (SPP), and the cone over $Q^{1,1,1}$.
The construction of flavored quivers and their transverse
geometries is discussed in \cite{Benini:2009qs,Jafferis:2009th}.
For each theory, Section~\ref{sec:QCexamples} specifies the
Fermi-gas operator and the allowed source parameters, evaluates
$J_0$, and extracts the constant and instanton terms.
We follow the source conventions of \cite{Hristov:2026zjh,Kubo:2024qhq,Kubo:2025dot}.
The phase-space normalization is explained in
Section~\ref{sec:spectral-method}. Appendix~\ref{app:quantum-curves}
collects the Wigner--Weyl conventions and integral identities
used in these calculations.

For illustration, consider the pairwise ABJM locus
$\Delta_1+\Delta_2=\Delta_3+\Delta_4=1$ at level $k$,
and the triangular locus $\Delta_4=1$,
$\Delta_1+\Delta_2+\Delta_3=1$ at level one.
For generic allowed $\Delta$, their exponential terms are
\begin{equation}
 \begin{aligned}
 J_{0,\mathrm{exp},\square}(\mu;\Delta)
 &\sim\sum_{i=1}^{4}\sum_{n\geq1}
 d_{i,n}(\Delta)e^{-n\mu/\Delta_i},\\
 J_{0,\mathrm{exp},\triangle}(\mu;\Delta)
 &\sim\sum_{i=1}^{3}\sum_{n\geq1}
 \widetilde d_{i,n}(\Delta)e^{-n\mu/\Delta_i}.
 \end{aligned}
 \label{eq:intro-qc-families}
\end{equation}
The coefficients are explicit Mellin residues, given in
Section~\ref{sec:QCexamples}.
When poles coincide, the corresponding residue contributions
combine into polynomials in $\mu$ multiplying an exponential.
In particular, the undeformed pairwise choice $\Delta_i=1/2$
recovers the membrane weights $e^{-2\ell\mu}$ discussed earlier.

\subsection{Geometric match and predictions}
\label{sec:intro-geometry}
\label{sec:intro-prediction}

The geometric input is the equivariant GLSM disk partition
function $\mathbb D_X(\lambda,\eps;z)$
\cite{Cassia:2022lfj}, defined in \eqref{eq:disk}.
It is the partition function of a two-dimensional
supersymmetric theory on a disk, with target space $X$,
and provides a quantum generalization of the equivariant
volume. Localization expresses it through Gamma functions and,
when compact curves are present, sums over map degrees
\cite{Honda:2013uca,Hori:2013ika,Sugishita:2013jca,Cassia:2022lfj}.
For $\C^4$, it takes the simple form
\begin{equation}
 \mathbb D_{\C^4}(\lambda,\eps;z)
 =z^{-4}e^{\sum_{i=1}^4\lambda^i\eps_i}
 \prod_{i=1}^4\Gamma(\eps_i/z).
 \label{eq:intro-C4-disk}
\end{equation}
Here $\eps_i$ are the weights of torus rotations of $X = \C^4$, $\lambda^i$ are redundant
equivariant K\"ahler parameters associated with the homogeneous
coordinates, and $z$ is the weight of disk rotation. For the round-sphere comparison, we normalize the
geometric weights by $\sum_i\eps_i=2$.

As explained in Section~\ref{sec:QC}, the leading WKB grand potential
can be reconstructed by a Mellin--Barnes contour integral over an
auxiliary complex parameter, which we denote by $s$.
Our geometric prescription identifies this integration variable with
the inverse disk rotation weight, $z=s^{-1}$. We then define the geometric counterpart of the grand potential,
\begin{equation}
 \boxed{\displaystyle
 J_{\mathrm{disk}}(\mu;\eps)
 :=-\frac{1}{4\pi^2}
 \int_{c-i\infty}^{c+i\infty}\frac{\mathrm ds}{2\pi i}\,
 \frac{\Gamma(-s)}{s^4\Gamma(s)}\,
 \mathbb D_X\!\left(
 s(\frac{\mu}{2}\mathbf 1+\lambda_\star),
 \eps;s^{-1}\right).}
 \label{eq:intro-master}
\end{equation}
The disk is evaluated at fixed weights $\eps$, rotation
weight $s^{-1}$, and K\"ahler arguments
$s(\lambda_\star+\mu\mathbf1/2)$.
The fixed offsets $\lambda_\star$ select the quantum counterpart
of the mesonic point, where the flat K\"ahler coordinates
vanish. These coordinates incorporate the mirror-map
corrections to the bare GLSM parameters, as explained in
Section~\ref{sec:flat-mesonic}.
For $\C^4$, we take $\lambda_\star=0$, so $\lambda^i=s\mu/2$.
In all cases, the common shift produces $e^{s\mu}$
directly through the disk exponential, using
$\sum_i\eps_i=2$.

The contour lies in a common convergence strip, initially
with $0<c<1$, and the result is continued where necessary.
Section~\ref{sec:disk} explains the Gamma normalization
and the details of the prescription.
Its intrinsic geometric origin and a derivation from
a bulk path integral remain to be understood.
Earlier spectral and topological-string uses of
Mellin--Barnes integrals provide useful context
\cite{Hatsuda:2015oaa,Krefl:2015qva}.

On the round-sphere QC loci, we identify the geometric
weights with the field-theory charges through the dictionaries below,
and establish the holographic match
\begin{equation}
 \boxed{\displaystyle
 J_0(\mu;\Delta)
 =J_{\mathrm{disk}}(\mu;\eps) \Big|_{\epsilon_i = \Delta_i}.}
 \label{eq:intro-holographic-match}
\end{equation}
The left-hand side is computed from the field-theory Fermi
gas, and the right-hand side from the transverse geometry,
with the disk normalized to the coefficient $J_0$ in
$J=\hbar^{-1}J_0+\cdots$.\footnote{
For a residual $\mathbb Z_k$ quotient with $\hbar=k$, this
comparison uses the disk before the residual quotient.
Including its factor $1/k$ instead gives the leading
contribution $J_0/k$ to the grand potential.}
Section~\ref{sec:diskexamples} verifies their equality
for every example studied here, with its specified
source restrictions and normalization.
The geometric description of the perturbative Airy structure
was established in \cite{Cassia:2025aus,Cassia:2025jkr},
up to an overall factor independent of $N$.
Here the disk transform determines the complete leading WKB
coefficient, including its constant term $A_0$ and its
membrane-instanton series.
These examples support a general holographic relation,
whose proof beyond the cases studied here remains open.

The transform explains how instanton terms arise even
for $\C^4$. A surviving pole of its integrand at
$s=-\alpha$ gives a contribution $e^{-\alpha\mu}$;
higher-order poles also produce polynomial factors in $\mu$.
The Gamma functions in \eqref{eq:intro-C4-disk} supply
the required poles despite the absence of compact curves.
When compact curves are present, the disk partition
function also contains contributions from maps of positive
degree. These must be summed before evaluating it at
vanishing flat K\"ahler coordinates, using analytic
continuation where needed.
The Mellin transform acts on this complete answer.
Thus, in all cases, it is the pole structure of the
transform that produces the membrane-instanton series,
rather than a direct identification of instanton terms
with individual curve degrees.

Importantly, the definition of $J_{\mathrm{disk}}$ also applies beyond
the QC loci, wherever the prescribed disk partition function
and its transform are defined.
For $\C^4$, arbitrary positive weights satisfying
$\sum_i\eps_i=2$ give
\begin{align}
 J_{\mathrm{disk,exp}}(\mu;\eps)
 &\sim\sum_{i=1}^4\sum_{n\geq1}
 c^{\mathrm{disk}}_{i,n}(\eps)e^{-n\mu/\eps_i},
 \nonumber\\
 c^{\mathrm{disk}}_{i,n}(\eps)
 &=\frac{(-1)^{n+1}}{4\pi^2n!\eps_i}
 \frac{\Gamma(n/\eps_i)}{\Gamma(-n/\eps_i)}
 \prod_{j\ne i}\Gamma(-n\eps_j/\eps_i).
 \label{eq:intro-general-instantons}
\end{align}
This expression assumes separated poles; coincident poles
are treated through their combined residues.
On the pairwise QC locus, it reproduces the four families
in \eqref{eq:intro-qc-families}.
On the triangular locus at level one, the unit weight in \eqref{eq:intro-C4-disk}
cancels one family and leaves the other three.

Away from these loci, the same expression provides
a geometric prediction for the membrane-instanton sector.
Its field-theory interpretation remains to be tested:
we do not assume that a spectral WKB expansion exists
there. The further information needed
to obtain the full finite-coupling answer is discussed
in Section~\ref{sec:outlook}.
\section{Quantum curves and the complete leading WKB grand potential}
\label{sec:QC}

Supersymmetric localization reduces the $S^3$ partition
functions of a broad class of three-dimensional supersymmetric
gauge theories to finite-dimensional matrix integrals
\cite{Kapustin:2009kz,Jafferis:2010un,Hama:2010av,Willett:2016adv}.
For certain theories, these integrals admit an exact
reformulation as canonical partition functions of an ideal
Fermi gas, whose one-particle density operator $\widehat\rho$
determines the grand partition function through
$\Xi=\det(1+e^\mu\widehat\rho)$
\cite{Marino:2011eh}.
Exact finite-rank computations and TBA methods provided early tests of this
spectral formulation and its nonperturbative corrections
\cite{Okuyama:2011su,Hatsuda:2012hm,Putrov:2012zi}.
In the examples studied here, the inverse operator
$\widehat O=\widehat\rho^{-1}$ also has an explicit
presentation in exponentials of conjugate operators:
this is the quantum-curve presentation.
These are distinct properties.
We call the source restrictions under which both are
established the QC loci
\cite{Hristov:2026zjh,Kubo:2024qhq,Kubo:2025dot}.

The WKB expansion is a subsequent semiclassical expansion
in the noncommutativity parameter $\hbar$.
As emphasized in Section~\ref{sec:intro-qc}, it is distinct
from the large-$N$ expansion: even its leading coefficient
$J_0(\mu;\Delta)$ contains exponentially small terms at
large chemical potential.
The Mellin--Barnes representation of the grand potential
in terms of the spectral zeta function makes these terms
accessible through the poles of its integrand
\cite{Hatsuda:2015oaa,Nosaka:2015iiw,Hatsuda:2015lpa,Moriyama:2014gxa}.

We build on the spectral results for deformed M2-brane
theories in
\cite{Kubo:2024qhq,Kubo:2025dot,Hristov:2026zjh}.
The comparison of the Airy coefficients $C$ and $B$
with the geometric predictions of \cite{Cassia:2025jkr}
was established in \cite{Hristov:2026zjh}.
Our aim is to determine the complete leading WKB
coefficient, including the membrane-instanton series
discarded by a polynomial large-energy approximation.
The calculation also recovers $C_0$ and $B_0$ and
determines the constant term $A_0$.
We first explain the method and then evaluate the individual
examples. Section~\ref{sec:disk} compares these results
with the disk partition functions of the corresponding
transverse fourfolds.

\subsection{From the Fermi gas to the leading WKB grand potential}
\label{sec:spectral-method}

On the source loci considered below, the localized sphere
partition function admits an ideal Fermi-gas description
\cite{Marino:2011eh,Nosaka:2015iiw,Hatsuda:2015lpa,Moriyama:2014gxa,Hristov:2026zjh,Kubo:2024qhq,Kubo:2025dot}.
The one-particle density operator
$\widehat\rho(\Delta,\hbar)$ determines the grand partition
function through
\begin{equation}
 \Xi(\mu;\Delta,\hbar)
 =\det\!\left(1+e^\mu\widehat\rho(\Delta,\hbar)\right).
 \label{eq:fermi-Fredholm}
\end{equation}
Consequently, the grand potential is
\begin{equation}
 \begin{aligned}
 J(\mu;\Delta,\hbar)
 =\operatorname{Tr}\log\!\left(
 1+e^\mu\widehat\rho(\Delta,\hbar)\right)=\sum_a\log\!\left(1+e^{\mu-E_a(\Delta,\hbar)}\right),
 \end{aligned}
 \label{eq:trace-log}
\end{equation}
where $e^{-E_a}$ are the eigenvalues of $\widehat\rho$.
We initially work with a positive, trace-class realization
of the density operator. The ordered quantum-curve
presentations used below are spectrally equivalent to this
realization, but need not themselves be self-adjoint
\cite{Marino:2011eh,Nosaka:2015iiw}.
We subsequently continue in the deformation parameters
where appropriate.
Our aim is to compute the complete leading WKB coefficient
$J_0(\mu;\Delta)$, retaining its exponentially small terms
as well as its polynomial large-$\mu$ behaviour.

In ABJM and the spectral completions where it is available,
the modified grand potential, denoted $\mathcal J$, is conventionally
specified by $\Xi(\mu)=\sum_{m\in\mathbb Z}\exp\mathcal J(\mu+2\pi i m)$
\cite{Hatsuda:2012dt,Grassi:2014zfa}. It organizes the standard
finite-coupling instanton expansion. We retain $J=\log\Xi$;
the image contributions are beyond its formal WKB expansion in the
large-positive-$\mu$ regime, so they do not alter the $J_0$ studied here.

The inverse density operator
$\widehat O=\widehat\rho^{-1}$ is expressed in terms of
conjugate operators satisfying
\begin{equation}
 [\widehat x,\widehat p]=2\pi i\hbar.
 \label{eq:spectral-commutator}
\end{equation}
To pass from operators to phase-space functions, we use the
Weyl symbol $O_W(x,p;\Delta,\hbar)$ of $\widehat O$,
defined by the Wigner transform reviewed in
Appendix~\ref{app:wigner}.
Its leading semiclassical term,
\begin{equation}
 O_0(x,p;\Delta)
 :=\lim_{\hbar\to0}O_W(x,p;\Delta,\hbar),
 \qquad
 H_0(x,p;\Delta):=\log O_0(x,p;\Delta),
 \label{eq:classical-symbol}
\end{equation}
is obtained by replacing the operators by commuting variables
and retaining the leading term in $\hbar$.
Thus $H_0$ is the classical one-particle Hamiltonian.

At leading semiclassical order, the sum over quantum states
in \eqref{eq:trace-log} becomes an integral over classical
phase space.
With the commutator \eqref{eq:spectral-commutator}, each
state occupies a phase-space area $4\pi^2\hbar$.
Writing $J\sim\hbar^{-1}J_0$ as $\hbar\to0$, we obtain
\begin{equation}
 J_0(\mu;\Delta)
 =\frac{1}{4\pi^2}
 \int_{\mathbb R^2}\mathrm dx\,\mathrm dp\,
 \log\!\left(1+e^{\mu-H_0(x,p;\Delta)}\right).
 \label{eq:classical-grand-potential}
\end{equation} 
For theories whose physical spectral coupling is fixed,
we define this approximation by formally varying $\hbar$
in the specified operator presentation while keeping the
sources fixed. This defines $J_0$ without requiring a
continuous variation of a quantized field-theory coupling.

To evaluate \eqref{eq:classical-grand-potential}, it is useful
to introduce the spectral zeta function
\begin{equation}
 \mathcal Z(s;\Delta,\hbar)
:=\operatorname{Tr}\widehat\rho(\Delta,\hbar)^s
 =\sum_a e^{-sE_a(\Delta,\hbar)},
 \label{eq:spectral-zeta}
\end{equation}
initially for $\operatorname{Re}s$ sufficiently large.
At positive integers $s=\ell$, this gives the spectral traces
appearing in the small-fugacity expansion
\begin{equation}
 J(\mu;\Delta,\hbar)
 =\sum_{\ell=1}^{\infty}
 \frac{(-1)^{\ell-1}}{\ell}\,
 e^{\ell\mu}\mathcal Z(\ell;\Delta,\hbar),
 \label{eq:grand-potential-traces}
\end{equation}
valid when $e^\mu\|\widehat\rho\|<1$.
Allowing complex $s$ provides the variable needed for the
Mellin--Barnes representation below.

The same semiclassical replacement of the trace gives
\begin{equation}
 \begin{aligned}
 \mathcal Z(s;\Delta,\hbar)
 &\sim\frac{I_0(s;\Delta)}{4\pi^2\hbar},
 \qquad \hbar\to0,\\
 I_0(s;\Delta)
 &:=\int_{\mathbb R^2}\mathrm dx\,\mathrm dp\,
 e^{-sH_0(x,p;\Delta)}
 =\int_{\mathbb R^2}\mathrm dx\,\mathrm dp\,
 O_0(x,p;\Delta)^{-s}.
 \end{aligned}
 \label{eq:classical-trace}
\end{equation}
We call $I_0$ the unnormalized classical spectral trace:
the phase-space measure $1/(4\pi^2\hbar)$ is kept separate.
Appendix~\ref{app:wigner} explains this relation using
Wigner--Weyl calculus.

The Mellin--Barnes identity
\begin{equation}
 \log(1+u)
 =-\int_{c-i\infty}^{c+i\infty}
 \frac{\mathrm ds}{2\pi i}\,
 \Gamma(s)\Gamma(-s)\,u^s,
 \qquad 0<c<1,
 \label{eq:MB-log}
\end{equation}
initially for $u>0$, now converts
\eqref{eq:classical-grand-potential} into
\begin{equation}
 J_0(\mu;\Delta)
 =-\frac{1}{4\pi^2}
 \int_{c-i\infty}^{c+i\infty}
 \frac{\mathrm ds}{2\pi i}\,
 \Gamma(s)\Gamma(-s)\,
 I_0(s;\Delta)e^{s\mu}.
 \label{eq:leading-Mellin-Barnes}
\end{equation}
The contour lies in a common convergence strip of the
Mellin kernel and the phase-space integral.
The resulting expression can then be analytically continued
where necessary
\cite{Hatsuda:2015oaa,Nosaka:2015iiw,Hatsuda:2015lpa,Moriyama:2014gxa}.

The large-$\mu$ expansion follows by moving the contour
to the left and collecting residues of the full integrand.
A surviving pole at $s=-a$, with $\operatorname{Re}a>0$,
produces a term proportional to $e^{-a\mu}$.
A pole of order $r$ produces a polynomial of degree at most
$r-1$ in $\mu$ multiplying this exponential.
Zeros of other factors can cancel candidate poles, while
coincident poles must be evaluated together.
These residues determine the membrane-instanton series
at leading WKB order.

The polynomial part follows from the pole at the origin.
For the examples below, the classical trace has the expansion
\begin{equation}
 I_0(s;\Delta)
 =\frac{a_{-2}(\Delta)}{s^2}
  +a_0(\Delta)+a_1(\Delta)s+O(s^2),
 \label{eq:classical-trace-origin}
\end{equation}
with no $s^{-1}$ term.
Combining this with
\begin{equation}
 -\Gamma(s)\Gamma(-s)
 =\frac{1}{s^2}+\frac{\pi^2}{6}+O(s^2)
 \label{eq:Mellin-kernel-origin}
\end{equation}
gives
\begin{equation}
 \begin{aligned}
 &J_{0,\mathrm{pert}}(\mu;\Delta)
 =\frac{C_0(\Delta)}{3}\mu^3
   +B_0(\Delta)\mu+A_0(\Delta),\\
 C_0&=\frac{a_{-2}}{8\pi^2},\qquad
 B_0=\frac{a_0}{4\pi^2}+\frac{a_{-2}}{24},
 \qquad
 A_0=\frac{a_1}{4\pi^2}.
 \end{aligned}
 \label{eq:leading-polynomial-residue}
\end{equation}
The absence of a quadratic term follows from the absence
of an $s^{-1}$ term in \eqref{eq:classical-trace-origin};
it is a property of the traces considered here, rather
than of the Mellin kernel alone.

The classical trace therefore determines both the polynomial
and exponentially suppressed parts of $J_0$.
Keeping only its behaviour near $s=0$ would discard the
instanton information central to our comparison with geometry.
We retain the full function $I_0(s;\Delta)$ in each example.
Higher WKB orders supply further corrections, while sectors
nonperturbative in $\hbar$, such as the ABJM worldsheet-instanton
sector, lie beyond this leading calculation
\cite{Calvo:2012du,Hatsuda:2012dt,Hatsuda:2015oaa}.

\subsection{Round-sphere quantum curves}
\label{sec:QCexamples}

We now compute the complete leading WKB grand potential
$J_0(\mu;\Delta)$ for the round-sphere quantum curves studied
in \cite{Hristov:2026zjh,Kubo:2024qhq,Kubo:2025dot}. Our main interest is its membrane-instanton
series: the exponential weights, their coefficients, and the
cancellations or enhancements that occur when Mellin poles coincide.
The same calculation determines the accompanying polynomial terms
$C_0\mu^3/3+B_0\mu+A_0$.
The leading contributions to $C$ and $B$ were obtained in
\cite{Hristov:2026zjh,Kubo:2024qhq,Kubo:2025dot}; results for the constant and instanton
terms in ABJM and related theories can be found in
\cite{Marino:2011eh,Nosaka:2015iiw,Hatsuda:2015lpa,Moriyama:2014gxa,Hatsuda:2014vsa}.
We compare the complete answers with the disk partition functions
in the following section.

For each example, we specify the operator and its allowed charges,
evaluate the classical trace $I_0$, and extract the instanton and
polynomial terms using \eqref{eq:leading-Mellin-Barnes}.
The required integral identities are collected in
Appendix~\ref{app:wkb-integrals}.
We attach an example label to the quantities defined in
Section~\ref{sec:spectral-method} and suppress their charge
arguments when no confusion can arise.
In particular, $J_0$ is the coefficient of $1/\hbar$, and
$I_0$ does not include the phase-space normalization.
We use $\Delta_{i_1\cdots i_r}:=\Delta_{i_1}+\cdots+\Delta_{i_r}$
for sums of charges. In every example, the charge variables satisfy
$\sum_i\Delta_i=2$, with the additional restrictions
defining each QC locus stated separately. The formulas are initially evaluated for
positive charges and $\operatorname{Re}s>0$, and then continued
meromorphically.
The sums below are understood as large-$\mu$ residue expansions;
the contour integral defines the complete function. Statements about
negative real poles refer initially to positive real charges.
All residue formulas apply to separated surviving poles.
At special charges, coincident poles should be evaluated
together and can produce polynomial factors in $\mu$.

\subsubsection{ABJM: pairwise locus}
\label{sec:ABJMpair}

We consider the pairwise R-charge deformation of ABJM,
whose density operator is given in
\cite[Eq.~(2.16), with $q=p=1$]{Nosaka:2015iiw}.
We use the ordered quantum curve of
\cite[Section~3.1, configuration (a), Eq.~(3.2)]{Hristov:2026zjh},
\begin{equation}
 \widehat O_\square
 =e^{\frac{D_{\rm m}}2\widehat x}
 \left(e^{(D-\frac12)\widehat p}+e^{(D+\frac12)\widehat p}\right)
 e^{\frac{D_{\rm m}}2\widehat x}
 \left(e^{-\widehat x/2}+e^{\widehat x/2}\right).
 \label{eq:pairO}
\end{equation}
The deformation dictionary is
\begin{equation}
 D=\Delta_1-\frac12=\frac12-\Delta_2,
 \qquad
 D_{\rm m}=\Delta_3-\frac12=\frac12-\Delta_4,
 \label{eq:pair-deformations}
\end{equation}
with
\begin{equation}
 \Delta_1+\Delta_2=\Delta_3+\Delta_4=1,
 \qquad \Delta_{1,2,3,4}>0.
 \label{eq:pairuv}
\end{equation}

Replacing the operators by commuting variables gives
\begin{equation}
 O_{0,\square}(x,p)=e^{-\Delta_4x-\Delta_2p}(1+e^x)(1+e^p).
 \label{eq:pair-classical-symbol}
\end{equation}
The classical trace is
\begin{equation}
 I_{0,\square}(s) =\int_{\R^2}\dd x\,\dd p O_{0,\square}(x,p)^{-s} =\frac{\prod_{i=1}^4\Gamma(s\Delta_i)}{\Gamma(s)^2},
 \label{eq:pair-leading-trace}
\end{equation}
using the Euler beta integral, \eqref{eq:general-beta}.

The negative Mellin poles give
\begin{align}
 J_{0,\mathrm{exp},\square}
 &\sim\sum_{i=1}^4\sum_{n=1}^\infty
 d_{i,n}\,e^{-n\mu/\Delta_i},\nonumber\\
 d_{i,n}
 &=\frac{(-1)^{n+1}}{4\pi^2 n!\Delta_i}
 \frac{\Gamma(n/\Delta_i)}{\Gamma(-n/\Delta_i)}
 \prod_{j\ne i}\Gamma(-n\Delta_j/\Delta_i).
 \label{eq:pairinstantons}
\end{align}

The pole at the origin gives the accompanying polynomial coefficients
\begin{equation}
 C_{0,\square} =\frac{1}{8\pi^2\prod_{i=1}^4\Delta_i}, \qquad B_{0,\square} =\frac{1-\Delta_1\Delta_2-\Delta_3\Delta_4} {24\prod_{i=1}^4\Delta_i}, \qquad A_{0,\square} =\frac{\zeta(3)}{4\pi^2} \sum_{i=1}^4\frac{1}{\Delta_i}.
 \label{eq:pairAcl}
\end{equation}
At the superconformal point $\Delta_i=1/2$, the combined
residues recover the undeformed leading-WKB instanton series, while
\eqref{eq:pairAcl} gives
$C_0=2/\pi^2$, $B_0=1/3$ and $A_0=2\zeta(3)/\pi^2$
\cite{Marino:2011eh,Nosaka:2015iiw,Hatsuda:2015lpa,Moriyama:2014gxa}.
The leading contribution to the grand potential is $J_0/k$.

\subsubsection{ABJM: triangular locus}
\label{sec:ABJMtriangle}

We next consider the triangular R-charge locus of level-one ABJM
theory. The flavored-SYM curve follows from
\cite[Eqs.~(2.22) and (A.23)]{Hristov:2026zjh} with
$q=1$ and $(F_1,F_2,F_3)=(1,0,0)$; its symbol is displayed
in Section~3.1, configuration (b), Eq.~(3.15) of that reference.
In the original deformation parameters, the operator is
\begin{equation}
 \widehat O_\triangle =e^{(\frac34-\frac D2+D_{\rm m})\widehat x +(D-\frac12)\widehat p} +e^{(-\frac14-\frac D2+D_{\rm m})\widehat x +(D-\frac12)\widehat p} +e^{(-\frac14-\frac D2+D_{\rm m})\widehat x +(D+\frac12)\widehat p}.
 \label{eq:triangleO}
\end{equation}
The deformation dictionary is
\begin{equation}
 D=\Delta_1+\Delta_3-\frac12=\frac12-\Delta_2,
 \qquad
 D_{\rm m}=\frac12(\Delta_1-\Delta_3),
 \label{eq:triangle-deformations}
\end{equation}
with
\begin{equation}
 \Delta_1+\Delta_2+\Delta_3=1,
 \qquad \Delta_4=1,
 \qquad \Delta_{1,2,3,4}>0.
 \label{eq:triangle-locus}
\end{equation}
Substituting this dictionary gives
\begin{equation}
 \widehat O_\triangle =e^{-\Delta_3\widehat x-\Delta_2\widehat p} +e^{(\Delta_1+\Delta_2)\widehat x-\Delta_2\widehat p} +e^{-\Delta_3\widehat x+(\Delta_1+\Delta_3)\widehat p}.
 \label{eq:triangleO-Delta}
\end{equation}

Replacing the operators by commuting variables gives
\begin{equation}
 O_{0,\triangle}
 =e^{-\Delta_3x-\Delta_2p}(1+e^x+e^p).
 \label{eq:triangleOW}
\end{equation}
The classical trace is
\begin{equation}
 I_{0,\triangle}(s) =\int_{\R^2}\dd x\,\dd p \left[e^{-\Delta_3x-\Delta_2p}(1+e^x+e^p)\right]^{-s} = \frac{\prod_{i=1}^3\Gamma(s\Delta_i)}{\Gamma(s)},
 \label{eq:triangle-leading-trace}
\end{equation}
where we used the two-variable Dirichlet integral
\eqref{eq:general-Dirichlet},
\begin{equation}
 \int_0^\infty\!\!\int_0^\infty
 \frac{X^{a-1}P^{b-1}}{(1+X+P)^{a+b+c}}
 \,\dd X\,\dd P
 =\frac{\Gamma(a)\Gamma(b)\Gamma(c)}{\Gamma(a+b+c)},
 \qquad \operatorname{Re} (a), \operatorname{Re} (b), \operatorname{Re} (c)>0,
 \label{eq:QC-Dirichlet}
\end{equation}
with $X=e^x$, $P=e^p$ and
$(a,b,c)=s(\Delta_3,\Delta_2,\Delta_1)$.
The negative Mellin poles give
\begin{align}
 J_{0,\mathrm{exp},\triangle}
 &\sim\sum_{i=1}^3\sum_{n=1}^\infty
 \widetilde d_{i,n}\,e^{-n\mu/\Delta_i},\nonumber\\
 \widetilde d_{i,n}
 &=\frac{(-1)^{n+1}}{4\pi^2 n!\Delta_i}
 \Gamma\left(\frac{n}{\Delta_i}\right)
 \prod_{j\ne i}
 \Gamma\left(-\frac{n\Delta_j}{\Delta_i}\right).
 \label{eq:triangleinstantons}
\end{align}

The perturbative contribution follows from the residue at $s=0$,
yielding the coefficients
\begin{equation}
 \begin{aligned}
 C_{0,\triangle}
 &=\frac{1}{8\pi^2\prod_{i=1}^3\Delta_i}, \qquad
 B_{0,\triangle}
 =\frac{(1-\Delta_1\Delta_2-\Delta_2\Delta_3
 -\Delta_3\Delta_1)}
 {24\prod_{i=1}^3\Delta_i},\\[4pt]
 A_{0,\triangle}
 &=\frac{\zeta(3)}{4\pi^2}
 \left(\sum_{i=1}^3\frac{1}{\Delta_i}-1\right).
 \end{aligned}
 \label{eq:triangleAcl}
\end{equation}

The quantum-curve operator \eqref{eq:triangleO} can be mapped
to the three-term family studied in
\cite[Section~2.6, Eq.~(74)]{Kashaev:2015kha} through the linear
change of variables
\begin{equation}
 \widehat X=(1-\Delta_3)\widehat x-\Delta_2\widehat p,
 \qquad
 \widehat P=-\Delta_3\widehat x+(1-\Delta_2)\widehat p,
 \label{eq:trianglechange}
\end{equation}
which gives
\begin{equation}
 \widehat O_\triangle
 =e^{\widehat X}+e^{\widehat P}+e^{-m\widehat X-n\widehat P},
 \label{eq:triangle-mirror-operator}
\end{equation}
where
\begin{equation}
 m=\frac{\Delta_3}{\Delta_1},
 \qquad
 n=\frac{\Delta_2}{\Delta_1},
 \qquad
 [\widehat X,\widehat P]=2\pi i\Delta_1\hbar.
 \label{eq:mn}
\end{equation}
For generic R-charges, $m,n$ are positive real parameters,
while for positive integer values the operator corresponds
to the mirror curve of local $\PP(1,m,n)$.
In particular, at $\Delta_1=\Delta_2=\Delta_3=1/3$,
we recover the local-$\PP^2$ operator, with
$[\widehat X,\widehat P]=2\pi i\hbar/3$.~\footnote{A relation between ABJM theory and local $\PP^2$ also
appears for the squashed sphere at $b^2=3$ and $k=1$
\cite{Hatsuda:2016uqa}.}
At the physical level-one ABJM point, $\hbar=1$,
the effective Planck constant of this operator is therefore
$2\pi/3$, in the convention where the commutator is
$i$ times the Planck constant.
The identification with local $\PP^2$ includes this
rescaling of the spectral coupling.

\subsubsection{ADHM: a distinguished locus}
\label{sec:ADHMdist-round}

We now turn to the distinguished R-charge locus of ADHM theory
with $f$ fundamental hypermultiplets. The flavored-SYM
construction of \cite[Eqs.~(2.22) and (A.23)]{Hristov:2026zjh}, specialized to $q=f$ and
$(F_1,F_2,F_3)=(f,0,0)$, gives the quantum curve
\begin{equation}
\begin{aligned}
\widehat O_{\mathrm{ADHM}}
={}&e^{\frac{D_{\rm m}}{2}\widehat x}
\Bigg[
 e^{(\frac14+\frac D2)\widehat p}
 e^{-f(\frac14+\frac D2)\widehat x}
 e^{(\frac14+\frac D2)\widehat p}
\\
&\quad+
 e^{-(\frac14-\frac D2)\widehat p}
 e^{f(\frac14-\frac D2)\widehat x}
 \left(e^{\widehat x/2}+e^{-\widehat x/2}\right)^f
 e^{-(\frac14-\frac D2)\widehat p}
\Bigg]
 e^{\frac{D_{\rm m}}{2}\widehat x}.
\end{aligned}
\label{eq:ADHM-operator}
\end{equation}
Here,
\begin{equation}
 D=\frac12-\Delta_2,
 \qquad
 D_{\rm m}=\frac f2(\Delta_4-\Delta_3),
 \label{eq:ADHM-deformations}
\end{equation}
with
\begin{equation}
 \Delta_1=1,
 \qquad
 \Delta_2+\Delta_3+\Delta_4=1,
 \qquad
 \Delta_{2,3,4}>0.
 \label{eq:ADHM-dist-locus}
\end{equation}

Replacing the operators by commuting variables gives
\begin{equation}
 O_{0,\mathrm{ADHM}}
 =e^{-f\Delta_3x-\Delta_2p}
 \left[(1+e^x)^f+e^p\right].
 \label{eq:ADHM-Wigner}
\end{equation}
The classical trace is
\begin{equation}
 \begin{aligned}
 I_{0,\mathrm{ADHM}}(s)
 &=\int_{\R^2}\dd x\,\dd p
   \left[e^{-f\Delta_3x-\Delta_2p} \left((1+e^x)^f+e^p\right)\right]^{-s}\\
 &=\int_{\R}\dd x\;
   e^{s f \Delta_3 x} (1+e^x)^{-s f}
   \int_{\R}\dd p\;
   e^{s \Delta_2 p}
   \left(1+\frac{e^p}{(1+e^x)^f}\right)^{-s}\\
    &= \int_{\R}\dd x\; e^{s f \Delta_3 x} (1+e^x)^{-s f (\Delta_3+\Delta_4)}\int_{\R}\dd p\; e^{s\Delta_2 p} (1+e^{p})^{-s}\\
 &=\frac{\Gamma(sf\Delta_3)\Gamma(sf\Delta_4)
   \Gamma(s\Delta_2)\Gamma(s\Delta_{34})}
   {\Gamma(s)\Gamma(sf\Delta_{34})}.
 \end{aligned}
 \label{eq:ADHM-leading-trace}
\end{equation}
The change of variables
$p_{\mathrm{old}}=p+f\log(1+e^x)$ separates the integrals;
each is then evaluated using the Euler beta integral.

In the Mellin--Barnes representation,
\eqref{eq:ADHM-leading-trace} gives the poles at
$s=-n/r_i$, where
$r=(f\Delta_3,f\Delta_4,\Delta_2,\Delta_{34})$ and
$n\in\mathbb Z_{>0}$.
Their residues yield the leading-WKB instanton series,
\begin{align}
 J_{0,\mathrm{exp},{\mathrm{ADHM}}}
 &\sim\sum_{i=1}^4\sum_{n=1}^\infty
 d^{\mathrm{ADHM}}_{i,n}\,e^{-n\mu/r_i},
 \nonumber\\
 d^{\mathrm{ADHM}}_{i,n}
 &=\frac{(-1)^{n+1}}{4\pi^2 n!r_i}
 \frac{\Gamma(n/r_i)}
 {\Gamma(-nf\Delta_{34}/r_i)}
 \prod_{j\ne i}\Gamma(-nr_j/r_i).
 \label{eq:dist-instantons}
\end{align}
The additional denominator allows cancellations between pole
families. In particular, for integer $f$, it cancels the poles
associated with $r_4=\Delta_{34}$, leaving the generic instanton
actions $n/(f\Delta_3)$, $n/(f\Delta_4)$, and $n/\Delta_2$. 

The pole at the origin gives
\begin{equation}
 \begin{aligned}
 C_{0,\mathrm{ADHM}}
 &=\frac{1}
 {8\pi^2 f\Delta_2\Delta_3\Delta_4},
 \qquad
 B_{0,\mathrm{ADHM}}
 =\frac{\left(1-\Delta_2\Delta_{34}
 -f^2\Delta_3\Delta_4\right)}
 {24 f\Delta_2\Delta_3\Delta_4},
 \\[4pt]
 A_{0,\mathrm{ADHM}}
 &=\frac{\zeta(3)}{4\pi^2}
 \left[
 \frac{1}{f\Delta_3}
 +\frac{1}{f\Delta_4}
 +f^2\left(\frac{1}{\Delta_2}-1\right)
 \right].
 \end{aligned}
 \label{eq:ADHM-leading-constant}
\end{equation}

At $f=1$, the quantum curve and the classical trace
reduce to those of the triangular ABJM locus in
Section~\ref{sec:ABJMtriangle}, upon
$\Delta_1 \longleftrightarrow \Delta_4$.
The corresponding pairwise ADHM locus is considered below.
\subsubsection{Grouped necklaces and pairwise ADHM}
\label{sec:necklaceQC}

The pairwise construction extends to the $\mathcal N=4$
necklace theories with $U(N)$ gauge nodes, bifundamental
hypermultiplets, and Chern--Simons levels
$k_a=k(s_a-s_{a-1})/2$, where $s_a=\pm1$.
Their transverse branch is
$(\C^2/\Z_\xi\times\C^2/\Z_\zeta)/\Z_k$
\cite{Hosomichi:2008jd,Imamura:2008dt,Imamura:2008nn}.
We assign $\Delta_1,\Delta_2$ to the first covering
surface and $\Delta_3,\Delta_4$ to the second.
We focus on the grouped ordering of $\xi$ position
factors and $\zeta$ momentum factors, with
$\xi,\zeta$ positive integers. The density operator is given in
\cite[Eq.~(2.16)]{Nosaka:2015iiw}. Inverting it and using an
ordered form gives
\begin{equation}
 \widehat O_{\mathrm{neck}} =F(\widehat x)G(\widehat p), \qquad F(x)=e^{-ux}(1+e^x)^\xi, \qquad G(p)=e^{-vp}(1+e^p)^\zeta.
 \label{eq:pqO}
\end{equation}
The parameters $(q,p,\xi,\eta)$ of that reference correspond to
$(\xi,\zeta,\xi-2u,\zeta-2v)$ here.
Our parameters are related to the R-charges by
\begin{equation}
 u=\xi\Delta_1,\qquad
 \xi-u=\xi\Delta_2,\qquad
 v=\zeta\Delta_3,\qquad
 \zeta-v=\zeta\Delta_4,
 \label{eq:necklace-deformations}
\end{equation}
with
\begin{equation}
 \Delta_1+\Delta_2=1,\qquad
 \Delta_3+\Delta_4=1,\qquad
 \Delta_{1,2,3,4}>0.
 \label{eq:necklace-locus}
\end{equation}

Replacing the operators by commuting variables gives
\begin{equation}
 O_{0,\mathrm{neck}}(x,p)=F(x)G(p)
 =e^{-ux-vp}(1+e^x)^\xi(1+e^p)^\zeta.
 \label{eq:necklace-classical}
\end{equation}
The classical trace therefore separates into two
Euler beta function integrals \cite{Nosaka:2015iiw,Hatsuda:2015lpa,Moriyama:2014gxa},
\begin{equation}
 \begin{aligned}
 I_{0,\mathrm{neck}}(s)
 &=\int_{\R^2}\dd x\,\dd p
   \left[
   e^{-ux-vp}(1+e^x)^\xi(1+e^p)^\zeta
   \right]^{-s}\\
 &=
   \mathrm B(us,(\xi-u)s)
   \mathrm B(vs,(\zeta-v)s)\\
 &=
   \frac{
   \Gamma(us)\Gamma((\xi-u)s)
   \Gamma(vs)\Gamma((\zeta-v)s)
   }{
   \Gamma(\xi s)\Gamma(\zeta s)
   } = \frac{
   \Gamma(\xi s \Delta_1)\Gamma(\xi s \Delta_2)
   \Gamma(\zeta s \Delta_3)\Gamma(\zeta s \Delta_4 )
   }{
   \Gamma(\xi s)\Gamma(\zeta s)
   }.
 \end{aligned}
 \label{eq:pqI}
\end{equation}

After inserting \eqref{eq:pqI} into the Mellin--Barnes
representation, the poles on the negative real $s$-axis
are organized by $r=(u,\xi-u,v,\zeta-v)$.
Their residues give the leading-WKB instanton series,
\begin{align}
 J_{0,\mathrm{exp},{\mathrm{neck}}}
 &\sim\sum_{i=1}^4\sum_{n=1}^\infty
   d^{\mathrm{neck}}_{i,n}\,e^{-n\mu/r_i},
 \nonumber\\
 d^{\mathrm{neck}}_{i,n}
 &=\frac{(-1)^{n+1}}{4\pi^2 n!r_i}
   \frac{
   \Gamma(n/r_i)\Gamma(-n/r_i)
   \prod_{j\ne i}\Gamma(-nr_j/r_i)
   }{
   \Gamma(-n\xi/r_i)\Gamma(-n\zeta/r_i)
   }.
 \label{eq:necklace-instantons}
\end{align}

The residue at $s=0$ fixes the accompanying perturbative
coefficients,
\begin{equation}
 \begin{aligned}
 C_{0,\mathrm{neck}}
 &=\frac{\xi\zeta}
   {8\pi^2 u(\xi-u)v(\zeta-v)},
 \qquad
 B_{0,\mathrm{neck}}
 =\frac{
   \xi\zeta
   \left[1-u(\xi-u)-v(\zeta-v)\right]
   }{
   24 u(\xi-u)v(\zeta-v)
   },
 \\[4pt]
 A_{0,\mathrm{neck}}
 &=\frac{\zeta(3)}{4\pi^2}
   \left[
   \zeta^2\left(\frac{1}{u}+\frac{1}{\xi-u}\right)
   +\xi^2\left(\frac{1}{v}+\frac{1}{\zeta-v}\right)
   \right].
 \end{aligned}
 \label{eq:pq-leading-constant}
\end{equation}

At $\xi=\zeta=1$, the classical trace reduces directly
to the pairwise ABJM expression in
Section~\ref{sec:ABJMpair}, with the same charge labels.

The pairwise ADHM trace is obtained by setting
$\xi=1$, $\zeta=f$, $u=\Delta_1$ and $v=f\Delta_3$.
Thus $\Delta_1,\Delta_2$ remain the free-factor weights,
while $\Delta_3,\Delta_4$ are the ALE covering weights.
An ordered quantum-curve presentation of this theory is
\begin{equation}
 \widehat O_{\mathrm{ADHM,pair}}
 =e^{-f\Delta_3\widehat x}(1+e^{\widehat x})^f
  e^{-\Delta_1\widehat p}(1+e^{\widehat p}).
 \label{eq:ADHM-pair-operator}
\end{equation}
Thus its classical trace is
\begin{equation}
 I_{0,\mathrm{ADHM,pair}}(s)
 =\frac{\Gamma(sf\Delta_3)\Gamma(sf\Delta_4)
         \Gamma(s\Delta_1)\Gamma(s\Delta_2)}
        {\Gamma(sf)\Gamma(s)}.
 \label{eq:ADHM-pair-trace}
\end{equation}
This is precisely the specialization of
\eqref{eq:pqI} stated above.
The position and momentum factors appear in a different
order in \eqref{eq:ADHM-pair-operator}, but the separated
classical trace is the same.
Its instanton coefficients and polynomial terms follow
from \eqref{eq:necklace-instantons} and
\eqref{eq:pq-leading-constant} with
$\xi=1$, $\zeta=f$, $u=\Delta_1$ and $v=f\Delta_3$.
At $\Delta_i=1/2$, this is the standard ADHM, or $N_f=f$,
Fermi-gas operator in an ordered form; the equivalent density
kernel is given in \cite[Eq.~(2.4)]{Hatsuda:2014vsa}; related sphere free-energy calculations appear in \cite{Mezei:2013gqa}.
Here the WKB expansion varies the formal spectral coupling
$\hbar$ at fixed $f$, rather than the number of flavors.
Our use of ``membrane-instanton'' refers to the exponential
terms captured by this expansion; it should therefore be
distinguished from the sector classification associated
with the small-$N_f$ expansion in \cite{Hatsuda:2014vsa}.

\subsubsection{Conifold: Weyl and ordered loci}
\label{sec:conifoldWeyl}
\label{sec:conifolda}

We now consider the one- and two-node presentations of
$\C\times\mathcal C$, where $\mathcal C$ is the conifold,
studied in \cite[Section~3.2]{Hristov:2026zjh}.
For configuration (c), with $q=0$ and $(F_1,F_2,F_3)=(1,1,0)$,
Eqs.~(2.22) and (A.23) of that reference give the operator below;
its symbol appears in Eq.~(3.42).
\begin{equation}
\begin{aligned}
\widehat O_{\mathrm{con,W}}
={}&e^{\frac{D_{\rm m}}2\widehat x}
\Bigg[
 e^{(\frac14+\frac D2)\widehat p}
 \left(e^{\widehat x/2}+e^{-\widehat x/2}\right)
 e^{(\frac14+\frac D2)\widehat p}
\\
&\quad+
 e^{-(\frac14-\frac D2)\widehat p}
 \left(e^{\widehat x/2}+e^{-\widehat x/2}\right)
 e^{-(\frac14-\frac D2)\widehat p}
\Bigg]
 e^{\frac{D_{\rm m}}2\widehat x}.
\end{aligned}
\label{eq:conifold-Weyl-operator}
\end{equation}
The deformation dictionary reads
\begin{equation}
 D=\frac12-\Delta_{14},
 \qquad
 D_{\rm m}=\frac12-\Delta_{13},
 \label{eq:conifold-Weyl-deformations}
\end{equation}
with
\begin{equation}
 \Delta_5=\Delta_{1234}=1,
 \qquad
 \Delta_i>0.
 \label{eq:conifold-Weyl-locus}
\end{equation}

Replacing the operators by commuting variables gives
\begin{equation}
 \begin{aligned}
 O_{0,\mathrm{con,W}}
 &=e^{D_{\rm m}x+Dp}
   \left(e^{x/2}+e^{-x/2}\right)
   \left(e^{p/2}+e^{-p/2}\right)\\
 &=e^{-\Delta_{13}x-\Delta_{14}p}
   (1+e^x)(1+e^p).
 \end{aligned}
 \label{eq:conifold-Weyl-symbol}
\end{equation}
Writing
$r=(\Delta_{13},\Delta_{24},\Delta_{14},\Delta_{23})$,
so that $r_1+r_2=r_3+r_4=1$, we find
\begin{equation}
 \begin{aligned}
 I_{0,\mathrm{con,W}}(s)
 &=\int_{\R^2}\dd x\,\dd p
   \left[
   e^{-r_1x-r_3p}(1+e^x)(1+e^p)
   \right]^{-s}\\
 &=
   \mathrm B(sr_1,sr_2)\mathrm B(sr_3,sr_4)\\
 &=
   \frac{\prod_{i=1}^4\Gamma(sr_i)}{\Gamma(s)^2}.
 \end{aligned}
 \label{eq:explicit-I0-con-w}
\end{equation}
This coincides with the classical trace of pairwise
ABJM under $\Delta_i\mapsto r_i$. The corresponding
Mellin--Barnes residues therefore give
\begin{align}
 J_{0,\mathrm{exp},{\mathrm{con,W}}}
 &\sim\sum_{i=1}^4\sum_{n=1}^\infty
   d^{\mathrm{con,W}}_{i,n}\,e^{-n\mu/r_i},
 \nonumber\\
 d^{\mathrm{con,W}}_{i,n}
 &=\frac{(-1)^{n+1}}{4\pi^2 n!r_i}
   \frac{\Gamma(n/r_i)}{\Gamma(-n/r_i)}
   \prod_{j\ne i}\Gamma(-nr_j/r_i).
 \label{eq:conifold-Weyl-instantons}
\end{align}

The perturbative coefficients likewise follow from
the pairwise expressions,
\begin{equation}
 \begin{aligned}
 C_{0,\mathrm{con,W}}
 &=\frac{1}{8\pi^2\prod_{i=1}^4r_i},
 \qquad
 B_{0,\mathrm{con,W}}
 =\frac{1-r_1r_2-r_3r_4}
   {24\prod_{i=1}^4r_i},
 \qquad
 A_{0,\mathrm{con,W}}
 =\frac{\zeta(3)}{4\pi^2}
   \sum_{i=1}^4\frac{1}{r_i}.
 \end{aligned}
 \label{eq:conifold-Weyl-constant}
\end{equation}
This agreement concerns the classical traces and does not
identify the ordered quantum operators.

At $r_i=1/2$, the linear change of variables
\begin{equation}
 \widehat X=\frac{\widehat x+\widehat p}{2},
 \qquad
 \widehat P=\frac{\widehat p-\widehat x}{2}
\end{equation}
puts the quantum curve in the mass-one
local-$\PP^1\times\PP^1$ form \cite[Section~5.2, Eq.~(5.21)]{Grassi:2014zfa},
\begin{equation}
 \widehat O_{\mathrm{con,W}}
 =e^{\widehat X}+e^{-\widehat X}+e^{\widehat P}+e^{-\widehat P},
 \qquad
 [\widehat X,\widehat P]=\pi i\hbar.
\end{equation}

A different R-charge restriction is realized by
configuration (b) of \cite[Section~3.2, Eqs.~(2.22), (A.23) and (A.24)]{Hristov:2026zjh}, with $q=1$ and
$(F_1,F_2,F_3)=(1,0,1)$. In this presentation, the
quantum curve contains an additional ordered line factor
(the full symbol is given in Eq.~(3.37) of that reference),
\begin{equation}
\begin{aligned}
\widehat O_{\mathrm{con,ord}}
={}&e^{\frac{D_{\rm m}}2\widehat x}
\Bigg[
 e^{(\frac14+\frac D2)\widehat p}
 e^{-(\frac14+\frac D2)\widehat x}
 e^{(\frac14+\frac D2)\widehat p}
\\
&\quad+
 e^{-(\frac14-\frac D2)\widehat p}
 e^{(\frac14-\frac D2)\widehat x}
 \left(e^{\widehat x/2}+e^{-\widehat x/2}\right)
 e^{-(\frac14-\frac D2)\widehat p}
\Bigg]
 e^{\frac{D_{\rm m}}2\widehat x}
 \left(e^{\widehat x/2}+e^{-\widehat x/2}\right).
\end{aligned}
\label{eq:conifold-ordered-operator}
\end{equation}
The deformation parameters now take the form
\begin{equation}
 D=\Delta_{14}-\frac12=\frac12-\Delta_5,
 \qquad
 D_{\rm m}
 =\frac12(\Delta_1-\Delta_2+\Delta_3-\Delta_4),
 \label{eq:conifold-ordered-deformations}
\end{equation}
with
\begin{equation}
 \Delta_{23}=1,
 \qquad
 \Delta_{14}+\Delta_5=1,
 \qquad
 \Delta_i>0.
 \label{eq:conifoldbUV}
\end{equation}

The relation to the triangular construction becomes
explicit in the exact factorization
\begin{equation}
 \widehat O_{\mathrm{con,ord}}=\widehat T\widehat L,
 \label{eq:conifold-TL-factorization}
\end{equation}
where
\begin{equation}
 \widehat T = e^{(\frac12-u)\widehat x-v\widehat p} +e^{(\frac32-u)\widehat x-v\widehat p} +e^{(\frac12-u)\widehat x+(1-v)\widehat p}, \qquad \widehat L =e^{-\widehat x/2}+e^{\widehat x/2},
 \label{eq:conifold-TL-operators}
\end{equation}
and $u=\Delta_{24}$, $v=\Delta_5$.
Replacing the operators by commuting variables gives
\begin{equation}
 O_{0,\mathrm{con,ord}}(x,p)
 =e^{-ux-vp}(1+e^x)(1+e^x+e^p).
 \label{eq:conifold-ordered-classical}
\end{equation}

For this locus, define
\begin{equation}
 r=(\Delta_{24},\Delta_{13},\Delta_5,\Delta_{14}),
 \qquad
 \tau=\Delta_{1234}=2-\Delta_5,
\end{equation}
which satisfy $r_1+r_2=\tau$ and $r_3+r_4=1$.
The translation $p_{\rm old}=p+\log(1+e^x)$ separates the two integrals.
Integrating first over $e^p$ and then over $e^x$ gives two beta integrals (\ref{eq:general-beta}),
\begin{equation}
 \begin{aligned}
 I_{0,\mathrm{con,ord}}(s)
 &=\int_{\R^2}\dd x\,\dd p
   \left[
   e^{-ux-vp}(1+e^x)(1+e^x+e^p)
   \right]^{-s}\\
 &=\int_{\R}\dd x\; e^{sux}(1+e^x)^{-s}\int_{\R} \dd p\,e^{svp} (1+e^x+e^p)^{-s}\\
 &= \int_{\R} \dd x\; e^{sux} (1+e^x)^{s(v-2)}\int_{\R} \dd p e^{svp} (1+e^p)^{-s} \\
 &=
   \mathrm B(sr_3,sr_4)\mathrm B(sr_1,sr_2)\\
 &=
   \frac{\prod_{i=1}^4\Gamma(sr_i)}
   {\Gamma(s)\Gamma(s\tau)}.
 \end{aligned}
 \label{eq:explicit-I0-con-ord}
\end{equation}
Compared with \eqref{eq:explicit-I0-con-w}, one denominator
factor $\Gamma(s)$ is replaced by $\Gamma(s\tau)$.
The leading-WKB instanton coefficients change accordingly,
\begin{align}
 J_{0,\mathrm{exp},{\mathrm{con,ord}}}
 &\sim\sum_{i=1}^4\sum_{n=1}^\infty
   d^{\mathrm{con,ord}}_{i,n}\,e^{-n\mu/r_i},
 \nonumber\\
 d^{\mathrm{con,ord}}_{i,n}
 &=\frac{(-1)^{n+1}}{4\pi^2 n!r_i}
   \frac{\Gamma(n/r_i)}{\Gamma(-n\tau/r_i)}
   \prod_{j\ne i}\Gamma(-nr_j/r_i).
 \label{eq:coninstantons}
\end{align}
The residue at the origin gives
\begin{equation}
 \begin{aligned}
 C_{0,\mathrm{con,ord}}
 &=\frac{\tau}{8\pi^2\prod_{i=1}^4r_i},
 \qquad
 B_{0,\mathrm{con,ord}}
 =\frac{
   \tau\left(1+\sum_{i=1}^4r_i^2-\tau^2\right)
   }{
   48\prod_{i=1}^4r_i
   },
 \\[4pt]
 A_{0,\mathrm{con,ord}}
 &=\frac{\zeta(3)}{4\pi^2}
   \left(
   \frac{1}{r_1}+\frac{1}{r_2}
   +\frac{4}{r_3}+\frac{1}{r_4}-1
   \right).
 \end{aligned}
 \label{eq:conifold-ordered-constant}
\end{equation}

Configuration (a) of \cite[Section~3.2, Eqs.~(2.23) and (3.31)]{Hristov:2026zjh} instead imposes
$\Delta_{24}=1$ and $\Delta_{13}+\Delta_5=1$.
With $r=(\Delta_{23},\Delta_{14},\Delta_5,\Delta_{13})$
and $\tau=\Delta_{1234}$, its classical formulas follow from the preceding
expressions by exchanging $\Delta_3$ and $\Delta_4$.

The three classical traces are restrictions of the common
meromorphic expression
\begin{equation}
 I_{0, \mathrm{con}}(s)
 =\frac{\Gamma(s\Delta_5)
 \Gamma(s\Delta_{13})\Gamma(s\Delta_{14})
 \Gamma(s\Delta_{23})\Gamma(s\Delta_{24})}
 {\Gamma(s)^2\Gamma(s\Delta_{1234})}.
 \label{eq:conIeps}
\end{equation}
Section~\ref{sec:diskexamples} identifies this common
expression with the corresponding mesonic disk.
The different operator orderings remain relevant beyond
leading WKB order.

\subsubsection{SPP: Weyl locus}
\label{sec:SPPWeyl}
\label{sec:SPP}

For the SPP Weyl locus, the flavored-SYM construction of
\cite[Section~3.4, Eqs.~(2.22) and (A.23)]{Hristov:2026zjh}, with $q=1$ and flavor multiplicities
$(F_1,F_2,F_3)=(2,1,0)$, gives the following operator.
Its symbol is displayed in Eq.~(3.80) of that reference.
\begin{equation}
\begin{aligned}
\widehat O_{\mathrm{SPP}}
={}&e^{\frac{D_{\rm m}}2\widehat x}
\Bigg[
 e^{(\frac14+\frac D2)\widehat p}
 e^{-(\frac14+\frac D2)\widehat x}
 \left(e^{\widehat x/2}+e^{-\widehat x/2}\right)
 e^{(\frac14+\frac D2)\widehat p}
\\
&\quad+
 e^{-(\frac14-\frac D2)\widehat p}
 e^{(\frac14-\frac D2)\widehat x}
 \left(e^{\widehat x/2}+e^{-\widehat x/2}\right)^2
 e^{-(\frac14-\frac D2)\widehat p}
\Bigg]
 e^{\frac{D_{\rm m}}2\widehat x}.
\end{aligned}
\label{eq:SPP-operator}
\end{equation}
The deformation dictionary is
\begin{equation}
 D=\Delta_1+\Delta_4+\Delta_5-\frac12, \qquad D_{\rm m} =\frac12(2\Delta_1+\Delta_2-\Delta_3-2\Delta_4),
 \label{eq:SPP-deformations}
\end{equation}
with
\begin{equation}
 \Delta_1+\Delta_2+\Delta_3+\Delta_4+\Delta_5=1,
 \qquad
 \Delta_6=1,
 \qquad
 \Delta_{1,2,3,4,5}>0.
 \label{eq:SPP-locus}
\end{equation}

Replacing the operators by commuting variables gives
\begin{equation}
 O_{0,\mathrm{SPP}} = e^{(D_{\rm m}-\frac34-\frac D2)x+(D-\frac12)p} \left(1+2e^x+e^{2x}+e^p+e^{x+p}\right).
 \label{eq:SPP-Wigner}
\end{equation}
In terms of the R-charges, this factorizes as
\begin{equation}
 \begin{aligned}
 O_{0,\mathrm{SPP}}
 &=e^{-ux-vp}(1+e^x)(1+e^x+e^p),
 \\
 u=\Delta_3&+2\Delta_4+\Delta_5,
 \qquad
 v=\Delta_2+\Delta_3.
 \end{aligned}
 \label{eq:SPP-factorization}
\end{equation}
The line and triangular factors reproduce the leading
ordered-conifold symbol \eqref{eq:conifold-ordered-classical},
with a different identification of the R-charges.

To express the spectral trace in the same form, we introduce
\begin{equation}
 \begin{aligned}
 r_1&=\Delta_3+2\Delta_4+\Delta_5,
 &
 r_2&=2\Delta_1+\Delta_2+\Delta_5,
 \\
 r_3&=\Delta_2+\Delta_3,
 &
 r_4&=\Delta_1+\Delta_4+\Delta_5,
 \end{aligned}
 \label{eq:SPP-weights}
\end{equation}
and $\tau=1+\Delta_1+\Delta_4+\Delta_5$, so that
$r_1+r_2=\tau$ and $r_3+r_4=1$.
The same two Euler beta integrals then yield
\begin{equation}
 \begin{aligned}
 I_{0,\mathrm{SPP}}(s)
 &=\int_{\R^2}\dd x\,\dd p
   \left[
   e^{-ux-vp}(1+e^x)(1+e^x+e^p)
   \right]^{-s}\\
 &=
   \mathrm B(sr_3,sr_4)\mathrm B(sr_1,sr_2)\\
 &=
   \frac{\prod_{i=1}^4\Gamma(sr_i)}
   {\Gamma(s)\Gamma(s\tau)}.
 \end{aligned}
 \label{eq:SPP-leading-trace}
\end{equation}

The leading-WKB instanton sector therefore follows from
the ordered-conifold residue formula \eqref{eq:coninstantons}.
For generic R-charges, the Mellin--Barnes poles at
$s=-n/r_i$ give
\begin{align}
 J_{0,\mathrm{exp},{\mathrm{SPP}}}
 &\sim\sum_{i=1}^4\sum_{n=1}^\infty
   d^{\mathrm{SPP}}_{i,n}\,e^{-n\mu/r_i},
 \nonumber\\
 d^{\mathrm{SPP}}_{i,n}
 &=\frac{(-1)^{n+1}}{4\pi^2 n!r_i}
   \frac{\Gamma(n/r_i)}{\Gamma(-n\tau/r_i)}
   \prod_{j\ne i}\Gamma(-nr_j/r_i).
 \label{eq:SPP-instantons}
\end{align}

The perturbative coefficients likewise follow from
\eqref{eq:conifold-ordered-constant}. Using the relations
among the weights, they can be written as
\begin{equation}
 \begin{aligned}
 C_{0,\mathrm{SPP}}
 &=\frac{\tau}{8\pi^2\prod_{i=1}^4r_i},
 \qquad
 B_{0,\mathrm{SPP}}
 =\frac{
   \tau(1-r_1r_2-r_3r_4)
   }{
   24\prod_{i=1}^4r_i
   },
 \\[4pt]
 A_{0,\mathrm{SPP}}
 &=\frac{\zeta(3)}{4\pi^2}
   \left[
   \frac{1}{r_1}+\frac{1}{r_2}
   +\tau^2\left(\frac{1}{r_3}+\frac{1}{r_4}\right)
   \right].
 \end{aligned}
 \label{eq:SPP-leading-constant}
\end{equation}
The two loci thus share the same leading-WKB grand potential
under the corresponding identification of weights.
This equality does not extend automatically to higher WKB orders.

\subsubsection{\texorpdfstring{$Q^{1,1,1}$}{Q111}: two ordered triangles}
\label{sec:Qexample}

We now consider the two-node flavored-quiver presentation of the cone over
$Q^{1,1,1}$ studied in \cite[Section~3.3]{Hristov:2026zjh}.
This is the flavored modified $L^{020}$ presentation of that reference,
also described there as an ABJM-type generalization. We take
$q_1=1$, $q_2=0$, $F_1=F_2=1$, and
$\widetilde F_1=\widetilde F_2=0$.
Equations~(2.23) and (A.23) of that reference give
\begin{equation}
 \widehat O_Q = e^{\frac{D_{\rm m}}4\widehat x} \widehat{\mathcal O}^{(1,1)}_{0,1}(D_1) e^{\frac{D_{\rm m}}2\widehat x} \widehat{\mathcal O}^{(1,0)}_{1,0}(D_2) e^{\frac{D_{\rm m}}4\widehat x} =\widehat T_1\widehat T_2.
 \label{eq:Q-general-operator}
\end{equation}
Here $\widehat{\mathcal O}^{(1,1)}_{0,1}$ and
$\widehat{\mathcal O}^{(1,0)}_{1,0}$ are the one-node building
blocks defined in \cite[Eq.~(A.23)]{Hristov:2026zjh}.
For clarity, their explicit form is specified by the following
factors through
$\widehat T_a=e^{D_{\rm m}\widehat x/4}
\widehat{\mathcal O}_a e^{D_{\rm m}\widehat x/4}$,
with $\widehat{\mathcal O}_1=
\widehat{\mathcal O}^{(1,1)}_{0,1}(D_1)$ and
$\widehat{\mathcal O}_2=
\widehat{\mathcal O}^{(1,0)}_{1,0}(D_2)$.
The two factors are
\begin{align}
 \widehat T_1
 ={}&
 e^{(\frac{D_{\rm m}}2-\frac14-\frac{D_1}2)\widehat x
       +(D_1-\frac12)\widehat p}
 +e^{(\frac{D_{\rm m}}2+\frac34-\frac{D_1}2)\widehat x
       +(D_1-\frac12)\widehat p}
 \nonumber\\
 &+
 e^{(\frac{D_{\rm m}}2-\frac14-\frac{D_1}2)\widehat x
       +(D_1+\frac12)\widehat p},
 \nonumber\\[4pt]
 \widehat T_2
 ={}&
 e^{(\frac{D_{\rm m}}2+\frac14-\frac{D_2}2)\widehat x
       +(D_2-\frac12)\widehat p}
 +e^{(\frac{D_{\rm m}}2-\frac34-\frac{D_2}2)\widehat x
       +(D_2+\frac12)\widehat p}
 \nonumber\\
 &+
 e^{(\frac{D_{\rm m}}2+\frac14-\frac{D_2}2)\widehat x
       +(D_2+\frac12)\widehat p}.
 \label{eq:Qfactor}
\end{align}
Each factor is equivalent to the quantum curve at the triangular
ABJM locus. For $\widehat T_1$, the parameter identification is
$D\to D_1$ and $D_{\rm m}\to D_{\rm m}/2$.
For $\widehat T_2$, it is $D\to-D_2$ and
$D_{\rm m}\to-D_{\rm m}/2$, together with
$(\widehat x,\widehat p)\to(-\widehat x,-\widehat p)$.

The R-charges obey
\begin{equation}
 \Delta_2+\Delta_4+\Delta_5
 =
 \Delta_1+\Delta_3+\Delta_6
 =1,
 \qquad
 \Delta_i>0.
 \label{eq:Q-locus}
\end{equation}
We define the charge differences
\begin{equation}
 \Delta^-_{ij}=\Delta_i-\Delta_j,
 \qquad
 \Delta^-_{56}=\Delta^-_{12}+\Delta^-_{34}.
 \label{eq:Q-charge-differences}
\end{equation}
The deformation dictionary is
\begin{equation}
 \begin{aligned}
 D_1&=\Delta_2+\Delta_4-\frac12
     =\frac12-\Delta_5,
 \\
 D_2&=\Delta_6-\frac12
     =\frac12-\Delta_1-\Delta_3,
 \\
 D_{\rm m}
 &=\frac12\left(\Delta^-_{34}-\Delta^-_{12}\right).
 \end{aligned}
 \label{eq:Q-deformations}
\end{equation}
In particular, $D_1+D_2=-\Delta^-_{56}$.
The redundancy discussed in \cite{Hristov:2026zjh,Kubo:2024qhq,Kubo:2025dot} permits
$D_1=D_2$; we keep both parameters to display the original
dictionary. Only their sum enters the classical symbol below.

At leading order, the two factors multiply as ordinary functions.
Writing the old variables in terms of new canonical variables as
$x_{\rm old}=p-x$ and $p_{\rm old}=-x$ then gives
\begin{equation}
 O_{0,Q}
 =
 e^{\Delta^-_{12}x+\Delta^-_{34}p}
 (1+e^x+e^p)(1+e^{-x}+e^{-p}).
 \label{eq:Qsymbol}
\end{equation}
The classical trace is therefore
\begin{equation}
 I_{0,Q}(s)
 =\int_{\R^2}\dd x\,\dd p\,
 \frac{
 e^{-s\Delta^-_{12}x-s\Delta^-_{34}p}
 }{
 (1+e^x+e^p)^s(1+e^{-x}+e^{-p})^s
 }.
\end{equation}

Applying \eqref{eq:general-Barnes-two} to the second triangle factor gives
\begin{equation}
 (1+e^{-x}+e^{-p})^{-s}
 =
 \frac{1}{\Gamma(s)}
 \int\frac{\dd \omega_1\,\dd \omega_2}{(2\pi i)^2}\,
 \Gamma(\omega_1)\Gamma(\omega_2)\Gamma(s-\omega_1-\omega_2)e^{\omega_1 x+\omega_2 p}.
 \label{eq:Q-inverse-Mellin}
\end{equation}
For real $s>0$, we choose
$\operatorname{Re}\omega_1=s\Delta_1$ and
$\operatorname{Re}\omega_2=s\Delta_3$.
The six Gamma-function arguments in the resulting Barnes
integral then have positive real parts by \eqref{eq:Q-locus}.
For other $s$, the integral is defined by analytic continuation,
with contour deformations separating the pole families.
With $X=e^x$ and $P=e^p$, the remaining phase integral
is a Dirichlet integral:
\begin{equation}
 \begin{aligned}
 &\int_0^\infty\!\!\int_0^\infty
 \frac{
 X^{\omega_1-s\Delta^-_{12}-1}
 P^{\omega_2-s\Delta^-_{34}-1}
 }{(1+X+P)^s}\,\dd X\,\dd P
 \\
 &\quad=
 \mathrm B\!\left(
 \omega_1-s\Delta^-_{12},
 \omega_2-s\Delta^-_{34}
 \right)
 \\
 &\qquad\quad\times
 \mathrm B\!\left(
 \omega_1+\omega_2-s\Delta^-_{56},
 s(1+\Delta^-_{56})-\omega_1-\omega_2
 \right)
 \\
 &\quad=
 \frac{
 \Gamma(\omega_1-s\Delta^-_{12})
 \Gamma(\omega_2-s\Delta^-_{34})
 \Gamma(s(1+\Delta^-_{56})-\omega_1-\omega_2)
 }{\Gamma(s)}.
 \end{aligned}
 \label{eq:Q-Dirichlet}
\end{equation}
Thus we find
\begin{equation}
 \begin{aligned}
 I_{0,Q}(s)
 =\frac{1}{\Gamma(s)^2}
 &\int\frac{\dd \omega_1\,\dd \omega_2}{(2\pi i)^2}\,
 \Gamma(\omega_1)\Gamma(\omega_2)\Gamma(s-\omega_1-\omega_2)
 \\
 &\times
 \Gamma(\omega_1-s\Delta^-_{12})
 \Gamma(\omega_2-s\Delta^-_{34})
 \Gamma(s(1+\Delta^-_{56})-\omega_1-\omega_2).
 \end{aligned}
 \label{eq:Qbarnes-QC}
\end{equation}

The complete leading WKB grand potential is
\begin{equation}
 J_{0,Q}(\mu)
 =
 -\frac{1}{4\pi^2}\int_{c-i\infty}^{c+i\infty}
 \frac{\dd s}{2\pi i}\,
 \Gamma(s)\Gamma(-s)I_{0,Q}(s)e^{s\mu},
 \qquad 0<c<1.
 \label{eq:Q-grand-potential}
\end{equation}

The leading-WKB instanton sector follows from the negative
poles of the Mellin--Barnes integrand.
The candidate pole families are
\begin{equation}
 s=-\frac{n}{1\pm\Delta^-_{12}},
 \qquad
 s=-\frac{n}{1\pm\Delta^-_{34}},
 \qquad
 s=-\frac{n}{1\pm\Delta^-_{56}},
 \qquad n\geq1.
 \label{eq:Q-instanton-poles}
\end{equation}
Their contributions are
\begin{equation}
 J_{0,\mathrm{exp},Q}(\mu)
 \sim
 -\frac{1}{4\pi^2}\sum_{s_*<0}
 \operatorname*{Res}_{s=s_*}
 \left[
 \Gamma(s)\Gamma(-s)I_{0,Q}(s)e^{s\mu}
 \right],
 \label{eq:Q-instantons}
\end{equation}
where the sum runs over distinct surviving poles of the full
integrand after cancellations. The explicit finite sums for these residues are obtained
in the geometric calculation of Section~\ref{sec:diskexamples},
using the Barnes-pinch identity in Appendix~\ref{app:gamma-mellin}.

The accompanying polynomial terms can be obtained from the
expansion of the same trace at the origin.
To expand around $s=0$, we first take $s>0$ and rescale
$\omega_1=su$, $\omega_2=sv$. This gives
\begin{equation}
 \begin{aligned}
 I_{0,Q}(s)
 =\frac{s^2}{\Gamma(s)^2}
 &\int\frac{\dd u\,\dd v}{(2\pi i)^2}\,
 \Gamma(su)\Gamma(sv)\Gamma(s(1-u-v))
 \\
 &\times
 \Gamma(s(u-\Delta^-_{12}))
 \Gamma(s(v-\Delta^-_{34}))
 \Gamma(s(1+\Delta^-_{56}-u-v)).
 \end{aligned}
 \label{eq:Q-rescaled-Barnes}
\end{equation}
We choose the vertical contours
\begin{equation}
 \operatorname{Re} (u)=\Delta_1,\qquad \operatorname{Re} (v)=\Delta_3.
 \label{eq:Q-contours}
\end{equation}

Define
\begin{equation}
 (\ell_1,\ldots,\ell_6)
 =\bigl(u,v,1-u-v,u-\Delta^-_{12},v-\Delta^-_{34},
 1+\Delta^-_{56}-u-v\bigr).
 \label{eq:Q-Barnes-linear-factors}
\end{equation}
Their sum is two, so the terms linear in $s$ cancel in the
Gamma-function expansion. Appendix~\ref{app:Qgeneraljets} justifies the extraction
of the singular Laurent coefficients by analytic
continuation; the expansion below is understood
coefficient by coefficient in this sense. Evaluating
these coefficients by Barnes residues gives 
\begin{equation}
 \begin{aligned}
 I_{0,Q}(s)
 =\frac{1}{s^2}
 \int\frac{\dd u\,\dd v}{(2\pi i)^2}
 \frac{1}{\prod_{a=1}^6\ell_a}
 \Biggl[1
 &+\frac{\zeta(2)s^2}{2}
 \left(\sum_{a=1}^6\ell_a^2-2\right)-\frac{\zeta(3)s^3}{3}
 \left(\sum_{a=1}^6\ell_a^3-2\right)
 \Biggr]\!+O(s^2).
 \end{aligned}
 \label{eq:Q-integrand-expansion}
\end{equation}
Closing the $u$-contour to the left picks the poles
$u=0$ and $u=\Delta^-_{12}$.
The remaining $v$-integral similarly picks
$v=0$ and $v=\Delta^-_{34}$.

Adding the residues at each order yields
\begin{equation}
 \begin{aligned}
 I_{0,Q}(s)
 =\Biggl\{
 &\frac{
 6-(\Delta^-_{12})^2-(\Delta^-_{34})^2-(\Delta^-_{56})^2
 }{
 s^2
 [1-(\Delta^-_{12})^2]
 [1-(\Delta^-_{34})^2]
 [1-(\Delta^-_{56})^2]
 }
 \\
 &-2\zeta(2)
 \left[
 \frac{1}{1-(\Delta^-_{12})^2}
 +\frac{1}{1-(\Delta^-_{34})^2}
 +\frac{1}{1-(\Delta^-_{56})^2}
 \right]
 \\
 &+2\zeta(3)s
 \left[
 \frac{1}{1-(\Delta^-_{12})^2}
 +\frac{1}{1-(\Delta^-_{34})^2}
 +\frac{1}{1-(\Delta^-_{56})^2}
 -1
 \right]
 +O(s^2)
 \Biggr\}.
 \end{aligned}
 \label{eq:Q-leading-trace-expansion}
\end{equation}

The residue at $s=0$ gives

\begin{equation}
 \begin{aligned}
 C_{0,Q}
 &=
 \frac{
 6-(\Delta^-_{12})^2-(\Delta^-_{34})^2-(\Delta^-_{56})^2
 }{
 8\pi^2
 [1-(\Delta^-_{12})^2]
 [1-(\Delta^-_{34})^2]
 [1-(\Delta^-_{56})^2]
 },
 \\[6pt]
 B_{0,Q}
 &=
 \frac{\pi^2}{6}
 \left[
 (\Delta^-_{12})^2
 +(\Delta^-_{34})^2
 +(\Delta^-_{56})^2
 \right]C_{0,Q},
 \\[6pt]
 A_{0,Q}
 &=
 \frac{\zeta(3)}{2\pi^2}
 \left[
 \frac{1}{1-(\Delta^-_{12})^2}
 +\frac{1}{1-(\Delta^-_{34})^2}
 +\frac{1}{1-(\Delta^-_{56})^2}
 -1
 \right].
 \end{aligned}
 \label{eq:Q-leading-constant}
\end{equation}

\section{Disk periods and the Mellin transform}
\label{sec:disk}

We now recover the complete classical spectral traces of
Section~\ref{sec:QC} from the transverse fourfolds.
Our geometric input is the equivariant disk partition function.
Its volume limit accounts for the leading large-rank behavior,
while its full dependence on the equivariant parameters supplies
the information needed for the subleading, constant and exponentially
suppressed terms in $J_0$. 
We first define this observable and explain its geometric
meaning, then specify the K\"ahler parameters used in the
comparison and establish the matches example by example.

A two-dimensional gauged linear sigma model (GLSM) presents
a toric target as a gauge quotient
\cite{Witten:1993yc,Morrison:1994fr}.
Supersymmetric localization on a disk or hemisphere reduces
its partition function to an integral over constant
vector-multiplet scalars.
The chiral multiplets contribute Gamma functions, while
the boundary condition determines the boundary factors
and the admissible integration contour
\cite{Hori:2013ika,Honda:2013uca,Sugishita:2013jca}.
The connections with Gamma classes and quantum cohomology
are developed in
\cite{Hori:2013ika,Iritani:2009qgl,Iritani:2023ngp},
and analytic continuation of hemisphere amplitudes is
studied in \cite{Knapp:2016rec}.
Here we use the equivariant disk observable of
\cite{Cassia:2022lfj}.

Consider a toric Calabi--Yau quotient
$X=\C^n//U(1)^r$, with charge matrix $Q_i^a$,
$n-r=4$ and $\sum_iQ_i^a=0$ for every $a$.
The equivariant weights $\eps_i$ describe torus rotations
of the homogeneous coordinates.
The integration variables $\phi_a$ are the complex
vector-multiplet scalars; geometrically, they represent
the degree-two classes associated with the quotient
$U(1)$ factors.
The combinations
\begin{equation}
 x_i=\eps_i+\sum_{a=1}^r Q_i^a\phi_a,
 \qquad
 t^a=\sum_{i=1}^n Q_i^a\lambda^i
 \label{eq:glsmweights}
\end{equation}
then have a direct geometric interpretation.
The $x_i$ represent the equivariant first Chern classes
of the line bundles associated with the homogeneous
coordinates
\cite{Cassia:2025aus,Cassia:2025jkr};
they are distinct from the phase-space coordinate $x$
of Section~\ref{sec:QC}.
The $t^a$ are the GLSM Fayet--Iliopoulos parameters,
which describe the K\"ahler class of the quotient:
in a geometric chamber, their real parts control
the sizes of its compact curves.
They are the bare K\"ahler parameters, whose relation
to flat coordinates will enter the prescription below.

The $\lambda^i$ provide a redundant description of these
K\"ahler parameters.
Indeed,
$\sum_i\lambda^i x_i
=\sum_i\lambda^i\eps_i+\sum_a t^a\phi_a$,
so combinations of the $\lambda^i$ that leave $t^a$
unchanged contribute only an overall equivariant
exponential.
We retain this dependence because it will supply
the chemical potential in the comparison with field theory.

Let $z$ be the equivariant weight of disk rotation.
With the normalization of
\cite[Definition~3.1]{Cassia:2022lfj}
and the equivariant exponential included, we define
\begin{equation}
 \mathbb D_X(\lambda,\eps;z)
 :=z^{-n}\oint_{\mathrm{QJK}}
 \prod_{a=1}^r\frac{\dd\phi_a}{2\pi i}\,
 e^{\sum_i\lambda^i x_i}
 \prod_{i=1}^n\Gamma(x_i/z).
 \label{eq:disk}
\end{equation}
Thus
$\mathbb D_X(\lambda,\eps;z)
=e^{\sum_i\lambda^i\eps_i}\mathcal F^D(t,\eps;z)$,
with $t=Q\lambda$, in the notation of
\cite{Cassia:2022lfj}.\footnote{
The scalar parameter denoted by $\lambda$ in
\cite{Cassia:2022lfj} is our $z$; it is distinct from
our parameters $\lambda^i$.
Their degree-counting variables $z_a$ are denoted
by $q_a$ here.}

The quantum Jeffrey--Kirwan prescription, abbreviated QJK,
extends the ordinary Jeffrey--Kirwan residue prescription
by including the towers of Gamma-function poles
$x_i=-mz$, with $m\geq0$, selected by the geometric chamber.
The residues organize a sum over curve degrees $d$,
weighted by
$\exp(-z\sum_a d_a t^a)$.
Appendix~\ref{app:QJK} explains the positive integral
representation of this contour and its normalization.
We use this prescribed contour without an additional
boundary insertion and analytically continue the resulting
function whenever its degree expansion does not converge
at the point of interest.

We also call $\mathbb D_X$ the disk period.
A geometric period is an integral of a differential form
over a specified cycle, viewed as a function of moduli.
In mirror symmetry, disk amplitudes admit a description
as oscillatory integrals, whose integrands contain an
exponential of the mirror superpotential
\cite[Eq.~(7.35)]{Hori:2000kt};
their relation to hemisphere partition functions is discussed
in \cite[Eq.~(2), Section~8 and Remark~15]{Iritani:2023cxf}.
In our setting, the dependence on the K\"ahler parameters
satisfies the equivariant Picard--Fuchs equations,
determined by the charge matrix
\cite[Section~7.1]{Cassia:2022lfj},
and the contour selects the solution used in our calculations.
Throughout this paper, ``disk period'' and
``disk partition function'' denote the same observable.

The relation to equivariant volume follows by taking
$z\to+\infty$ with $\lambda$ and $\eps$ fixed and
$t=Q\lambda$ in the interior of a geometric chamber.
Since $z^{-1}\Gamma(x_i/z)\to x_i^{-1}$,
\begin{equation}
 \lim_{z\to+\infty}\mathbb D_X(\lambda,\eps;z)
 =\mathbb V_X(\lambda,\eps)
 :=\oint_{\mathrm{JK}}
 \prod_{a=1}^r\frac{\dd\phi_a}{2\pi i}\,
 \frac{e^{\sum_i\lambda^i x_i}}{\prod_i x_i}.
 \label{eq:disk-volume-limit}
\end{equation}
This is the equivariant volume studied in
\cite{Nekrasov:2021ked,Cassia:2022lfj}, written in the conventions of
\cite{Cassia:2025aus}.
The degree-zero residues become the ordinary JK residues,
while nonzero effective degrees are suppressed by
$\exp(-z\sum_a d_a t^a)$
\cite[Eqs.~(5.9)--(5.10)]{Cassia:2022lfj}.
The disk partition function is thus a quantum
generalization of equivariant volume.
A change of integration variables also gives the exact
scaling identity
\begin{equation}
 \mathbb D_X(\lambda,\eps;z)
 =z^{-4}\mathbb D_X(z\lambda,\eps/z;1).
 \label{eq:main-disk-scaling}
\end{equation}

The degree expansion connects the disk partition function
to genus-zero equivariant Gromov--Witten theory.
The GLSM charges determine the toric $I$-function,
a generating series that encodes Gromov--Witten data
after the mirror map from bare to flat K\"ahler
coordinates
\cite{Givental1998,Cassia:2022lfj}.
The foundations of equivariant Gromov--Witten theory
and localization on moduli spaces of stable maps are
developed in \cite{Givental:1996heu,Graber:1997tvy}.
Mirror theorems relate these invariants to hypergeometric
series \cite{Givental1998,Lian:1997fkq}, while quantum
Riemann--Roch describes characteristic-class twists
\cite{Coates:2001ewh}.
Related GLSM constructions use sphere and vortex
partition functions
\cite{Benini:2012ui,Doroud:2012xw,Jockers:2012dk,Bonelli:2013mma}.
For our comparison, the essential point is that the full
disk includes the positive-degree contributions whenever
they are present.
We evaluate their sum through \eqref{eq:disk}, without
extracting individual Gromov--Witten invariants.
The Gamma factors themselves also admit a regularized
disk-map interpretation \cite{Gerasimov:2009eb},
although this does not explain the additional Mellin
transform used below.

Even without compact curves, the disk contains more
information than equivariant volume.
For $\C^4$, the Gamma product in
\eqref{eq:intro-C4-disk} is the complete answer.
Replacing each Gamma factor by its leading pole recovers
the volume but discards its remaining dependence on the
equivariant weights.
After the Gamma normalization and Mellin transform
in \eqref{eq:intro-master}, the full expression reproduces
$J_0$ on the stated QC loci, including its constant term
and membrane-instanton series.
For resolved singularities, the comparison likewise
uses the complete disk, with the additional prescription that follows.

\subsection{The geometric prescription}
\label{sec:flat-mesonic}

We first specify the quantum mesonic restriction on the
K\"ahler parameters and call the disk evaluated there
the mesonic disk.
We then define its Mellin transform, which determines
both polynomial and exponentially small terms.
The construction is geometric: the restrictions defining
a QC locus enter only in the comparison with field theory.

\paragraph{The quantum mesonic restriction.}

The disk integral is expressed in terms of the bare
GLSM parameters $t^a=\sum_iQ_i^a\lambda^i$.
The flat K\"ahler coordinates $T^a$ are determined by
logarithmic periods of the mirror geometry, normalized
to agree with the bare coordinates at large radius
\cite{Givental1998,Cassia:2022lfj}.
Their relation is the mirror map, which includes
corrections from positive-degree maps.
Keeping the disk weight $z$ explicit, we write
\begin{equation}
 zT^a=zt^a+f^a(q_1,\ldots,q_r),
 \qquad q_a=e^{-zt^a},
 \qquad f^a(0,\ldots,0)=0.
 \label{eq:flat-coordinate-definition}
\end{equation}
The corresponding flat degree variables are
$\widetilde q_a=e^{-zT^a}$.
These conventions follow
\cite[Sections~3.1 and 9.1]{Cassia:2022lfj}.

The ALE mirror maps are described in
\cite{Lau:2010jf}, while a flat-coordinate
parametrization of the SPP mirror curve is given in
\cite[Section~3.2]{Ooguri:2010yk}.
The disk calculations and mirror maps for the other
resolved geometries considered here are presented in
\cite[Sections~10.1--10.2 and 11.1--11.2]{Cassia:2022lfj};
see also \cite{Chan:2013dma} for the general
enumerative interpretation.
We state the required maps alongside the examples
and use them to locate the mesonic specialization.

We impose the \emph{quantum mesonic restriction}
by setting all flat K\"ahler coordinates to zero,
on the branch obtained by analytic continuation
from large radius:
\begin{equation}
 T^a=0,\qquad a=1,\ldots,r.
 \label{eq:intro-flat-condition}
\end{equation}
This is the counterpart, in flat coordinates, of the
classical mesonic restriction used for equivariant volume
in \cite{Cassia:2025aus,Cassia:2025jkr}, following the
mesonic-twist construction of \cite{Hosseini:2019ddy}.
Here ``quantum'' refers to the mirror map, independently
of the spectral WKB expansion.
In particular, \eqref{eq:intro-flat-condition} need not
set the bare parameters $t^a$ to zero.

Let $t_\star^a$ denote the bare parameters corresponding
to $T=0$ at unit disk weight, $z = 1$, on the chosen branch.
Choose a fixed representative $\lambda_\star$ satisfying
\begin{equation}
 \sum_iQ_i^a\lambda_\star^i=t_\star^a.
 \label{eq:Pi}
\end{equation}
The remaining freedom in $\lambda_\star$ fixes the
equivariant exponential and, consequently, the origin
of the chemical potential. We specify this choice in
each example.\footnote{
A change $\lambda_\star\mapsto\lambda_\star+\delta\lambda$
with $Q\delta\lambda=0$ multiplies the mesonic disk by
$e^{s\,\delta\lambda\cdot\eps}$ and shifts the transformed
function to
$J_{\mathrm{disk}}(\mu+\delta\lambda\cdot\eps;\eps)$.
The representative is therefore part of the normalization
used in the spectral comparison.}
At general disk weight, the same coefficient point is
represented by $\lambda=\lambda_\star/z$.
Indeed, this keeps $zt^a=t_\star^a$ and hence
$q_a=e^{-t_\star^a}$ fixed.

We now introduce a complex Mellin variable $s$.
Motivated by the spectral calculation in
Section~\ref{sec:QC} and the matches below, we identify
the inverse disk rotation weight with this variable:
\begin{equation}
 z=s^{-1}.
 \label{eq:disk-mellin-variable}
\end{equation}
Keeping the coefficient point fixed then requires
$\lambda=s\lambda_\star$, while the target weights
$\eps$ remain unchanged.
We therefore define the mesonic disk by
\begin{equation}
 \mathbb D_X^{\text{mes}}(s,\eps)
 :=\mathbb D_X(s\lambda_\star,\eps;s^{-1})
 =s^4\mathbb D_X(\lambda_\star,s\eps;1).
 \label{eq:mesonic-new-definition}
\end{equation}
The last equality follows from
\eqref{eq:main-disk-scaling}.
The specialization is first made for positive real $s$
and then analytically continued with the chosen disk cycle.
The flat degree variables are $e^{-zT^a}$, so setting
$T=0$ does not suppress positive degrees.
The complete function is analytically continued when
its degree expansion does not converge at that point.

\paragraph{The Mellin transform.}

Following the single-chemical-potential prescription
motivated and developed in \cite{Cassia:2025jkr},
we shift every component of $\lambda_\star$ by $\mu/2$
and impose $\sum_i\eps_i=2$.
The Calabi--Yau condition $Q\mathbf1=0$ preserves the
K\"ahler specialization, while $\sum_i x_i=2$ gives
\begin{equation}
 \mathbb D_X\left(s(\lambda_\star+\tfrac\mu2\mathbf1),\eps;s^{-1}\right)
 =e^{s\mu}\mathbb D_X^{\text{mes}}(s,\eps).
 \label{eq:mesonic}
\end{equation}
The factor $e^{s\mu}$ therefore follows directly
from the shifted disk.

The geometric transform proposed in
\eqref{eq:intro-master} can therefore be written as
\begin{equation}
 J_{\mathrm{disk}}(\mu;\eps)
 :=-\frac1{4\pi^2}\int_{c-i\infty}^{c+i\infty}
 \frac{\dd s}{2\pi i}\,
 \frac{\Gamma(-s)}{s^4\Gamma(s)}
 \mathbb D_X^{\text{mes}}(s,\eps)e^{s\mu}.
 \label{eq:geometric-J-definition}
\end{equation}
The contour is initially chosen in a common convergence
strip with $0<c<1$, and the result is continued where
necessary.
Equivalently, the disk is dressed by
$[s^4\Gamma(s)^2]^{-1}$ and integrated against the
spectral Mellin kernel $\Gamma(s)\Gamma(-s)$.
This prescription is motivated by the spectral calculation.
Its intrinsic geometric origin remains an open question:
the identities proved here do not yet provide a geometric
or bulk derivation of the integration kernel.

The limit $s\to0$ recovers the classical mesonic volume
in the conventions of \cite{Cassia:2025aus}:
\begin{equation}
 \lim_{s\to0^+}\mathbb D_X^{\text{mes}}(s,\eps)
 =\mathbb V_X(0,\eps)
 =:C_X(\eps).
 \label{eq:volume-CX}
\end{equation}
The positive integral representation in App.\
\ref{app:QJK} establishes this limit
in its convergence domain, with $\lambda=s\lambda_\star$
and $z=s^{-1}$. It concerns the complete mesonic disk,
with its degree contributions already summed.

On the QC loci, we use the covering-space normalization
when a residual finite quotient is present.
The comparison with Section~\ref{sec:QC} then takes the form
\begin{equation}
 \boxed{\displaystyle
 \left.
 \frac{\mathbb D_X^{\text{mes}}(s,\eps)}
 {s^4\Gamma(s)^2}
 \right|_{\eps_i=\Delta_i}
 =I_0(s;\Delta).
 }
 \label{eq:disk-trace-match}
\end{equation}
This is the spectral-trace form of
\eqref{eq:intro-holographic-match}, so checking it
suffices to establish that match.
With this normalization, the common contour and measure
imply $J_{\mathrm{disk}}=J_0$.
The equality includes the constant and exponentially
small terms of the complete leading WKB functions.

\paragraph{Geometric predictions at general weights.}

The definition \eqref{eq:geometric-J-definition}
does not require a QC locus.
For several geometries, the mesonic disk evaluates
to a finite quotient of Gamma functions, yielding
explicit predictions for the polynomial and exponentially
small terms of $J_{\mathrm{disk}}$.
Suppose the result takes the form
\begin{equation}
 \mathbb D_X^{\text{mes}}(s,\eps)
 =C_X(\eps)
\frac{\prod_{I=1}^{M}\Gamma(1+s\nu_I)}
 {\prod_{A=1}^{M-4}\Gamma(1+s\kappa_A)}.
 \label{eq:general-geometric-period}
\end{equation}
Here $\nu_I$ and $\kappa_A$ are linear combinations
of the equivariant weights, to be derived for each example. In the examples admitting this factorization,
they are the weights of the specified invariant
generators and their relations.

These weights obey
\begin{equation}
 \sum_{I=1}^{M}\nu_I-\sum_{A=1}^{M-4}\kappa_A
 =\sum_i\eps_i=2.
 \label{eq:disk-weight-balance}
\end{equation}
Using the expansion of $\log\Gamma(1+x)$, this identity
cancels the linear Euler-constant contribution
from the disk against that of the Mellin kernel.
After removing $e^{s\mu}$, the integrand in \eqref{eq:geometric-J-definition} therefore
has no $s^{-3}$ term, and $J_{\mathrm{disk}}$
has no quadratic term in $\mu$.
The resulting decomposition is
\begin{equation}
 J_{\mathrm{disk}}(\mu;\eps)
 =\frac{C_{\mathrm{disk}}(\eps)}{3}\mu^3
  +B_{\mathrm{disk}}(\eps)\mu+A_{\mathrm{disk}}(\eps)
  +J_{\mathrm{disk},\mathrm{exp}}(\mu;\eps),
 \label{eq:disk-polynomial-decomposition}
\end{equation}
with
\begin{equation}
 \begin{aligned}
 C_{\mathrm{disk}}(\eps)
 &=\frac{C_X(\eps)}{8\pi^2},\qquad
 B_{\mathrm{disk}}(\eps)
 =\frac{C_X(\eps)}{48}
 \left(\sum_I\nu_I^2-\sum_A\kappa_A^2\right),\\
 A_{\mathrm{disk}}(\eps)
 &=\frac{\zeta(3)C_X(\eps)}{12\pi^2}
 \left(2-\sum_I\nu_I^3+\sum_A\kappa_A^3\right).
 \end{aligned}
 \label{eq:geometric-polynomial}
\end{equation}
These are coefficients of the geometric transform
at general weights.
On a QC locus, with the normalization of
\eqref{eq:disk-trace-match}, they equal $C_0,B_0,A_0$.

For positive real $\nu_I$, a surviving pole at
$s=-n/\nu_I$ contributes an exponentially small term
$e^{-n\mu/\nu_I}$.
For separated simple poles,
\begin{align}
 J_{\mathrm{disk},\mathrm{exp}}(\mu;\eps)
 &\sim\sum_{I=1}^{M}\sum_{n\geq1}
 c^X_{I,n}(\eps)e^{-n\mu/\nu_I},
 \nonumber\\
 c^X_{I,n}(\eps)
 &=\frac{C_X(\eps)\prod_J\nu_J}{\prod_A\kappa_A}
 \frac{(-1)^{n+1}}{4\pi^2n!\nu_I}
 \frac{\Gamma(n/\nu_I)}{\Gamma(-n/\nu_I)}
 \frac{\prod_{J\ne I}\Gamma(-n\nu_J/\nu_I)}
 {\prod_A\Gamma(-n\kappa_A/\nu_I)}.
 \label{eq:general-geometric-residues}
\end{align}
Empty products equal one.
The symbol $\sim$ denotes the large-$\mu$ residue
expansion; the contour integral
\eqref{eq:geometric-J-definition} defines the
function itself. The resulting expressions are continued in the weights
where appropriate.
Zeros of reciprocal Gamma functions can remove
candidate poles.
When poles coincide, their combined residue can
produce a polynomial in $\mu$ multiplying the
exponential, according to \eqref{eq:collisionrule}.

The disk calculations below establish
\eqref{eq:general-geometric-period} in every example
except the resolved cone over $Q^{1,1,1}$.
For that geometry, we retain a Barnes integral
and extract its polynomial and exponentially small
terms directly from that representation.

\subsection{The fourfold examples}
\label{sec:diskexamples}

Our calculations rely extensively on the systematic
treatment of equivariant disk partition functions
by Cassia, Piazzalunga and Zabzine
\cite{Cassia:2022lfj}.
Their explicit residue expansions and mirror maps
provide much of the geometric input below.
We use these results in our normalization and
concentrate on the additional steps needed here:
evaluating the complete disk at the quantum mesonic
point, extracting its Mellin-transform predictions,
and establishing the spectral matches.

Each fourfold example follows this order.
We first describe the geometry and disk integral,
then give the mesonic disk and its predictions
at general equivariant weights, and finally impose
the QC restrictions.
For the ALE geometries, we first collect the surface
calculations and then form the relevant fourfold
products.

In analogy to the field-theory notation, we abbreviate sums of
weights by $\eps_{ij\cdots\ell}:=\eps_i+\eps_j+\cdots+\eps_\ell$.
For the general disk formulas we retain the disk rotation weight $z$,
$c_i=e^{-z\lambda^i}$ and $q_a=e^{-zt^a}$, with
$t^a=\sum_iQ_i^a\lambda^i$.
We use $\widetilde q_a=e^{-zT^a}$ for the flat
degree variables. We occasionally suppress the equivariant-weight arguments
of the volume and polynomial coefficients.

\subsubsection{$\C^4/\Z_k$}
\label{sec:c4geometry}

\paragraph{Geometry and disk integral.}

The covering space $\C^4$ has no gauge contour,
no effective K\"ahler parameters and no
compact-curve degree sum.
Its general disk
\eqref{eq:intro-C4-disk} is
\begin{equation}
 \mathbb D_{\C^4}(\lambda,\eps;z)
 =z^{-4}e^{\sum_{i=1}^4\lambda^i\eps_i}
  \prod_{i=1}^4\Gamma(\eps_i/z),
 \qquad
 \mathbb D_{\C^4/\Z_k}(\lambda,\eps;z)
 =\frac1k\mathbb D_{\C^4}(\lambda,\eps;z).
 \label{eq:C4-general-disk}
\end{equation}
For the quotient, we use the untwisted equivariant
pushforward, which multiplies the covering-space
answer by $1/k$ with the covering weights held fixed.
This quotient normalization is explained in \eqref{eq:finite-quotient-measure}.

\paragraph{Mesonic disk and geometric prediction.}

There are no effective K\"ahler parameters to
specialize, and we choose $\lambda_\star=0$.
The covering-space and quotient mesonic disks are
\begin{equation}
 \mathbb D_{\C^4}^{\text{mes}}(s,\eps)
 =s^4\prod_{i=1}^4\Gamma(s\eps_i),
 \qquad
 \mathbb D_{\C^4/\Z_k}^{\text{mes}}(s,\eps)
 =\frac{s^4}{k}\prod_{i=1}^4\Gamma(s\eps_i).
 \label{eq:C4-quotient-disk}
\end{equation}
For positive weights with $\eps_{1234}=2$,
\begin{equation}
 J_{\mathrm{disk},\C^4/\Z_k}(\mu;\eps)
 =-\frac1{4\pi^2}
 \int_{c-i\infty}^{c+i\infty}\frac{\dd s}{2\pi i}\,
 \frac{\Gamma(-s)}{s^4\Gamma(s)}
 \mathbb D_{\C^4/\Z_k}^{\text{mes}}(s,\eps)e^{s\mu}.
 \label{eq:C4-Mellin}
\end{equation}
Using the mesonic equivariant volume
$C_{\C^4/\Z_k}=1/(k\prod_i\eps_i)$,
the polynomial coefficients are
\begin{equation}
 \begin{aligned}
 C_{\mathrm{disk}}
 =\frac{C_{\C^4/\Z_k}}{8\pi^2},
 \quad
 B_{\mathrm{disk}}
 =\frac{C_{\C^4/\Z_k}}{48}\sum_i\eps_i^2,\quad
 A_{\mathrm{disk}}
 =\frac{\zeta(3)C_{\C^4/\Z_k}}{12\pi^2}
 \left(2-\sum_i\eps_i^3\right).
 \end{aligned}
 \label{eq:C4jets}
\end{equation}
For separated surviving poles,
\begin{align}
 J_{\mathrm{disk},\C^4/\Z_k,\mathrm{exp}}
 &\sim\sum_{i=1}^4\sum_{n\geq1}
 c_{i,n}^{\C^4/\Z_k}e^{-n\mu/\eps_i},
 \nonumber\\
 c_{i,n}^{\C^4/\Z_k}
 &=\frac{(-1)^{n+1}}{4\pi^2k\,n!\eps_i}
 \frac{\Gamma(n/\eps_i)}{\Gamma(-n/\eps_i)}
 \prod_{j\ne i}\Gamma(-n\eps_j/\eps_i).
 \label{eq:C4-disk-exp}
\end{align}
The exponentially small terms arise from the
Gamma factors despite the absence of compact curves.

\paragraph{Match on the QC loci.}

On the pairwise ABJM locus,
$\eps_i=\Delta_i$ and
$\eps_{12}=\eps_{34}=1$.
The covering-space disk satisfies
\begin{equation}
 \frac{\mathbb D_{\C^4}^{\text{mes}}(s,\eps)}{s^4\Gamma(s)^2}\Big|_{\mathrm{QC}}
 =\frac{\prod_{i=1}^4\Gamma(s\Delta_i)}{\Gamma(s)^2}
 =I_{0,\square}(s).
 \label{eq:C4-pairwise-match}
\end{equation}
Consequently $J_{\mathrm{disk},\C^4}\Big|_{\mathrm{QC}}=J_{0,\square}$.
Since $\hbar=k$ and $J=\hbar^{-1}J_0+\cdots$,
the quotient-normalized disk instead gives
$J_{\mathrm{disk},\C^4/\Z_k}\Big|_{\mathrm{QC}}=J_{0,\square}/k$.

At the undeformed point $\eps_i=1/2$,
\begin{align}
 J_{0,\square}(\mu)
 ={}&\frac{2\mu^3}{3\pi^2}+\frac{\mu}{3}
 +\frac{2\zeta(3)}{\pi^2}
 \nonumber\\
 &+\left[
 \frac{-4\mu^2+4\mu+4}{\pi^2}-\frac23
 \right]e^{-2\mu}
 +O(\mu^2e^{-4\mu}).
 \label{eq:c4firstinst}
\end{align}
The pole at $s=-2$ has order three after the
reciprocal-Gamma cancellation, giving the
quadratic instanton prefactor.

On the triangular locus at level one,
$\eps_4=1$ and $\eps_{123}=1$.
With $\eps_i=\Delta_i$,
\begin{equation}
 \frac{\mathbb D_{\C^4}^{\text{mes}}(s,\eps)}{s^4\Gamma(s)^2}\Big|_{\mathrm{QC}}
 =\frac{\prod_{i=1}^3\Gamma(s\Delta_i)}{\Gamma(s)}
 =I_{0,\triangle}(s).
 \label{eq:C4-triangular-match}
\end{equation}
The unit-weight family cancels, leaving the
three families in \eqref{eq:intro-qc-families}.
Thus $J_{\mathrm{disk},\C^4}\Big|_{\mathrm{QC}}=J_{0,\triangle}$.

\subsubsection{ALE resolutions and their products}
\label{sec:ALE-disk}
Let $A_{p-1}$ denote the minimal crepant resolution of $\C^2/\Z_p$, in particular $A_1=\operatorname{Tot}(\mathcal O_{\mathbb P^1}(-2))$. We first evaluate the surface disk and express its mesonic specialization as the Gamma quotient \eqref{eq:ALE-mesonic-surface}. We then use this result to construct the fourfold disks for $A_{p-1}\times\C^2$ and $(A_{\xi-1}\times A_{\zeta-1})/\Z_k$. The former involves an additional free $\C^2$ factor, while the latter combines two ALE surface disks. We finally compare these fourfold results with their field-theory counterparts.

\paragraph{Geometry and disk integral.}
Write $u,v$ for the two covering-space weights
of the surface.
The homogeneous-coordinate weights
$\widetilde\eps_j$, $j=0,\ldots,p$, obey
\begin{equation}
 \sum_{j=0}^{p}\widetilde\eps_j=u+v,
 \qquad
 \sum_{j=0}^{p}j\widetilde\eps_j=pu.
 \label{eq:ALE-weight-dictionary}
\end{equation}
We express the surface disk directly in these
covering weights and write
$\mathbb D_{A_{p-1}}(\lambda,(u,v);z)$,
where $\lambda=(\lambda^0,\ldots,\lambda^p)$.
Its quantum mesonic specialization is denoted
by $\mathbb D_{A_{p-1}}^{\text{mes}}(s,(u,v))$. For these complex surfaces, the scaling relation in
\eqref{eq:mesonic-new-definition} carries $s^2$ in place
of $s^4$.

\paragraph{$\boldsymbol{A_1}$ resolution.} For $A_1$, the charge row is
\begin{equation}
 Q_{A_1}=(1,-2,1).
 \label{eq:A1-charge}
\end{equation}
In the positive chamber, the QJK prescription selects the first and
third coordinates.
The Coulomb integral and its complete residue expansion
are given in \cite[Eqs.~(10.8) and (10.11)]{Cassia:2022lfj}.
Here
\begin{equation}
 t=\lambda^0-2\lambda^1+\lambda^2,
 \qquad
 q=e^{-zt}=\frac{c_0c_2}{c_1^2}.
\end{equation}
Applying the positive integral representation
\eqref{eq:disk-positive-integral}, solving the charge
constraint and integrating the radial variable gives
the equivalent expression \footnote{For comparison with the homogeneous-coordinate formulas,
a shift of the Coulomb variables allows us to choose
the weight representative $(v,0,\ldots,0,u)$.
The equivariant exponential is then
$e^{v\lambda^0+u\lambda^p}$.
Our coordinate index $j$ corresponds to index $j+1$
in \cite{Cassia:2022lfj}.}
\begin{equation}
 \mathbb D_{A_1}(\lambda,(u,v);z)
 =z^{-2}e^{v\lambda^0+u\lambda^2}\Gamma((u+v)/z)
 \int_0^\infty
 \frac{U^{2u/z-1}\,\dd U}{(1+bU+U^2)^{(u+v)/z}},
 \label{eq:A1-general-disk}
\end{equation}
where $b=\frac{c_1}{\sqrt{c_0c_2}}=e^{zt/2}$.\\

The polynomial $P_2(U):=1+bU+U^2$ appearing in the disk integrand also defines the mirror surface $\xi\eta=P_2(U)$.
This correspondence extends to the general ALE resolutions.

We now express the disk partition function in terms of the flat K\"ahler coordinate. In our conventions, the mirror map of \cite[Eqs.~(10.33)--(10.34)]{Cassia:2022lfj} takes the form
\begin{equation}
 \widetilde q=e^{-zT}
 =\frac{1-\sqrt{1-4q}}{1+\sqrt{1-4q}},
 \qquad
 q=\frac{\widetilde q}{(1+\widetilde q)^2},
 \qquad
 b=\frac{1+\widetilde q}{\sqrt{\widetilde q}}.
 \label{eq:A1-main-map}
\end{equation}

The residue expansion of the same disk partition function, given in \cite[Eq.~(10.11)]{Cassia:2022lfj}, can then be expressed in flat coordinates using the identities \eqref{eq:general-quadratic-transformation} and \eqref{eq:A1-quadratic-transformation}. This gives
\begin{align}
 \mathbb D_{A_1}(\lambda,(u,v);z)
 ={}&z^{-2}e^{v\lambda^0+u\lambda^2}
 \Biggl[
 \Gamma(2u/z)\Gamma((v-u)/z)\widetilde q^{u/z}
 \sum_{d=0}^{\infty}
 \frac{(2u/z)_d((u+v)/z)_d}{d!\,(1+(u-v)/z)_d}
 \widetilde q^{\,d}
 \nonumber\\
 &\hspace{45mm}+(u\leftrightarrow v)
 \Biggr].
 \label{eq:A1-flat-degree-series}
\end{align}
Here $(a)_d=\Gamma(a+d)/\Gamma(a)$, and the exchange acts only inside the brackets. The contribution proportional to $\widetilde q^{u/z}$ begins as
\begin{equation}
 \widetilde q^{u/z}
 \left[
 1+\frac{2u(u+v)}{z(z+u-v)}e^{-zT}
 +O(e^{-2zT})
 \right].
 \label{eq:A1-first-degree}
\end{equation}

The degree sums in \eqref{eq:A1-flat-degree-series} are Gauss
hypergeometric functions. Applying the connection formula
\eqref{eq:hypergeometric-connection}, the two chart
contributions combine into
\begin{align}
 \mathbb D_{A_1}(\lambda,(u,v);z)
 ={}&z^{-2}e^{v\lambda^0+u\lambda^2}
 \frac{
 \Gamma((u+v)/z)\Gamma(2u/z)\Gamma(2v/z)}
 {\Gamma(2(u+v)/z)}
 \nonumber\\
 &\times
 \widetilde q^{u/z}
 {}_2F_1\!\left(
 \frac{2u}{z},\frac{u+v}{z};
 \frac{2(u+v)}{z};1-\widetilde q
 \right).
 \label{eq:A1-combined-hypergeometric}
\end{align}

The restriction $T=0$ sets $\widetilde q=1$.
Consequently,
\begin{equation}
 \left.
 \mathbb D_{A_1}(\lambda,(u,v);z)
 \right|_{T=0}
 =
 z^{-2}e^{v\lambda^0+u\lambda^2}
 \frac{
 \Gamma((u+v)/z)\Gamma(2u/z)\Gamma(2v/z)}
 {\Gamma(2(u+v)/z)}.
 \label{eq:A1-mesonic-evaluation}
\end{equation}

\paragraph{$\boldsymbol{A_2}$ resolution.}
For $A_2$, the charges are
\begin{equation}
 Q_{A_2}
 =\begin{pmatrix}
 1&-2&1&0\\
 0&1&-2&1
 \end{pmatrix}.
 \label{eq:A2-charge}
\end{equation}
The positive chamber has three fixed-point charts.
The cohomological disk is expressed as a degree
operator acting on its three chart contributions in
\cite[Eqs.~(10.49)--(10.52)]{Cassia:2022lfj};
the logarithmic periods are given in
\cite[Eqs.~(10.54)--(10.56)]{Cassia:2022lfj}.
We use the equivalent positive integral obtained
from \eqref{eq:disk-positive-integral}.
The bare parameters are
$t^1=\lambda^0-2\lambda^1+\lambda^2$ and
$t^2=\lambda^1-2\lambda^2+\lambda^3$.
Define the normalized coefficients
\begin{equation}
 b_1=\frac{c_1}{c_0^{2/3}c_3^{1/3}}
     =e^{z(2t^1+t^2)/3},
 \qquad
 b_2=\frac{c_2}{c_0^{1/3}c_3^{2/3}}
     =e^{z(t^1+2t^2)/3}.
\end{equation}
With $P_3(U)=1+b_1U+b_2U^2+U^3$,
\begin{equation}\begin{aligned}
 \mathbb D_{A_2}(\lambda,(u,v);z)
&=z^{-2}e^{v\lambda^0+u\lambda^3}\Gamma((u+v)/z)
 \int_0^\infty
 \frac{U^{3u/z-1}\,\dd U}{P_3(U)^{(u+v)/z}},
 \\ &\qquad q_1=\frac{b_2}{b_1^2},\quad
 q_2=\frac{b_1}{b_2^2}.
 \label{eq:A2-general-disk}
\end{aligned}\end{equation}
Factoring $P_3(U)=\prod_{j=1}^3(1+a_jU)$,
with $a_1a_2a_3=1$, gives
$\widetilde q_1=a_2/a_1$ and
$\widetilde q_2=a_3/a_2$.
Equivalently, the inverse mirror map is
\begin{equation}
 \begin{aligned}
 q_1&=
 \frac{\widetilde q_1
 (1+\widetilde q_2+\widetilde q_1\widetilde q_2)}
 {(1+\widetilde q_1+\widetilde q_1\widetilde q_2)^2},
 \qquad
 q_2=
 \frac{\widetilde q_2
 (1+\widetilde q_1+\widetilde q_1\widetilde q_2)}
 {(1+\widetilde q_2+\widetilde q_1\widetilde q_2)^2}.
 \end{aligned}
 \label{eq:A2-inverse-mirror-map}
\end{equation}
These root ratios give the flat coordinates determined by the
logarithmic periods of
\cite[Eq.~(10.56)]{Cassia:2022lfj}.
They express \eqref{eq:A2-general-disk} in flat
coordinates, while its remaining $\lambda$
dependence stays in the exponential prefactor.
It can be expanded near
$\widetilde q_1=\widetilde q_2=0$,
but the resulting double degree series is less
economical than the Euler representation.

\paragraph{The general $\boldsymbol{A_{p-1}}$ resolution.}
The charge rows are
\begin{equation}
 Q_j^a=\delta_{j,a-1}-2\delta_{j,a}
       +\delta_{j,a+1},
 \qquad
 j=0,\ldots,p,\quad a=1,\ldots,p-1.
 \label{eq:ALE-charges}
\end{equation}
The positive chamber has $p$ charts,
corresponding to complements of adjacent
pairs of surface coordinates.
For arbitrary $\lambda$, normalize the polynomial
coefficients by
\begin{equation}
 b_j=\frac{c_j}{c_0^{1-j/p}c_p^{j/p}},
 \qquad
 P_p(U)=\sum_{j=0}^{p}b_jU^j
       =\prod_{j=1}^{p}(1+a_jU),
 \qquad b_0=b_p=1.
 \label{eq:ALE-normalized-coefficients}
\end{equation}

The polynomial $P_p(U)$ also defines the mirror ALE
surface
\begin{equation}
 \check A_{p-1}
 =\left\{(\xi,\eta,U)\in\C^2\times\C^*:
 \xi\eta=P_p(U)\right\}.
\end{equation}
The $\C^*$ fiber degenerates at the roots
$U=-a_j^{-1}$. Compact two-spheres are obtained by
fibering its circle over paths joining adjacent roots,
and their logarithmic periods determine the flat
K\"ahler coordinates \cite{Lau:2010jf,Chan:2013dma}.
With the normalization fixed at large radius, these
periods give the root ratios in (\ref{eq:ALE-flat-map}).
The mesonic restriction $T^j=0$ thus corresponds to
coincident roots, giving $P_p(U)=(1+U)^p$. Before imposing this specialization, the complete disk is
\begin{equation}
 \mathbb D_{A_{p-1}}(\lambda,(u,v);z)
=z^{-2}e^{v\lambda^0+u\lambda^p}\Gamma((u+v)/z)
 \int_0^\infty
 \frac{U^{pu/z-1}\,\dd U}{P_p(U)^{(u+v)/z}}.
 \label{eq:ALE-general-disk}
\end{equation}
The bare variables and flat coordinates are
\begin{equation}
 t^j=\lambda^{j-1}-2\lambda^j+\lambda^{j+1},
 \qquad
 q_j=\frac{c_{j-1}c_{j+1}}{c_j^2}
     =\frac{b_{j-1}b_{j+1}}{b_j^2},
 \qquad
 zT^j=\log\frac{a_j}{a_{j+1}}.
 \label{eq:ALE-flat-map}
\end{equation}
The root ordering is fixed by continuation
from large radius. The mirror map determines the normalized
coefficients $b_j$ from the flat coordinates.
This gives a flat-coordinate integral for every
$p$, without requiring a compact formula for
each coefficient of its degree expansion.

\paragraph{Mesonic ALE disks.}
We specify the fixed coefficient representative at unit disk weight $z=1$;
the actual mesonic arguments are then $s\lambda_\star$.

\begin{equation}
\begin{aligned}
A_1:\qquad
&\lambda_\star=(0,-\log 2,0),
&&t_\star=\log 4,\\
A_2:\qquad
&\lambda_\star=(0,-\log 3,-\log 3,0),
&&(t_\star^1,t_\star^2)=(\log 3,\log 3),\\
A_{p-1}:\qquad
&\lambda_\star^j=-\log\binom pj,
&&t_\star^j=
\log\frac{(j+1)(p-j+1)}{j(p-j)}.
\end{aligned}
\label{eq:ALE-flat-point}
\end{equation}
Here $\lambda_\star^j$ is defined for $j=0,\ldots,p$,
while $t_\star^j$ is defined for $j=1,\ldots,p-1$,
corresponding to the $p-1$ independent K\"ahler parameters.

At the mesonic specialization, the denominator polynomial becomes
$(1+U)^p$. The Euler beta integral \eqref{eq:general-beta}
then gives
\begin{equation}
 \begin{aligned}
 \mathbb D_{A_1}^{\text{mes}}(s,(u,v))
 &=s^2\frac{\Gamma(s(u+v))\Gamma(2su)\Gamma(2sv)}{\Gamma(2s(u+v))},\\
 \mathbb D_{A_2}^{\text{mes}}(s,(u,v))
 &=s^2\frac{\Gamma(s(u+v))\Gamma(3su)\Gamma(3sv)}{\Gamma(3s(u+v))},\\
 \mathbb D_{A_{p-1}}^{\text{mes}}(s,(u,v))
 &=s^2\frac{\Gamma(s(u+v))\Gamma(psu)\Gamma(psv)}{\Gamma(ps(u+v))}.
 \end{aligned}
 \label{eq:ALE-mesonic-surface}
\end{equation}

\paragraph{ADHM geometry and mesonic disk.}

The ADHM fourfold is the product of a free
$\C^2$ and $\C^2/\Z_f$, with $f=N_f$.
We use its resolution $A_{f-1}\times\C^2$.
The charge matrix consists of
\eqref{eq:ALE-charges}, with $p=f$, and two
zero columns.
Assign covering weights $\eps_1,\eps_2$ to
the free factor and $\eps_3,\eps_4$ to the
surface, with $\eps_{1234}=2$.

Write
$\lambda=(\lambda_{\mathrm{free}}^1,
\lambda_{\mathrm{free}}^2,\lambda^{(A)})$,
where $\lambda^{(A)}$ contains the surface
parameters.
The disk at arbitrary $\lambda$ is
\begin{equation}
 \mathbb D_{A_{f-1}\times\C^2}(\lambda,\eps;z)
 =e^{\lambda_{\mathrm{free}}^1\eps_1+
      \lambda_{\mathrm{free}}^2\eps_2}
  z^{-2}\Gamma(\eps_1/z)\Gamma(\eps_2/z)
  \mathbb D_{A_{f-1}}(\lambda^{(A)},(\eps_3,\eps_4);z).
 \label{eq:ADHM-general-disk}
\end{equation}
The surface factor is the full expression
\eqref{eq:ALE-general-disk}.
Here and below the disk arguments for ALE
products are expressed in covering weights,
using \eqref{eq:ALE-weight-dictionary}.

Set the two free-coordinate parameters to zero
and impose the surface specialization
\eqref{eq:ALE-flat-point}.
This gives
\begin{align}
 \mathbb D_{A_{f-1}\times\C^2}^{\text{mes}}(s,\eps)
 &=s^2\Gamma(s\eps_1)\Gamma(s\eps_2)
 \mathbb D_{A_{f-1}}^{\text{mes}}(s,(\eps_3,\eps_4))\nonumber\\
 &=s^4\Gamma(s\eps_1)\Gamma(s\eps_2)
 \frac{\Gamma(s\eps_{34})\Gamma(sf\eps_3)\Gamma(sf\eps_4)}
 {\Gamma(sf\eps_{34})}.
 \label{eq:adhmdisk}
\end{align}
The mesonic volume is
$C_{A_{f-1}\times\C^2}=1/(f\prod_i\eps_i)$,
while the generator and relation weights are, cf.\ \eqref{eq:general-geometric-period}
\begin{equation}
 (\nu_1,\ldots,\nu_5)
 =(\eps_1,\eps_2,\eps_{34},f\eps_3,f\eps_4),
 \qquad
 \kappa_1=f\eps_{34}.
 \label{eq:adhm-geometric-poles}
\end{equation}
The geometric prediction for the perturbative coefficients then follows from \eqref{eq:geometric-polynomial}. In the non-perturbative sector, the $\nu_3=\eps_{34}$ pole family cancels
against the relation factor for integer $f$.
The remaining instanton expansion is
\begin{equation}
 J_{\mathrm{disk},A_{f-1}\times\C^2,\mathrm{exp}}
 \sim\sum_{I\in\{1,2,4,5\}}\sum_{n\geq1}
 c_{I,n}^{A_{f-1}\times\C^2}e^{-n\mu/\nu_I}.
 \label{eq:ALE-disk-exp}
\end{equation}
Its coefficients are
\eqref{eq:general-geometric-residues} with
\eqref{eq:adhm-geometric-poles}.

\paragraph{Match on the ADHM QC loci.}

For the distinguished locus, set
$\eps_1=1$, $\eps_{234}=1$
and $\eps_i=\Delta_i$.
Then
\begin{equation}
 \frac{\mathbb D_{A_{f-1}\times\C^2}^{\text{mes}}(s,\eps)}{s^4\Gamma(s)^2}\Big|_{\mathrm{QC}}
 =
 \frac{\Gamma(s\Delta_2)\Gamma(s\Delta_{34})
       \Gamma(sf\Delta_3)\Gamma(sf\Delta_4)}
      {\Gamma(s)\Gamma(sf\Delta_{34})}
 =I_{0,\mathrm{ADHM}}(s).
 \label{eq:ADHM-distinguished-disk-match}
\end{equation}
Thus $J_{\mathrm{disk},A_{f-1}\times\C^2}\Big|_{\mathrm{QC}}
=J_{0,\mathrm{ADHM}}$.

For the pairwise locus, set $\eps_{12}=\eps_{34}=1$ and
$\eps_i=\Delta_i$.
The mesonic disk then gives
\begin{equation}
 \frac{\mathbb D_{A_{f-1}\times\C^2}^{\text{mes}}(s,\eps)}{s^4\Gamma(s)^2}\Big|_{\mathrm{QC}}
 =
 \frac{\Gamma(sf\Delta_3)\Gamma(sf\Delta_4)
       \Gamma(s\Delta_1)\Gamma(s\Delta_2)}
      {\Gamma(sf)\Gamma(s)}
 =I_{0,\mathrm{ADHM,pair}}(s).
 \label{eq:ADHM-pairwise-disk-match}
\end{equation}
This is \eqref{eq:ADHM-pair-trace}, establishing
the corresponding equality with
$J_{0,\mathrm{ADHM,pair}}$. At $f=1$, the two matches reduce to the triangular and
pairwise ABJM cases, respectively, with the appropriate
relabeling of the unit-weight coordinate.

\paragraph{Product geometry and disk integral.}
\label{sec:necklace-disk}

For the grouped necklaces, the transverse
space is
$(\C^2/\Z_\xi\times\C^2/\Z_\zeta)/\Z_k$.
We resolve the two surfaces and retain the
residual quotient normalization.
The GLSM charge matrix is
\begin{equation}
 Q=
 \begin{pmatrix}
 Q_{A_{\xi-1}}&0\\
 0&Q_{A_{\zeta-1}}
 \end{pmatrix}.
 \label{eq:necklace-charges}
\end{equation}
The positive chamber in each factor gives
$\xi\zeta$ fixed-point charts.
A length-one block has no gauge charges.

Assign $\eps_1,\eps_2$ to the first covering
surface and $\eps_3,\eps_4$ to the second.
At arbitrary $\lambda$,
\begin{equation}
 \mathbb D_{(A_{\xi-1}\times A_{\zeta-1})/\Z_k}
 (\lambda,\eps;z)
 =\frac1k
 \mathbb D_{A_{\xi-1}}(\lambda^{(1)},(\eps_1,\eps_2);z)
 \mathbb D_{A_{\zeta-1}}(\lambda^{(2)},(\eps_3,\eps_4);z).
 \label{eq:necklace-general-disk}
\end{equation}
Here $\lambda^{(1)},\lambda^{(2)}$ are the two
blocks of $\lambda$.
The factorization follows from the product
charges and integration cycle.

\paragraph{Mesonic disk and geometric prediction for the product.}

The two mirror maps are independent.
We impose \eqref{eq:ALE-flat-point} in each factor.
The mesonic disk is
\begin{align}
 \mathbb D_{(A_{\xi-1}\times A_{\zeta-1})/\Z_k}^{\text{mes}}(s,\eps)
 &=\frac1k\mathbb D_{A_{\xi-1}}^{\text{mes}}(s,(\eps_1,\eps_2))
 \mathbb D_{A_{\zeta-1}}^{\text{mes}}(s,(\eps_3,\eps_4))\nonumber\\
 &=\frac{s^4}{k}
 \frac{\Gamma(s\eps_{12})\Gamma(\xi s\eps_1)\Gamma(\xi s\eps_2)}{\Gamma(\xi s\eps_{12})}
 \frac{\Gamma(s\eps_{34})\Gamma(\zeta s\eps_3)\Gamma(\zeta s\eps_4)}{\Gamma(\zeta s\eps_{34})}.
 \label{eq:necklace-disk-main}
\end{align}
The mesonic volume is
$C_{(A_{\xi-1}\times A_{\zeta-1})/\Z_k} = 1/(k\xi\zeta\prod_i\eps_i)$, and the Gamma
weights are, cf.\ \eqref{eq:general-geometric-period}
\begin{equation}
 \begin{aligned}
 (\nu_1,\ldots,\nu_6)
 &=(\eps_{12},\xi\eps_1,\xi\eps_2,
          \eps_{34},\zeta\eps_3,\zeta\eps_4),\\
 (\kappa_1,\kappa_2)&=(\xi\eps_{12},\zeta\eps_{34}).
 \end{aligned}
 \label{eq:necklace-geometric-weights}
\end{equation}
The geometric prediction for the perturbative coefficients then follows from \eqref{eq:geometric-polynomial}. In the non-perturbative sector, the $\eps_{12}$ and $\eps_{34}$ pole families
cancel, leaving
\begin{equation}
 J_{\mathrm{disk},
 (A_{\xi-1}\times A_{\zeta-1})/\Z_k,\mathrm{exp}}
 \sim
 \sum_{I\in\{2,3,5,6\}}\sum_{n\geq1}
 c_{I,n}^{(A_{\xi-1}\times A_{\zeta-1})/\Z_k}
 e^{-n\mu/\nu_I}.
 \label{eq:necklace-disk-exp}
\end{equation}
The coefficients follow from
\eqref{eq:general-geometric-residues} with
\eqref{eq:necklace-geometric-weights}.

\paragraph{Match on the necklace QC locus.}

Set $\eps_{12}=\eps_{34}=1$ and $\eps_i=\Delta_i$,
with the ordering of \eqref{eq:necklace-deformations}.
Then
\begin{equation}
 \frac{
 \mathbb D_{(A_{\xi-1}\times A_{\zeta-1})/\Z_k}^{\text{mes}}
 (s,\eps)}
 {s^4\Gamma(s)^2}\Big|_{\mathrm{QC}}
 =
 \frac{\Gamma(s\xi\Delta_1)\Gamma(s\xi\Delta_2)
       \Gamma(s\zeta\Delta_3)\Gamma(s\zeta\Delta_4)}
      {k\Gamma(s\xi)\Gamma(s\zeta)}
 =\frac{I_{0,\mathrm{neck}}(s)}{k}.
 \label{eq:necklace-disk-match}
\end{equation}
The covering-space disk computes $J_{0,\mathrm{neck}}$;
the residual quotient gives $J_{0,\mathrm{neck}}/k$,
the leading contribution to $J$ when $\hbar=k$.
The choices $(\xi,\zeta,k)=(1,f,1)$ and $(1,1,k)$
recover the pairwise ADHM and ABJM results,
respectively, with the same weight labels.

\subsubsection{The conifold}

\paragraph{Geometry and disk integral.}

We consider the fourfold $\C\times\mathcal C$, where
$\mathcal C=\{uv=xy\}\subset\C^4$ is the conifold.
We resolve the singularity of $\mathcal C$ at the origin by the small
resolution $\operatorname{Tot}[\mathcal O(-1)\oplus\mathcal O(-1)
\to\PP^1]$.
The resulting GLSM has charge row
\begin{equation}
 Q=(1,1,-1,-1,0).
 \label{eq:conifold-charges}
\end{equation}
The fifth coordinate parametrizes the additional free $\C$ factor.
The bare K\"ahler parameter and degree variable are
\begin{equation}
 t=\lambda^1+\lambda^2-\lambda^3-\lambda^4,
 \qquad
 q=e^{-zt}=\frac{c_1c_2}{c_3c_4}.
\end{equation}
For these charges, the disk partition function takes the following form

\begin{align}
 \mathbb D_{\C\times\mathcal C}(\lambda,\eps;z)
 ={}&z^{-5}e^{\lambda\cdot\eps}\Gamma(\eps_5/z)
 \oint_{\mathrm{QJK}}\frac{\dd\phi}{2\pi i}\,e^{t\phi} \nonumber
 \\
 &\times\Gamma\!\left(\frac{\eps_1+\phi}{z}\right)
 \Gamma\!\left(\frac{\eps_2+\phi}{z}\right)
 \Gamma\!\left(\frac{\eps_3-\phi}{z}\right)
 \Gamma\!\left(\frac{\eps_4-\phi}{z}\right).
\end{align}

In the positive geometric chamber $t>0$, the QJK prescription
selects the two towers of poles $\phi=-\eps_i-zd$,
where $i=1,2$ and $d\geq0$.
Evaluating the residues of \cite[Eq.~(11.14)]{Cassia:2022lfj},
or applying its degree operator \cite[Eq.~(11.16)]{Cassia:2022lfj}
to the two chart contributions, gives
\begin{align}
 \mathbb D_{\C\times\mathcal C}(\lambda,\eps;z)
 ={}&z^{-4}e^{\lambda\cdot\eps}\Gamma(\eps_5/z)
 \sum_{i=1}^{2}e^{-\eps_i t}
 \Gamma\!\left(\frac{\eps_j-\eps_i}{z}\right)
 \Gamma(\eps_{i3}/z)\Gamma(\eps_{i4}/z)
 \nonumber\\
 &\times{}_2F_1\!\left(
 \frac{\eps_{i3}}{z},\frac{\eps_{i4}}{z};
 1+\frac{\eps_i-\eps_j}{z};q\right),
 \qquad \{i,j\}=\{1,2\}.
 \label{eq:conifold-QJK-series}
\end{align}
Here $i$ labels the fixed-point chart and $j$ is the other element
of $\{1,2\}$. The free-coordinate determinant is
$z^{-1}\Gamma(\eps_5/z)$; its equivariant exponential is
included in $e^{\lambda\cdot\eps}$.

The conifold mirror map is trivial \cite[Eq.~(11.22)]{Cassia:2022lfj}:
\begin{equation}
 T=t,\qquad \widetilde q=e^{-zT}=q,
 \qquad T=0\ \Longrightarrow\ q=1.
 \label{eq:conifold-flat-map}
\end{equation}
Thus the hypergeometric functions in \eqref{eq:conifold-QJK-series}
are already expanded in the flat degree variable:
\begin{equation}
 {}_2F_1(a,b;c;e^{-zT})
 =\sum_{d\geq0}\frac{(a)_d(b)_d}{d!\,(c)_d}e^{-dzT}
 =1+\frac{ab}{c}e^{-zT}+O(e^{-2zT}).
 \label{eq:conifold-first-degree}
\end{equation}

\paragraph{Mesonic disk and geometric prediction.}

We impose $T=0$ and choose $\lambda_\star=0$.
The two chart contributions in \eqref{eq:conifold-QJK-series}
carry the opposite tangent weights $\eps_2-\eps_1$ and
$\eps_1-\eps_2$. Their complete sum is continued from the
large-radius chamber to $q=1$.
Rather than performing this continuation on the two
hypergeometric functions separately, we evaluate the equivalent
Coulomb integral at $t=0$.
With $z=s^{-1}$ and $\omega=s\phi$, it becomes

\begin{align}
 \mathbb D_{\C\times\mathcal C}^{\mathrm{mes}}(s,\eps)
 ={}&s^4\Gamma(s\eps_5)
 \int_{-i\infty}^{i\infty}\frac{\dd\omega}{2\pi i}
 \Gamma(s\eps_1+\omega)\Gamma(s\eps_2+\omega) \nonumber
 \\
 &\times\Gamma(s\eps_3-\omega)\Gamma(s\eps_4-\omega).
\end{align}

The vertical contour separates the two families of Gamma-function
poles. Barnes' first lemma, \eqref{eq:Barnes-first-lemma},
immediately gives
\begin{equation}
 \mathbb D_{\C\times\mathcal C}^{\text{mes}}(s,\eps)
 =s^4\frac{\Gamma(s\eps_5)
 \displaystyle\prod_{i=1}^2\prod_{j=3}^4
 \Gamma(s\eps_{ij})}
 {\Gamma(s\eps_{1234})}.
 \label{eq:conifoldDisk}
\end{equation}
Equivalently, Gauss summation of the combined two-chart expression,
or two Euler beta integrals in the positive-integral representation,
produces the same result. 

The volume and Gamma weights are, cf.\ \eqref{eq:general-geometric-period},
\begin{align}
 C_{\C\times\mathcal C}
 &=\frac{\eps_{1234}}
 {\eps_5\displaystyle\prod_{i=1}^2\prod_{j=3}^4
 \eps_{ij}},
 \nonumber\\
 (\nu_1,\ldots,\nu_5)
 &=(\eps_5,\eps_{13},\eps_{14},\eps_{23},\eps_{24}),
 \qquad
 \kappa_1=\eps_{1234}.
 \label{eq:conifold-geometric-poles}
\end{align}
The geometric prediction for the perturbative coefficients then
follows from \eqref{eq:geometric-polynomial}.
In the non-perturbative sector,
\begin{equation}
 J_{\mathrm{disk},\C\times\mathcal C,\mathrm{exp}}
 \sim\sum_{I=1}^5\sum_{n\geq1}
 c_{I,n}^{\C\times\mathcal C}e^{-n\mu/\nu_I}.
 \label{eq:conifold-disk-exp}
\end{equation}
The coefficients follow from
\eqref{eq:general-geometric-residues} with
\eqref{eq:conifold-geometric-poles}.

\paragraph{Match on the conifold QC loci.}

With $\eps_i=\Delta_i$,
\begin{equation}
\left.
\frac{\mathbb D^{\mathrm{mes}}_{\C\times\mathcal C}(s,\eps)}
{s^4\Gamma(s)^2}
\right|_{\mathrm{QC}}
=I_{0,\mathrm{con}}(s;\Delta).
 \label{eq:conifold-disk-trace}
\end{equation}
Here $I_{0,\mathrm{con}}$ is \eqref{eq:conIeps}, whose three
restrictions are
\begin{equation}
 \begin{aligned}
 \text{Weyl:}\qquad
 &\Delta_5=1,\qquad \Delta_{1234}=1,\\
 \text{one-node ordered:}\qquad
 &\Delta_{23}=1,\qquad \Delta_{145}=1,\\
 \text{two-node ordered:}\qquad
 &\Delta_{24}=1,\qquad \Delta_{135}=1.
 \end{aligned}
 \label{eq:conifold-disk-loci}
\end{equation}
In the first case the free-coordinate and relation factors
cancel. In each ordered case a unit-weight generator cancels
one factor of $\Gamma(s)$.
The resulting expressions are the complete classical traces
of Section~\ref{sec:QCexamples}.
Their Mellin transforms therefore reproduce the corresponding
$J_0$.

\subsubsection{The suspended pinch point}

\paragraph{Geometry and disk integral.}

We consider the fourfold $\C\times\mathrm{SPP}$, where
$\mathrm{SPP}=\{uv=xy^2\}\subset\C^4$ is the suspended pinch point.
Resolving the transverse $A_1$ singularity along
$\{u=v=y=0\}$ and then the residual conifold point gives
a toric crepant resolution with three affine charts.
The resulting GLSM has charge matrix
\begin{equation}
 Q_{\mathrm{SPP}}
 =\begin{pmatrix}
 1&0&0&1&-2&0\\
 -1&1&-1&0&1&0
 \end{pmatrix}.
 \label{eq:sppcharges}
\end{equation}
The ordering agrees with
\cite[Eqs.~(3.85)--(3.86)]{Hristov:2026zjh};
the sixth coordinate parametrizes the additional free $\C$ factor.
Unlike the preceding surface and conifold examples, this disk
is not tabulated in \cite{Cassia:2022lfj};
we apply their general prescription to these charges.

After quotienting the two gauge factors, the residual torus action
is described by the weights of the invariant generators,
\begin{equation}
 \begin{aligned}
 \nu_1&=\eps_{35}+2\eps_4,
 &\nu_2&=2\eps_1+\eps_{25},\\
 \nu_3&=\eps_{23},
 &\nu_4&=\eps_{145},
 \end{aligned}
 \qquad
 \nu_1+\nu_2=2\nu_4+\nu_3,
 \label{eq:spp-geometric-poles}
\end{equation}
together with the free-coordinate weight $\eps_6$.
We will express the disk partition function in terms of these
gauge-invariant combinations as usual.
The bare parameters and degree variables are
\begin{equation}
 \begin{aligned}
 t^1&=\lambda^1+\lambda^4-2\lambda^5,
 &q_1&=\frac{c_1c_4}{c_5^2},\\
 t^2&=-\lambda^1+\lambda^2-\lambda^3+\lambda^5,
 &q_2&=\frac{c_2c_5}{c_1c_3}.
 \end{aligned}
 \label{eq:SPP-bare-parameters}
\end{equation}

For these charges, the disk partition function is
\begin{align}
 \mathbb D_{\C\times\mathrm{SPP}}(\lambda,\eps;z)
 ={}&z^{-6}e^{\lambda\cdot\eps}\Gamma(\eps_6/z)
 \oint_{\mathrm{QJK}}
 \frac{\dd\phi_1\,\dd\phi_2}{(2\pi i)^2}\,
 e^{t^1\phi_1+t^2\phi_2}
 \nonumber\\
 &\times
 \Gamma\!\left(\frac{\eps_1+\phi_1-\phi_2}{z}\right)
 \Gamma\!\left(\frac{\eps_2+\phi_2}{z}\right)
 \Gamma\!\left(\frac{\eps_3-\phi_2}{z}\right)
 \nonumber\\
 &\times
 \Gamma\!\left(\frac{\eps_4+\phi_1}{z}\right)
 \Gamma\!\left(\frac{\eps_5-2\phi_1+\phi_2}{z}\right).
 \label{eq:SPP-Coulomb-disk}
\end{align}
The contour is appropriately deformed in accordance with the QJK prescription, in the chamber containing
$\eta_{\mathrm{JK}}=(1,1/3)$.
The selected pairs $(1,2),(2,4),(4,5)$ of charge vectors label intersections
of the pole families giving the three fixed-point contributions.
Using the torus parametrization of \eqref{eq:app-toric-parametrization} gives~\footnote{The substitution
$(y_1,y_2,y_3,y_4,y_5)=(R,S,SU,RU^2,RU)$
solves the two charge constraints in \eqref{eq:disk-positive-integral}.
Integrating $R$ and $S$ gives
$\Gamma(\nu_4/z)\Gamma(\nu_3/z)$.}
\begin{equation}
 \begin{aligned}
 \mathbb D_{\C\times\mathrm{SPP}}(\lambda,\eps;z)
 ={}&z^{-4}c_6^{-\eps_6/z}
 \Gamma(\eps_6/z)\Gamma(\nu_4/z)\Gamma(\nu_3/z)\\
 &\times\int_0^\infty
 \frac{U^{\nu_1/z-1}\,\dd U}
 {(c_1+c_5U+c_4U^2)^{\nu_4/z}
  (c_2+c_3U)^{\nu_3/z}}.
 \end{aligned}
 \label{eq:SPP-general-disk}
\end{equation}

Rescaling $U\to(c_2/c_3)U$, the denominator polynomials
determine the normalized mirror curve
\begin{equation}
 \begin{aligned}
 P_\text{SPP} (U,V)
 &=1+q_2U+q_1q_2^2U^2+V(1+U)\\
 &=(1+a_1U)(1+a_2U)+V(1+U)=0,
 \end{aligned}
 \label{eq:SPP-mirror-curve}
\end{equation}
where $a_1+a_2=q_2$ and $a_1a_2=q_1q_2^2$.
This agrees with the SPP mirror curve 
\cite{Ooguri:2010yk},
written here in bare coordinates.\footnote{
The parameters $(Q,\mu)$ of that reference correspond to
$(a_1^{-1},a_2)$, with $e^y=U$ and $e^x=V/a_1$.
Their $\mu$ is a K\"ahler modulus, not the chemical potential
used here.}

The factor ordering is fixed by continuation from large radius,
where $a_1\sim q_2$ and $a_2\sim q_1q_2$.
The logarithmic periods then determine the mirror map
\begin{equation}
 \begin{aligned}
 zT^1&=\log(a_1/a_2), \qquad
 zT^2=-\log a_1,  \qquad
 \widetilde q_a=e^{-zT^a},\\
 q_1&=\frac{\widetilde q_1}{(1+\widetilde q_1)^2}, \qquad
 q_2=\widetilde q_2(1+\widetilde q_1).
 \end{aligned}
 \label{eq:SPP-inverse-mirror-map}
\end{equation}

\paragraph{Mesonic disk and geometric prediction.}

The quantum mesonic restriction $T^1=T^2=0$ gives
$a_1=a_2=1$, or equivalently $(q_1,q_2)=(1/4,2)$.
At unit disk weight, we choose the representative
\begin{equation}
 \begin{aligned}
 c^\star&=(1,1,1,1,2,1),\\
 (t_\star^1,t_\star^2)&=(\log4,-\log2),\\
 \lambda_\star&=(0,0,0,0,-\log2,0).
 \end{aligned}
 \label{eq:spppoint}
\end{equation}
Although $t_\star^2<0$, this specialization is obtained by
analytic continuation of the disk from the chosen large-radius
chamber, retaining its integration cycle.
For this representative, the coefficient prefactor
in \eqref{eq:SPP-general-disk} is one.
The quadratic and linear denominator factors become
$(1+U)^2$ and $1+U$, respectively, so the beta identity
\eqref{eq:general-beta} gives
\begin{equation}
 \begin{aligned}
 \mathbb D_{\C\times\mathrm{SPP}}^{\text{mes}}(s,\eps)
 &=s^4
 \frac{\Gamma(s\eps_6)\prod_{i=1}^4\Gamma(s\nu_i)}
 {\Gamma(s(\nu_1+\nu_2))},\\
 C_{\C\times\mathrm{SPP}}
 &=\frac{\nu_1+\nu_2}{\eps_6\prod_{i=1}^4\nu_i}.
 \end{aligned}
 \label{eq:spp-full-period}
\end{equation}

Let $\nu_5=\eps_6$ and $\kappa_1=\nu_1+\nu_2$,
cf.\ \eqref{eq:general-geometric-period}.
The geometric prediction for the perturbative coefficients
then follows from \eqref{eq:geometric-polynomial}.
In the non-perturbative sector,
\begin{equation}
 J_{\mathrm{disk},\C\times\mathrm{SPP},\mathrm{exp}}
 \sim\sum_{I=1}^5\sum_{n\geq1}
 c_{I,n}^{\C\times\mathrm{SPP}}e^{-n\mu/\nu_I}.
 \label{eq:SPP-disk-exp}
\end{equation}
The coefficients are given by
\eqref{eq:general-geometric-residues}
with these generator and relation weights.

\paragraph{Match on the SPP QC locus.}

Impose $\eps_6=\nu_3+\nu_4=1$ and
$\eps_i=\Delta_i$.
The four $\nu_i$ then equal the $r_i$ of
\eqref{eq:SPP-weights}, and
\begin{equation}
 \frac{
 \mathbb D_{\C\times\mathrm{SPP}}^{\text{mes}}(s,\eps)}
 {s^4\Gamma(s)^2} \Big|_{\mathrm{QC}}
 =
 \frac{\prod_{i=1}^4\Gamma(sr_i)}
 {\Gamma(s)\Gamma(s(r_1+r_2))}
 =I_{0,\mathrm{SPP}}(s).
 \label{eq:SPP-disk-match}
\end{equation}
This proves
$J_{\mathrm{disk},\C\times\mathrm{SPP}}\Big|_{\mathrm{QC}}=J_{0,\mathrm{SPP}}$.

\subsubsection{The cone over \texorpdfstring{$Q^{1,1,1}$}{Q111}}

\paragraph{Geometry and disk integral.}

We consider the Calabi--Yau fourfold $C(Q^{1,1,1})$,
the cone over the seven-dimensional Sasaki--Einstein
manifold $Q^{1,1,1}$, and its resolution
$\operatorname{Tot}[
\mathcal O(-1,-1)\oplus\mathcal O(-1,-1)
\to\PP^1\times\PP^1]$.
The GLSM charge matrix is
\begin{equation}
 Q_{Q^{1,1,1}}
 =\begin{pmatrix}
 1&1&0&0&-1&-1\\
 0&0&1&1&-1&-1
 \end{pmatrix}.
 \label{eq:qcharges}
\end{equation}
The bare K\"ahler parameters and degree variables are
\begin{equation}
 \begin{aligned}
 t^1&=\lambda^1+\lambda^2-\lambda^5-\lambda^6,
 &q_1&=e^{-zt^1},\\
 t^2&=\lambda^3+\lambda^4-\lambda^5-\lambda^6,
 &q_2&=e^{-zt^2}.
 \end{aligned}
 \label{eq:Q-bare-parameters}
\end{equation}
The disk integral of
\cite[Eq.~(11.30)]{Cassia:2022lfj}, including our
equivariant exponential, is
\begin{align}
 \mathbb D_{C(Q^{1,1,1})}(\lambda,\eps;z)
 ={}&z^{-6}e^{\lambda\cdot\eps}
 \oint_{\mathrm{QJK}}
 \frac{\dd\phi_1\,\dd\phi_2}{(2\pi i)^2}\,
 e^{t^1\phi_1+t^2\phi_2}
 \nonumber\\
 &\times
 \Gamma\!\left(\frac{\eps_1+\phi_1}{z}\right)
 \Gamma\!\left(\frac{\eps_2+\phi_1}{z}\right)
 \Gamma\!\left(\frac{\eps_3+\phi_2}{z}\right)
 \Gamma\!\left(\frac{\eps_4+\phi_2}{z}\right)
 \nonumber\\
 &\times
 \Gamma\!\left(\frac{\eps_5-\phi_1-\phi_2}{z}\right)
 \Gamma\!\left(\frac{\eps_6-\phi_1-\phi_2}{z}\right).
 \label{eq:Q-Coulomb-disk}
\end{align}

In the chamber $t^1,t^2>0$, the QJK prescription selects
four charts, indexed by $i\in\{1,2\}$ and $j\in\{3,4\}$.
Their Gamma-pole towers are
\begin{equation}
 \phi_1=-\eps_i-zm,\qquad
 \phi_2=-\eps_j-zn,\qquad m,n\geq0.
 \label{eq:Q-chart-poles}
\end{equation}
The residues involve the Appell function
\begin{equation}
 F_4(a,b;c,d;q_1,q_2)
 =\sum_{m,n\geq0}
 \frac{(a)_{m+n}(b)_{m+n}}
 {(c)_m(d)_n\,m!\,n!}\,q_1^m q_2^n.
 \label{eq:Q-Appell-definition}
\end{equation}
Writing $i'$ and $j'$ for the other members of the
respective pairs, the complete chart sum is
\begin{align}
 \mathbb D_{C(Q^{1,1,1})}(\lambda,\eps;z)
 ={}&z^{-4}e^{\lambda\cdot\eps}
 \sum_{i=1,2}\sum_{j=3,4}
 e^{-\eps_i t^1-\eps_j t^2}
 \Gamma\!\left(\frac{\eps_{i'}-\eps_i}{z}\right)
 \Gamma\!\left(\frac{\eps_{j'}-\eps_j}{z}\right)
 \nonumber\\
 &\times
 \Gamma(\eps_{ij5}/z)\Gamma(\eps_{ij6}/z)
 F_4\!\left(
 \frac{\eps_{ij5}}{z},\frac{\eps_{ij6}}{z};
 1+\frac{\eps_i-\eps_{i'}}{z},
 1+\frac{\eps_j-\eps_{j'}}{z};
 q_1,q_2
 \right).
 \label{eq:Q-general-QJK}
\end{align}
The terms with $m+n>0$ are the positive-degree
contributions.
The series converges for
$\sqrt{|q_1|}+\sqrt{|q_2|}<1$; elsewhere we analytically
continue the complete chart sum.
The expressions above are initially understood at
generic equivariant weights.

\paragraph{Mesonic disk and geometric prediction.}

The mirror map is trivial
\cite[Eqs.~(11.35)--(11.36)]{Cassia:2022lfj}:
\begin{equation}
 T^1=t^1,\qquad T^2=t^2,\qquad
 T^1=T^2=0\ \Longrightarrow\ q_1=q_2=1.
 \label{eq:Q-flat-map}
\end{equation}
Thus the Appell series already uses flat degree variables,
but its mesonic value requires analytic continuation.
We choose
\begin{equation}
 \lambda_\star=(0,0,0,0,0,0),\qquad
 (t_\star^1,t_\star^2)=(0,0).
 \label{eq:Q-mesonic-point}
\end{equation}
Both exponentials in \eqref{eq:Q-Coulomb-disk} then
equal one.
We evaluate the mesonic disk directly from this integral,
continuing its QJK cycle to the chosen point.

By \eqref{eq:mesonic-new-definition},
$\mathbb D_{C(Q^{1,1,1})}^{\text{mes}}(s,\eps)
=\mathbb D_{C(Q^{1,1,1})}(0,\eps;s^{-1})$.
Define
\begin{equation}
 \alpha=\eps_1-\eps_2,\qquad
 \beta=\eps_3-\eps_4.
 \label{eq:Q-geometric-differences}
\end{equation}
Setting $z=s^{-1}$ and changing variables to
$\omega_1=s(\eps_1+\phi_1)$ and
$\omega_2=s(\eps_3+\phi_2)$ gives
\begin{align}
 \mathbb D_{C(Q^{1,1,1})}^{\text{mes}}(s,\eps)
 =s^4\int\frac{\dd\omega_1\,\dd\omega_2}{(2\pi i)^2}\,
 &\Gamma(\omega_1)\Gamma(\omega_2)
 \Gamma(s\eps_{136}-\omega_1-\omega_2)
 \nonumber\\
 {}\times&
 \Gamma(\omega_1-s\alpha)
 \Gamma(\omega_2-s\beta)
 \Gamma(s\eps_{135}-\omega_1-\omega_2).
 \label{eq:Q-general-period}
\end{align}
For positive weights and real $s>0$, we take
$\operatorname{Re}\omega_1=s\eps_1$ and
$\operatorname{Re}\omega_2=s\eps_3$, so all six Gamma
arguments have positive real parts.
The expression is then analytically continued.

Keeping the weights general, with $\sum_i\eps_i=2$,
the Mellin transform \eqref{eq:geometric-J-definition}
gives geometric predictions for the polynomial and
exponentially small terms.
Although the full disk remains a Barnes integral,
Appendix~\ref{app:Qgeneraljets} shows that its polynomial
contribution is determined by the four fixed-point
weight sets
\begin{equation}
 w_{ij}=
 (\eps_{i'}-\eps_i,\eps_{j'}-\eps_j,
  \eps_{ij5},\eps_{ij6}),
 \qquad i\in\{1,2\},\quad j\in\{3,4\}.
 \label{eq:Q-chart-weights}
\end{equation}
The mesonic equivariant volume is
\begin{equation}
 C_{C(Q^{1,1,1})}(\eps)
 =\sum_{i=1,2}\sum_{j=3,4}
 \frac1{\prod_{a=1}^4w_{ij,a}},
 \label{eq:Q-mesonic-volume}
\end{equation}
and the polynomial coefficients are
\begin{equation}
 \begin{aligned}
 C_{\mathrm{disk}}(\eps)
 &=\frac{C_{C(Q^{1,1,1})}(\eps)}{8\pi^2},\qquad
 B_{\mathrm{disk}}(\eps)
 =\frac1{48}
 \sum_{i=1,2}\sum_{j=3,4}
 \frac{\sum_{a=1}^4w_{ij,a}^2}
 {\prod_{a=1}^4w_{ij,a}},\\
 A_{\mathrm{disk}}(\eps)
 &=\frac{\zeta(3)}{12\pi^2}
 \sum_{i=1,2}\sum_{j=3,4}
 \frac{2-\sum_{a=1}^4w_{ij,a}^3}
 {\prod_{a=1}^4w_{ij,a}}.
 \end{aligned}
 \label{eq:Q-geom-main-jets}
\end{equation}
The chart contributions must be combined before
taking coincident-weight limits.

The exponentially small terms require the complete
Barnes integral.
Its candidate negative poles are associated with the
eight invariant-generator weights
\begin{equation}
 \nu_{ijk}=\eps_{ijk},\qquad
 i\in\{1,2\},\quad j\in\{3,4\},\quad k\in\{5,6\}.
 \label{eq:Q-general-actions}
\end{equation}
For each family, set $s_n=-n/\nu_{ijk}$ and let $k'$
be the other member of $\{5,6\}$.
For separated simple poles,
\begin{align}
 J_{\mathrm{disk},C(Q^{1,1,1}),\mathrm{exp}}(\mu;\eps)
 &\sim
 \sum_{i=1,2}\sum_{j=3,4}\sum_{k=5,6}
 \sum_{n\geq1}c_{ijk,n}(\eps)e^{-n\mu/\nu_{ijk}},
 \nonumber\\
 c_{ijk,n}(\eps)
 &=
 \frac{(-1)^{n+1}\Gamma(-s_n)}
 {4\pi^2\nu_{ijk}\Gamma(s_n)}
 \sum_{u=0}^{n}\sum_{v=0}^{n-u}
 \frac{
 \Gamma(s_n(\eps_{i'}-\eps_i)-u)
 \Gamma(s_n(\eps_{j'}-\eps_j)-v)}
 {u!\,v!\,(n-u-v)!}
 \nonumber\\
 &\hspace{15mm}\times
 \Gamma(s_n\eps_{ijk'}+u+v).
 \label{eq:Q-general-residues}
\end{align}
The finite sum follows from the Barnes-pinch identity
\eqref{eq:Barnes-pinch-residue} in
Appendix~\ref{app:Qpinch}.
Its integers distribute Gamma-pole shifts and are
distinct from the curve degrees in
\eqref{eq:Q-general-QJK}.

\paragraph{Match on the QC locus.}

We identify the weights with the field-theory charges
and impose the QC restrictions,
\begin{equation}
 \eps_i=\Delta_i,\qquad
 \eps_{136}=\eps_{245}=1.
 \label{eq:Q-disk-QC-locus}
\end{equation}
Then $\alpha=\Delta^-_{12}$,
$\beta=\Delta^-_{34}$ and
$\eps_{135}=1+\Delta^-_{56}$.
The six Gamma factors in \eqref{eq:Q-general-period}
therefore coincide with those in \eqref{eq:Qbarnes-QC}.
With the normalization of \eqref{eq:disk-trace-match},
\begin{equation}
 \left.
 \frac{\mathbb D_{C(Q^{1,1,1})}^{\text{mes}}(s,\eps)}
 {s^4\Gamma(s)^2}
 \right|_{\mathrm{QC}}
 =I_{0,Q}(s).
 \label{eq:Q-disk-match}
\end{equation}
At generic weights on this locus, the two unit-weight
families $\nu_{136}=\nu_{245}=1$ are canceled by the
reciprocal Gamma factor in the Mellin kernel, as shown
in \eqref{eq:Q-unit-pole-cancellation}.
The remaining six families agree with
\eqref{eq:Q-instanton-poles}. The common Mellin transform therefore gives
$J_{\mathrm{disk},C(Q^{1,1,1})}|_{\mathrm{QC}}=J_{0,Q}$.
\section{Discussion}
\label{sec:outlook}

We have related the mesonic disk partition function of the transverse
fourfold to the complete classical spectral trace on the round-sphere
QC loci studied here. With the Gamma normalization and the covering/quotient
convention specified in Section~\ref{sec:flat-mesonic}, the match holds as an identity of meromorphic
functions. Its Mellin transform therefore reproduces the full
$\mu$-dependence of $J_0$, including the constant term and the
membrane-instanton series. This extends the geometric description
based on equivariant volumes and constant maps
\cite{Cassia:2025aus,Cassia:2025jkr} to exponentially small terms,
even for $\C^4$, where there is no compact-curve degree sum.

For resolved conical singularities, the prescription uses the complete
disk analytically continued to vanishing flat K\"ahler coordinates.
Positive-degree contributions are summed at this point and remain
part of the mesonic disk. Thus, also in the presence of compact
curves, the immediate source of the membrane expansion is the pole
structure of the Mellin integrand, rather than a term-by-term
identification with curve degree. The boundary cycle and equivariant
normalization specify the disk entering this construction
\cite{Cassia:2022lfj}. The match determines a complete leading WKB
coefficient; its physical derivation and extension to the finite-$\hbar$
answer remain open.

We conclude by outlining several related questions that we hope
to revisit.

\paragraph{A general identity and its deformations.}
The common integral structure of our examples suggests
a route to a broader proof: in suitable two-slice
presentations, radial integrations produce the
$\Gamma(s)^2$ factor relating the disk to the classical
trace. The equivariant
$\mathrm{CY}_4$--$\C\times\mathrm{CY}_3$ correspondence
of \cite{Hristov:2026zjh} provides a natural setting
for this argument.

At the quantum mesonic point, the paired classical
symbols agree on the QC loci when their coefficients,
equivariant weights and measures are identified.
This includes the multiplicities of coincident monomials,
not only the Newton polygon.
The correspondence therefore preserves the complete
$J_0$, including its constant and exponentially small
terms. Its scope is nevertheless restricted: away from
the QC loci the equivariant volumes already differ,
while the mismatch in $B_1$ for the paired theories
shows that the correspondence does not extend in
general to $J_1$
\cite[Section~4.2]{Hristov:2026zjh}.
It identifies a common leading semiclassical sector
of distinct quantum theories.

Away from the known QC loci, $J_{\mathrm{disk}}$ supplies
predictions for each geometry that could be tested by a
direct large-$N$ analysis of the source-deformed localization
integral.
Such a test should first compare the exponential scales;
matching amplitudes additionally requires identifying the
sector represented by the geometric transform when no WKB
description is known.
Varying the flat K\"ahler coordinates gives a separate
extension, changing polynomial coefficients rather than
equivariant weights.
Five-brane constructions may help identify the corresponding
mass and FI parameters \cite{Kubo:2025jxi}, and clarify
whether these deformations incorporate baryonic directions
\cite{Hosseini:2025mgf}.

\paragraph{Quantum corrections and topological strings.}
At special weights, the curves reduce to familiar local-threefold
spectral problems. The relations between ABJM theory, refined
topological strings and TS/ST therefore provide a useful guide
\cite{Marino:2009jd,Aganagic:2011mi,Hatsuda:2013oxa,Grassi:2014zfa,Gu:2015pda,Honda:2014npa,Huang:2014eha,Codesido:2015dia,Marino:2015nla}.
Explicit expressions for the full partition functions of maximally supersymmetric ABJ(M) theories provide a particularly close precedent \cite{Codesido:2014oua,Grassi:2014uua}.
These comparisons concern full spectral functions rather than a few
asymptotic coefficients. In our case the exact geometric identity fixes
the entire classical trace, while higher spectral WKB orders remain separate.
A precise identification must match the quantum operator and its
parameters, not merely its polygon. Higher WKB orders require
ordering and symplectic data absent from the classical disk match,
while worldsheet-type sectors are nonperturbative in $\hbar$.
Geometric insertions, boundary-brane data and the Gamma-class
integral structure offer possible ways to supply additional
information \cite{Hori:2013ika,Iritani:2009qgl,Iritani:2023cxf,Jockers:2018sfl,Beem:2012mb,Cassia:2023uvd,Cassia:2024txc}.
A follow-up study \cite{CompanionII} investigates higher WKB
terms, supersymmetric squashing and the all-orders constant
functions, including their relation to Hodge-integral structures.

\paragraph{Wrapped branes and the bulk interpretation.}
Wrapped-string and membrane calculations give independent access
to instanton actions and fluctuation prefactors
\cite{Drukker:2011zy,Cagnazzo:2009zh,Gautason:2023igo,
Beccaria:2023ujc,Gautason:2025per,Kurlyand:2026yke,Tseytlin:2026dhy,Kurlyand:2026gxi}.
Relating these results to the present Mellin residues first
requires identifying how the worldsheet sector computed by
these probes and the leading-WKB membrane sector fit into
a common geometric description. The proposed M-theory
grand-canonical ensemble also gives a bulk interpretation of
$\mu$ \cite{Gautason:2025plx,Bobev:2026gir}.
However, the change from fixed rank to fixed chemical potential
is distinct from the Mellin reconstruction of the spectral
logarithm. A bulk derivation must explain the latter operation
as well as the disk state and its contour.

\paragraph{Giant-graviton expansions.}
The giant-graviton expansion organizes
finite-$N$ index corrections through brane-worldvolume
contributions \cite{Gaiotto:2021xce,Hristov:2026tde}; related calculations for
M2-brane indices involve wrapped M5-branes \cite{Arai:2020uwd}.
The free-fermion, determinant and eigenvalue-instanton
descriptions of these expansions make a comparison with the
disk transform particularly interesting
\cite{Murthy:2022ien,Chen:2024cvf}.
One could ask whether a suitable disk amplitude supplies a
worldvolume contribution or a building block of such an
expansion. The indices and the round-sphere partition function
have different backgrounds and asymptotic regimes, so this
would require a dictionary between their observables and
brane sectors rather than an identification of their
exponential terms alone.

\paragraph{Other brane partition functions.}
The disk may also enter other brane observables as a building block or
through a different expansion. Related equivariant constructions include
CY$_4$ elliptic genera and D-brane bound-state partition functions
\cite{Franco:2017cjj,Nekrasov:2017cih,Nekrasov:2018xsb}; a common geometry
alone does not identify these observables with the disk.
A more direct setting is provided by M5-brane indices on $S^1\times L_5$,
where $L_5$ is Sasaki--Einstein. Their anomaly-based Cardy description
uses $X=\C\times Y$, with $Y$ a resolution of $C(L_5)$, and relates its
equivariant characteristic classes to M2-brane observables
\cite{Hristov:2026tde}.
It would be interesting to ask whether the disk on this same fourfold
enters an extension beyond that regime. Factorization into local
Nekrasov functions offers a possible route to such a comparison
\cite{Qiu:2014oqa,Qiu:2015rwp}.
An appropriate block or limit, its boundary data and its parameter
dictionary would have to be identified; neither the full M5 index nor
its integration kernel follows from the present disk match.

\paragraph{Compact threefolds and BPS integral representations.}
The product $\C\times Y$, with $Y$ a compact Calabi--Yau
threefold, offers a test beyond the conical setting.
For factorized boundary data, the free factor contributes a
Gamma function multiplying a disk amplitude of $Y$.
Transforming that factor at fixed compact moduli cannot by
itself supply the independent higher-genus data of the closed
topological string. A possible connection instead comes from
the Gopakumar--Vafa description in terms of M2-brane states
\cite{Gopakumar:1998ii,Gopakumar:1998jq,Dedushenko:2014nya} and the
Schwinger-contour proposal of
\cite{Hattab:2024ewk,Lockhart:2012vp,Krefl:2015qva}.
Comparing the pole structures may help explain how periods
enter different expansions of a BPS observable.
The kernels and physical quantities must first be related;
scaling compact K\"ahler parameters with the Mellin variable
would be an additional prescription, not a consequence of
the construction used here.

\paragraph{A sixfold observable and different expansions.}
Equivariant localization of gravitational actions and anomaly
inflow suggest a broader framework
\cite{BenettiGenolini:2023kxp,Martelli:2023oqk,
Colombo:2023fhu,Hristov:2026uxs}.
For M2-brane backgrounds, a possible local model for the twelve-dimensional filling is the complex sixfold $\C^2\times X$, retaining
both external and transverse equivariant data.
The recent localization proposal uses a tangent--cotangent
character, an equivariant index associated with $TW-T^*W$
and relative boundary conditions on the filling $W$, to describe
the bulk one-loop factor \cite{BenettiGenolini:2026kjy}.
Together with the anomaly-based description of constant maps
\cite{Hristov:2026uxs}, this motivates asking whether the
fourfold disk is a limit or specialization of a more complete
sixfold observable. Such a construction is not established,
and the rewriting on a QC locus need not select a unique
quantum completion. The disk could naturally belong to a
different expansion of that larger observable, in which terms
classified as instantons here are organized differently.
Identifying the observable and its physical boundary conditions
would give a clearer meaning to both the established matches
and their proposed extensions.

\section*{Acknowledgements}
We thank Luca Cassia, Sebastian Franco, Seyed Morteza Hosseini, Naotaka Kubo, Ji Hoon Lee, and Alberto Zaffaroni for helpful discussions, and especially Alba Grassi for the initial motivation and comments on a parallel project. We gratefully acknowledge the support of the European Union's Erasmus+ programme for research mobility. We acknowledge assistance from Mathematica (Wolfram) with analytic calculations and from ChatGPT (OpenAI) with helpful discussions, derivations and editing. K.H.\ is supported in part by the Bulgarian NSF grant KP-06-N68/3. K.H.\ gratefully acknowledges the hospitality and support of the Institute for Advanced Study, where this work was carried out, with funding provided by an anonymous donor.

\bigskip
\bigskip
\bigskip
\appendix
\section{Quantum curve identities}
\label{app:quantum-curves}
\label{app:wigner}

We collect the operator and integral identities used in
Sections~\ref{sec:QC} and \ref{sec:disk}.
Throughout, $[\widehat x,\widehat p]=2\pi i\hbar$,
$O_W$ denotes the full Weyl symbol of $\widehat O$,
and $O_0$ its leading semiclassical term.
The classical trace $I_0$ is unnormalized, as in
\eqref{eq:classical-trace}.

\subsection{Wigner--Weyl calculus}
\label{app:wigner-calculus}

With position states normalized by
$\langle x|y\rangle=\delta(x-y)$, the Weyl symbol of
$\widehat A$ is its Wigner transform,
\begin{equation}
 A_W(x,p)=\int_{\mathbb R}\mathrm dy\,
 e^{-ipy/(2\pi\hbar)}
 \left\langle x+\frac y2\middle|\widehat A
 \middle|x-\frac y2\right\rangle.
 \label{eq:wigner}
\end{equation}
The corresponding trace formula is
\begin{equation}
 \operatorname{Tr}\widehat A
 =\frac{1}{4\pi^2\hbar}
 \int_{\mathbb R^2}\mathrm dx\,\mathrm dp\,A_W(x,p).
 \label{eq:wigner-trace}
\end{equation}
These conventions agree with the spectral methods of
\cite{Marino:2011eh,Hatsuda:2015oaa} after accounting
for our normalization of $\hbar$.

Operator composition is represented by the Moyal product:
\begin{equation}
 (\widehat A\widehat B)_W=A_W\star B_W,
 \qquad
 A\star B=A\exp\!\left[i\pi\hbar
 \left(\overleftarrow{\partial_x}\overrightarrow{\partial_p}
 -\overleftarrow{\partial_p}\overrightarrow{\partial_x}\right)
 \right]B.
 \label{eq:wigner-moyal}
\end{equation}
Equivalently,
\begin{equation}
 A\star B=\sum_{n=0}^{\infty}\frac{(i\pi\hbar)^n}{n!}
 \sum_{r=0}^{n}(-1)^r\binom nr
 (\partial_x^{n-r}\partial_p^r A)
 (\partial_p^{n-r}\partial_x^r B).
 \label{eq:wigner-moyal-expansion}
\end{equation}
The product is associative, and reversing the sign of
$\hbar$ exchanges its arguments:
\begin{equation}
 (A\star B)\star C=A\star(B\star C),
 \label{eq:moyal-associativity}
\end{equation}
\begin{equation}
 A\star_\hbar B=B\star_{-\hbar}A.
 \label{eq:moyal-reversal}
\end{equation}
The first identity follows from operator associativity;
the second follows because the bidifferential operator
in \eqref{eq:wigner-moyal} changes sign when its
arguments are exchanged.

For $f(z)=\sum_{n\geq0}c_nz^n$, the corresponding
star-function is
\begin{equation}
 \bigl(f(\widehat O)\bigr)_W
 =f_\star(O_W):=\sum_{n=0}^{\infty}c_n O_W^{\star n},
 \qquad O_W^{\star0}=1,
 \label{eq:wigner-function}
\end{equation}
whenever the operator series and its symbol transform
are justified.
Here $O_W^{\star n}$ denotes the $n$-fold star product.
Inverses, noninteger powers and logarithms are defined
through operator functional calculus; their semiclassical
symbols can then be expanded using the same product.
In particular, for $O_W=O_0+O(\hbar)$,
\begin{equation}
 (\widehat O^{-s})_W=O_0^{-s}+O(\hbar),
 \label{eq:wigner-leading-power}
\end{equation}
within the domain of the semiclassical expansion.
Consequently,
\begin{equation}
 \mathcal Z(s;\Delta,\hbar)
 =\frac{1}{4\pi^2\hbar}
 \int_{\mathbb R^2}\mathrm dx\,\mathrm dp\,
 (\widehat O^{-s})_W
 \sim\frac{I_0(s;\Delta)}{4\pi^2\hbar},
 \qquad \hbar\to0.
 \label{eq:spectral_zeta_wigner}
\end{equation}
Even when $O_W=O_0$ exactly, the star powers generally
receive quantum corrections.
Exactness of the symbol therefore does not imply
exactness of the classical spectral trace.

Every operator considered here either has an
$\hbar$-independent Weyl symbol or is a product of two
blocks with this property.
By \eqref{eq:moyal-reversal}, reversing $\hbar$ leaves
star functions of a single such symbol unchanged,
while for two blocks it reverses their order.
Cyclicity then leaves the spectral traces unchanged.
After extracting the phase-space factor $1/\hbar$,
their formal expansion therefore contains only even
powers of $\hbar$, giving
$J\sim\hbar^{-1}J_0+\hbar J_1+\hbar^3J_2+\cdots$.
This statement concerns the traces; individual ordered
symbols need not be even in $\hbar$.

\subsection{Integral identities}
\label{app:wkb-integrals}
\label{app:dictionary}

The classical symbols in Section~\ref{sec:QCexamples}
can be written as
\begin{equation}
 O_0(x,p)=e^{-ux-vp}P(e^x,e^p),
\end{equation}
where $P$ is a polynomial and $u,v$ are determined
by the curve.
A Laurent polynomial can be put in this form by
extracting a monomial and shifting $u,v$ accordingly.
The substitution $t_1=e^x$, $t_2=e^p$ gives
\begin{equation}
 I_0(s)=\int_0^\infty\frac{\mathrm dt_1}{t_1}
 \int_0^\infty\frac{\mathrm dt_2}{t_2}\,
 t_1^{us}t_2^{vs}P(t_1,t_2)^{-s}.
 \label{eq:general-Euler-trace}
\end{equation}
The normalization $1/(4\pi^2\hbar)$ enters when
reconstructing $\mathcal Z$, not in the definition
of $I_0$.
We use the following identities first in their
convergence domains and then by analytic continuation.

The Euler beta integral is
\begin{equation}
 \int_0^\infty\mathrm dt\,t^{\alpha-1}(a+t)^{-\beta}
 =a^{\alpha-\beta}\mathrm B(\alpha,\beta-\alpha)
 =a^{\alpha-\beta}
 \frac{\Gamma(\alpha)\Gamma(\beta-\alpha)}{\Gamma(\beta)},
 \label{eq:general-beta}
\end{equation}
where $\mathrm B$ is the Euler beta function,
$a>0$ and
$0<\operatorname{Re}\alpha<\operatorname{Re}\beta$.
Its multidimensional counterpart is the Dirichlet integral,
\begin{equation}
 \int_{(0,\infty)^d}
 \frac{\prod_{j=1}^d(\mathrm dt_j\,t_j^{\alpha_j-1})}
 {(a+\sum_{j=1}^d t_j)^\beta}
 =a^{\sum_j\alpha_j-\beta}
 \frac{\Gamma(\beta-\sum_j\alpha_j)
       \prod_{j=1}^d\Gamma(\alpha_j)}{\Gamma(\beta)}.
 \label{eq:general-Dirichlet}
\end{equation}
It converges for $a>0$, $\operatorname{Re}\alpha_j>0$
and $\operatorname{Re}(\beta-\sum_j\alpha_j)>0$.
The two-variable case appears in \eqref{eq:QC-Dirichlet}.

For the Barnes representation of the $Q^{1,1,1}$ trace,
we use
\begin{equation}
 (a+b)^{-\beta}
 =\frac{1}{\Gamma(\beta)}
 \int_{c-i\infty}^{c+i\infty}
 \frac{\mathrm d\omega}{2\pi i}\,
 \Gamma(\omega)\Gamma(\beta-\omega)
 a^{-\omega}b^{\omega-\beta},
 \label{eq:general-Barnes}
\end{equation}
where $a,b>0$ and $0<c<\operatorname{Re}\beta$.
Applying this identity twice gives
\begin{equation}
 (1+a+b)^{-\beta}
 =\frac{1}{\Gamma(\beta)}
 \int\frac{\mathrm d\omega_1\,\mathrm d\omega_2}{(2\pi i)^2}\,
 \Gamma(\omega_1)\Gamma(\omega_2)
 \Gamma(\beta-\omega_1-\omega_2)
 a^{-\omega_1}b^{-\omega_2},
 \label{eq:general-Barnes-two}
\end{equation}
with vertical contours satisfying
$\operatorname{Re}\omega_1>0$,
$\operatorname{Re}\omega_2>0$ and
$\operatorname{Re}(\omega_1+\omega_2)<\operatorname{Re}\beta$.
Setting $a=e^{-x}$, $b=e^{-p}$ and $\beta=s$
yields \eqref{eq:Q-inverse-Mellin}.

The conifold calculation uses Barnes' first lemma
\cite[Section~5.13]{Olver:2010ouy}:
\begin{equation}
 \begin{aligned}
 &\int_{-i\infty}^{i\infty}\frac{\dd\omega}{2\pi i}\,
 \Gamma(\alpha+\omega)\Gamma(\beta+\omega)
 \Gamma(\gamma-\omega)\Gamma(\delta-\omega)
 \\
 &\qquad=
 \frac{\Gamma(\alpha+\gamma)\Gamma(\alpha+\delta)
       \Gamma(\beta+\gamma)\Gamma(\beta+\delta)}
 {\Gamma(\alpha+\beta+\gamma+\delta)}.
 \end{aligned}
 \label{eq:Barnes-first-lemma}
\end{equation}
Initially, all four parameters have positive real parts,
so the imaginary axis separates the two families of poles.
The identity extends by analytic continuation, with the
contour deformed to preserve this separation.

For the $A_1$ disk, we use two hypergeometric identities.
The quadratic transformation
\cite[Section~15.8]{Olver:2010ouy} reads
\begin{equation}
 {}_2F_1\!\left(a,a+\frac12;c;
 \frac{4x}{(1+x)^2}\right)
 =(1+x)^{2a}
 {}_2F_1(2a,2a-c+1;c;x).
 \label{eq:general-quadratic-transformation}
\end{equation}
Using the mirror map
$q=\widetilde q/(1+\widetilde q)^2$, this gives
\begin{equation}
 \begin{aligned}
 &q^{u/z}
 {}_2F_1\!\left(
 \frac uz,\frac uz+\frac12;
 1+\frac{u-v}{z};4q
 \right)\\
 &\qquad=
 \widetilde q^{u/z}
 {}_2F_1\!\left(
 \frac{2u}{z},\frac{u+v}{z};
 1+\frac{u-v}{z};\widetilde q
 \right).
 \end{aligned}
 \label{eq:A1-quadratic-transformation}
\end{equation}
The second chart follows by exchanging $u$ and $v$.
This yields the flat-coordinate degree expansion
\eqref{eq:A1-flat-degree-series}.

To combine the two charts, we use the connection formula
\cite[Section~15.10]{Olver:2010ouy},
\begin{align}
 {}_2F_1(\alpha,\beta;\gamma;x)
 ={}&
 \frac{\Gamma(\gamma)\Gamma(\gamma-\alpha-\beta)}
 {\Gamma(\gamma-\alpha)\Gamma(\gamma-\beta)}
 {}_2F_1\!\left(
 \alpha,\beta;\alpha+\beta-\gamma+1;1-x
 \right)
 \nonumber\\
 &+
 (1-x)^{\gamma-\alpha-\beta}
 \frac{\Gamma(\gamma)\Gamma(\alpha+\beta-\gamma)}
 {\Gamma(\alpha)\Gamma(\beta)}
 \nonumber\\
 &\hspace{12mm}\times
 {}_2F_1\!\left(
 \gamma-\alpha,\gamma-\beta;
 \gamma-\alpha-\beta+1;1-x
 \right).
 \label{eq:hypergeometric-connection}
\end{align}
Taking $\alpha=2u/z$, $\beta=(u+v)/z$,
$\gamma=2(u+v)/z$ and $x=1-\widetilde q$ relates
the combined expression
\eqref{eq:A1-combined-hypergeometric} to the chart sum
\eqref{eq:A1-flat-degree-series}.
After multiplication by the prefactor in the combined
expression, the two terms on the right reproduce the
two chart contributions, including their Gamma factors.

We initially use these hypergeometric identities for
$0<x<1$ and generic parameters, then continue with
consistent branches.
Exceptional parameter values are obtained by taking
limits of the complete expressions.
\section{Identities for the disk calculations}
\label{app:examples}

This appendix supplies the integral representations,
volume limit and Mellin residues used in
Section~\ref{sec:diskexamples}.
We build on the disk calculations of
\cite{Cassia:2022lfj}.

\subsection{Positive integral representations of the disk period}
\label{app:QJK}
\label{app:radial-proof}

We start from the disk partition function
\eqref{eq:disk} for a toric Calabi--Yau quotient
$X=\C^n//U(1)^r$ of complex dimension $d=n-r$,
with $\sum_iQ_i^a=0$:
\begin{equation}
 \mathbb D_X(\lambda,\eps;z)
 =z^{-n}\oint_{\mathrm{QJK}}
 \prod_{a=1}^r\frac{\dd\phi_a}{2\pi i}\,
 e^{\sum_i\lambda^i x_i}
 \prod_{i=1}^n\Gamma(x_i/z),
 \qquad x_i=\eps_i+\sum_aQ_i^a\phi_a.
 \label{eq:app-disk-definition}
\end{equation}
Using the Euler integral for each Gamma function
and $t^a=\sum_iQ_i^a\lambda^i$ gives
\begin{align}
 \mathbb D_X(\lambda,\eps;z)
 ={}&z^{-n}e^{\lambda\cdot\eps}
 \int_{\sigma+i\mathbb R^r}
 \prod_a\frac{\dd\phi_a}{2\pi i}\,
 e^{\sum_a t^a\phi_a}
 \nonumber\\
 &\times\int_{(0,\infty)^n}
 \left(\prod_i\frac{\dd y_i}{y_i}\,
 y_i^{\eps_i/z}\right)e^{-\sum_i y_i}
 \exp\!\left(
 \frac1z\sum_{a,i}\phi_aQ_i^a\log y_i
 \right).
 \label{eq:app-mixed-integral}
\end{align}
Here the Coulomb contour has been deformed to
a vertical contour on which the Gamma arguments
have positive real parts
\cite[Eq.~(2.45) and Remark~2.5]{Cassia:2023uvd}.
We initially take $z>0$ and work in a convergence
domain where closing this contour recovers the
QJK pole towers, then continue the resulting
function.

Fourier inversion in the imaginary Coulomb
variables imposes the charge constraints.
Each integration supplies a factor of $z$, giving
\begin{align}
 \mathbb D_X(\lambda,\eps;z)
 =z^{-d}e^{\lambda\cdot\eps}
 \int_{(0,\infty)^n}
 \left(\prod_i\frac{\dd y_i}{y_i}\,
 y_i^{\eps_i/z}\right)e^{-\sum_i y_i} \prod_a\delta\!\left(
 zt^a+\sum_iQ_i^a\log y_i
 \right).
 \label{eq:app-disk-delta-bare}
\end{align}
Equivalently, the constraints are
\begin{equation}
 \prod_i y_i^{Q_i^a}=q_a,
 \qquad q_a=e^{-zt^a}.
 \label{eq:app-HV-constraints}
\end{equation}
Set $c_i=e^{-z\lambda^i}$ and replace each old
integration variable by $c_i y_i$.
This moves the K\"ahler dependence into the
exponential and cancels $e^{\lambda\cdot\eps}$:
\begin{equation}
 \mathbb D_X(\lambda,\eps;z)
 =z^{-d}\int_{(0,\infty)^n}
 \left(\prod_i\frac{\dd y_i}{y_i}\,
 y_i^{\eps_i/z}\right)
 e^{-\sum_i c_i y_i}
 \prod_a\delta\!\left(\sum_iQ_i^a\log y_i\right).
 \label{eq:disk-positive-integral}
\end{equation}

The constraints can be solved using toric
vectors $v_i=(1,m_i)$, with
$m_i\in\mathbb Z^{d-1}$ and
$\sum_iQ_i^av_i=0$:
\begin{equation}
 y_i=R\,U^{m_i},
 \qquad
 U^{m_i}=\prod_{\alpha=1}^{d-1}
 U_\alpha^{m_{i\alpha}},
 \qquad R,U_\alpha>0.
 \label{eq:app-toric-parametrization}
\end{equation}
The potential becomes
\begin{equation}
 W_c(R,U)=R\,P_c(U),
 \qquad
 P_c(U)=\sum_i c_iU^{m_i},
 \label{eq:app-HV-potential}
\end{equation}
where $P_c$ is the Laurent polynomial entering the
Hori--Vafa mirror construction.
In primitive torus coordinates, with the
GLSM measure normalization,\footnote{
For the effective continuous quotients used here,
the charge map is surjective on the integral
charge lattice. An integral basis of its kernel
then gives unit Jacobian for the constrained
logarithmic measure. Residual finite quotients
are included separately.}
the disk is
\begin{equation}
 \mathbb D_X(\lambda,\eps;z)
 =z^{-d}\int_{(0,\infty)^d}
 \frac{\dd R}{R}
 \prod_{\alpha=1}^{d-1}\frac{\dd U_\alpha}{U_\alpha}\,
 R^{\sum_i\eps_i/z}
 \prod_{\alpha=1}^{d-1}
 U_\alpha^{\sum_i m_{i\alpha}\eps_i/z}
 e^{-R P_c(U)}.
 \label{eq:app-equivariant-HV-period}
\end{equation}
Integrating the common scale $R$ gives
\begin{equation}
 \mathbb D_X(\lambda,\eps;z)
 =z^{-d}\Gamma\!\left(\frac{\sum_i\eps_i}{z}\right)
 \int_{(0,\infty)^{d-1}}
 \prod_\alpha\frac{\dd U_\alpha}{U_\alpha}\,
 \frac{\displaystyle\prod_\alpha
 U_\alpha^{\sum_i m_{i\alpha}\eps_i/z}}
 {P_c(U)^{\sum_i\eps_i/z}}.
 \label{eq:app-radial-period}
\end{equation}

\paragraph{The mesonic-volume limit.}
\label{app:classical-limit}

For the mesonic disk, $\lambda=s\lambda_\star$
and $z=s^{-1}$, so the coefficients
$c_i^\star=e^{-\lambda_\star^i}$ remain fixed.
Setting $y_i=e^{-u_i/s}$ in
\eqref{eq:disk-positive-integral} cancels the
prefactor $s^d$ against the measure and
delta-function factors:
\begin{equation}
 \mathbb D_X^{\text{mes}}(s,\eps)
 =
 \int_{\mathbb R^n}
 \left(\prod_i\dd u_i\right)
 e^{-\sum_i\eps_i u_i}
 \exp\!\left[-\sum_i c_i^\star e^{-u_i/s}\right]
 \prod_a\delta\!\left(\sum_iQ_i^a u_i\right).
 \label{eq:reduced-positive-cycle}
\end{equation}
As $s\to0^+$, the second exponential tends
to one where all $u_i>0$, and to zero wherever
some $u_i<0$.
For positive real weights and coefficients,
dominated convergence therefore gives
\begin{equation}
 \lim_{s\to0^+}\mathbb D_X^{\text{mes}}(s,\eps)
 =
 \int_{[0,\infty)^n}
 \left(\prod_i\dd u_i\right)e^{-\sum_i\eps_i u_i}
 \prod_a\delta\!\left(\sum_iQ_i^a u_i\right)
 =\mathbb V_X(0,\eps)=C_X(\eps).
 \label{eq:app-mesonic-volume-limit}
\end{equation}
The last integral is the moment-map representation
of the mesonic equivariant volume.
This establishes \eqref{eq:volume-CX} for the
complete disk: its degree variables remain fixed,
and positive-degree terms are not discarded
individually.

\paragraph{Products and finite quotients.}

Disks of product geometries factorize when their
charge constraints are independent.
For an invariant equivariant class on a finite
cover, the untwisted normalization is
\begin{equation}
 \int_{[\widetilde X/G]}\alpha
 =\frac1{|G|}\int_{\widetilde X}\alpha.
 \label{eq:finite-quotient-measure}
\end{equation}
This supplies the residual factor $1/k$ with
covering weights held fixed.
It multiplies both the disk and its volume limit
and is included only once.
The factors $1/p$ associated with resolved ALE
surfaces already follow from their continuous
GLSM integrals and must not be inserted again.

\subsection{Mellin expansion and residues}
\label{app:gamma-mellin}
\label{app:charges}

Throughout this subsection, $X$ is a fourfold
and $\sum_i\eps_i=2$.

\paragraph{Polynomial terms.}

Starting from the Gamma-quotient expression
\eqref{eq:general-geometric-period}, the Mellin
kernel without its chemical exponential is
\begin{equation}
 K_X(s;\eps)
 :=-\frac{C_X(\eps)}{4\pi^2s^4}
 \frac{\Gamma(-s)}{\Gamma(s)}
 \frac{\prod_{I=1}^{M}\Gamma(1+s\nu_I)}
 {\prod_{A=1}^{M-4}\Gamma(1+s\kappa_A)}.
\end{equation}
Its residue at the origin, after multiplication
by $e^{s\mu}$, determines the polynomial part
of $J_{\mathrm{disk}}$.

Using
$-\Gamma(-s)/\Gamma(s)=\Gamma(1-s)/\Gamma(1+s)$
and
\begin{equation}
 \log\Gamma(1+x)
 =-\gamma_{\mathrm E}x
  +\frac{\zeta(2)}2x^2
  -\frac{\zeta(3)}3x^3+O(x^4),
 \qquad |x|<1,
 \label{eq:gamma-log-expansion}
\end{equation}
the weight balance
$\sum_I\nu_I-\sum_A\kappa_A=2$
cancels the linear Euler-constant terms.
Hence
\begin{align}
 K_X(s;\eps)
 =\frac{C_X(\eps)}{4\pi^2s^4}\Bigg[
 1
 &+\frac{\zeta(2)}2
 \left(\sum_I\nu_I^2-\sum_A\kappa_A^2\right)s^2
 \nonumber\\
 &+\frac{\zeta(3)}3
 \left(2-\sum_I\nu_I^3+\sum_A\kappa_A^3\right)s^3
 +O(s^4)\Bigg].
 \label{eq:full-kernel-jet}
\end{align}
Multiplying by $e^{s\mu}$ and taking the residue
yields
\begin{align}
 \operatorname*{Res}_{s=0}
 \bigl[K_X(s;\eps)e^{s\mu}\bigr]
 ={}&\frac{C_X(\eps)}{24\pi^2}\mu^3
 +\frac{C_X(\eps)}{48}
 \left(\sum_I\nu_I^2-\sum_A\kappa_A^2\right)\mu
 \nonumber\\
 &+\frac{\zeta(3)C_X(\eps)}{12\pi^2}
 \left(2-\sum_I\nu_I^3+\sum_A\kappa_A^3\right).
 \label{eq:zero-residue}
\end{align}
This gives \eqref{eq:geometric-polynomial},
including the absence of a quadratic term in $\mu$.

\paragraph{Exponentially small terms.}

Using $\Gamma(1+x)=x\Gamma(x)$, the same kernel
can be written as
\begin{equation}
 K_X(s;\eps)
 =-\frac{C_X(\eps)}{4\pi^2}
 \frac{\prod_I\nu_I}{\prod_A\kappa_A}
 \frac{\Gamma(-s)}{\Gamma(s)}
 \frac{\prod_I\Gamma(s\nu_I)}
 {\prod_A\Gamma(s\kappa_A)}.
\end{equation}
For positive $\nu_I$, its candidate negative poles
occur at $s=-n/\nu_I$, with $n\geq1$.
The identity
\begin{equation}
 \operatorname*{Res}_{x=-n}\Gamma(x)
 =\frac{(-1)^n}{n!},
 \qquad n=0,1,\ldots,
 \label{eq:gamma-poles-and-shifts}
\end{equation}
gives
$\operatorname*{Res}_{s=-n/\nu_I}\Gamma(s\nu_I)
=(-1)^n/(n!\nu_I)$.
For a surviving simple pole, evaluating the
remaining factors there gives
$c^X_{I,n}(\eps)$ in
\eqref{eq:general-geometric-residues},
while $e^{s\mu}$ supplies $e^{-n\mu/\nu_I}$.

At a surviving pole $s=s_\star$ of order $m$,
\begin{equation}
 \operatorname*{Res}_{s=s_\star}
 \bigl[K_X(s;\eps)e^{s\mu}\bigr]
 =
 \left.
 \frac{1}{(m-1)!}
 \frac{\dd^{m-1}}{\dd s^{m-1}}
 \left[(s-s_\star)^mK_X(s;\eps)e^{s\mu}\right]
 \right|_{s=s_\star}.
 \label{eq:collisionrule}
\end{equation}
The result is $e^{s_\star\mu}$ times a polynomial
in $\mu$ of degree at most $m-1$.
Reciprocal Gamma factors can remove poles or
lower their order.
When families collide, one must therefore take
the combined residue of the complete kernel,
rather than specialize their potentially
divergent coefficients separately.
This rule also applies to the Barnes case below.

\paragraph{Polynomial terms from the Barnes representation.}
\label{app:Qgeneraljets}
\label{app:Q-periods}

For the resolved cone over $Q^{1,1,1}$, the full
mesonic disk remains a Barnes integral.
Its polynomial contribution to $J_{\mathrm{disk}}$
can nevertheless be obtained from the degree-zero
chart sum.

For every nonzero effective degree $(d_1,d_2)$,
the divisors with negative degree pairing are
$D_5$ and $D_6$, whose intersection is the compact
zero section $\PP^1\times\PP^1$.
By \cite[Proposition~7.8 and Eq.~(11.32)]{Cassia:2022lfj},
the corresponding positive-degree contributions
are regular in the nonequivariant limit.
In the locally convergent large-radius expansion,
their sum is likewise regular near the nonequivariant limit.
The singular Laurent coefficients under
$\eps\mapsto s\eps$ therefore come from the
degree-zero chart sum, including its FI
exponentials.

We continue these coefficients to the mesonic
point using the complete disk.
Along positive coefficient paths, the polynomial
and its face polynomials have no zeros on the
positive integration domain, so Euler--Mellin
continuation applies
\cite[Section~2]{Berkesch_2014}.\footnote{
On a compact positive coefficient path, the
integration-by-parts continuation has locally
uniform estimates and possible polar hyperplanes
fixed by the monomial support.
The Laurent coefficients can therefore be
continued along the same path, without continuing
the degree series term by term.}
At the unit-coefficient mesonic representative
$\lambda_\star=0$, the mirror map gives $t=T=0$
and all equivariant and FI exponentials equal one.

The four charts have tangent weights
\begin{equation}
 w_{ij}
 =
 (\eps_{i'}-\eps_i,\eps_{j'}-\eps_j,
  \eps_{ij5},\eps_{ij6}),
 \qquad i=1,2,\quad j=3,4,
\end{equation}
where $i'$ and $j'$ are the other members of
$\{1,2\}$ and $\{3,4\}$.
Consequently,
\begin{equation}
 \mathbb D_{C(Q^{1,1,1})}^{\text{mes}}(s,\eps)
 =
 s^4\sum_{i=1,2}\sum_{j=3,4}
 \prod_{a=1}^4\Gamma(sw_{ij,a})
 +O(s^4).
 \label{eq:Q-singular-part}
\end{equation}
The chart contributions are combined before
taking coincident-weight limits.

Each chart satisfies $\sum_a w_{ij,a}=2$.
Expanding its Gamma product and the Mellin
factor through cubic order therefore gives
\eqref{eq:Q-geom-main-jets}.
The remainder in \eqref{eq:Q-singular-part}
is regular after multiplication by
$-\Gamma(-s)/(4\pi^2s^4\Gamma(s))$
and contributes no residue at the origin.
It can contribute at negative poles, however,
so the exponentially small terms require the
complete Barnes representation.

\paragraph{Exponentially small terms from the Barnes representation.}
\label{app:Qpinch}

Poles of a Barnes integral arise when poles
in its Coulomb variables approach the contours
from opposite sides and prevent their further
deformation.
The relevant local contour pinch has the form
\begin{equation}
 \mathcal I(s)
 =
 \int\frac{\dd\phi\,\dd\psi}{(2\pi i)^2}\,
 \Gamma(s\alpha_1+\phi)
 \Gamma(s\alpha_2+\psi)
 \Gamma(s\alpha_3-\phi-\psi)
 H(s;\phi,\psi).
 \label{eq:Barnes-pinch-integral}
\end{equation}
Initially, the upward-oriented vertical contours
make the displayed Gamma arguments have positive
real parts.
We continue in $s$ while keeping the pole families
separated until a pinch occurs.
Assume that $H$ is regular at the relevant pinch
points and that no additional families collide
there.

Set $\nu=\alpha_1+\alpha_2+\alpha_3>0$.
The pole conditions are
\begin{equation}
 s\alpha_1+\phi=-u,\qquad
 s\alpha_2+\psi=-v,\qquad
 s\alpha_3-\phi-\psi=-w,
 \qquad u,v,w\in\mathbb Z_{\geq0}.
\end{equation}
Their sum gives $s\nu=-(u+v+w)$.
Thus the negative candidate poles occur at
$s_n=-n/\nu$, with $n\geq1$, and receive
contributions from the triples $u+v+w=n$.

Taking residues in $\phi$ and $\psi$ leaves
\begin{equation}
 \frac{(-1)^{u+v}}{u!\,v!}\,
 \Gamma(s\nu+u+v)\,
 H(s;-s\alpha_1-u,-s\alpha_2-v).
\end{equation}
The remaining Gamma pole then gives
\begin{equation}
 \operatorname*{Res}_{s=s_n}\mathcal I(s)
 =
 \frac{(-1)^n}{\nu}
 \sum_{u=0}^{n}\sum_{v=0}^{n-u}
 \frac{
 H(s_n;-s_n\alpha_1-u,-s_n\alpha_2-v)}
 {u!\,v!\,(n-u-v)!}.
 \label{eq:Barnes-pinch-residue}
\end{equation}

For the $Q^{1,1,1}$ disk, choose one Gamma
factor from each of the three charge pairs.
Then
\begin{equation}
 \nu=\nu_{ijk}=\eps_{ijk},
 \qquad i\in\{1,2\},\quad j\in\{3,4\},\quad
 k\in\{5,6\},
\end{equation}
and the remaining three factors form $H$.
These choices give eight candidate families
$e^{-n\mu/\nu_{ijk}}$.
The factor $s^4$ in the mesonic disk cancels
$s^{-4}$ in the Mellin kernel, so multiplying
\eqref{eq:Barnes-pinch-residue} by
$-\Gamma(-s_n)/(4\pi^2\Gamma(s_n))$
gives \eqref{eq:Q-general-residues}
for generic weights.

On the QC locus, $\nu_{136}=\nu_{245}=1$.
The corresponding pinches occur at $s=-n$
but, for generic remaining weights, at distinct
Coulomb points.
Each produces at most a simple pole, and their
sum still has at most a simple pole.
It is canceled by the reciprocal Gamma factor:
\begin{equation}
 \frac1{\Gamma(s)}
 =(-1)^n n!(s+n)+O((s+n)^2),
 \qquad
 \frac{\mathbb D_{C(Q^{1,1,1})}^{\text{mes}}(s,\eps)}
 {\Gamma(s)}
 =O(1),
 \qquad s\to-n,\quad n\geq1.
 \label{eq:Q-unit-pole-cancellation}
\end{equation}
Since $s^{-4}\Gamma(-s)$ is regular there,
these two families give no Mellin residue.
The remaining six are those listed in
\eqref{eq:Q-instanton-poles}.

At special weights, additional families can
meet negative integers or collide with one
another, requiring the combined residue
\eqref{eq:collisionrule}.
The cancellation above removes the two generic
unit-weight families; it does not exclude
integer-action terms at resonant weights.
\clearpage
\bibliographystyle{JHEP}
\bibliography{references}
\end{document}